\documentclass[prb, aps,
twocolumn,
showpacs,amsmath,amssymb
,floatfix
]{revtex4}
\usepackage{epsf}
\usepackage{graphicx}
\usepackage{bm}
\usepackage{relsize}
\usepackage[normalem]{ulem}
\usepackage[usenames]{color}
\usepackage{float}

\def \be {\begin{equation}}
\def \ee {\end{equation}}
\def \bea {\begin{align}}
\def \eea {\end{align}}

\def\bee{\begin{eqnarray}}
\def\eee{\end{eqnarray}}
\def \BC {\begin{cases}}
\def \EC {\end{cases}}

\begin{document}

\title{Interference-induced magnetoresistance in HgTe quantum wells}

\author{I. V. Gornyi$^{1,2}$}
\author{V. Yu. Kachorovskii$^{1,2}$}
\author{P. M.  Ostrovsky$^{3,4}$}
\affiliation{$^{1}$ Institut f\"ur Nanotechnologie,  Karlsruhe Institute of Technology,
76021 Karlsruhe, Germany \\ $^{2}$ A .F. Ioffe Physico-Technical Institute,
194021 St.~Petersburg, Russia
\\
$^3$ Max-Planck-Institute for Solid State Research, D-70569 Stuttgart, Germany\\
$^{4}$ L.~D.~Landau Institute for Theoretical Physics RAS,
 119334 Moscow, Russia
}

\begin{abstract}
  We study the quantum interference correction to the conductivity in
  HgTe quantum wells using the Bernevig-Hughes-Zhang
  model. This model consists of two independent species (blocks) of
  massive Dirac fermions.  We describe the crossover between the
  orthogonal and symplectic classes with  increasing the carrier
  concentration and calculate, respectively, weak localization and
  antilocalization corrections in the absence of the block mixing and assuming the
  white-noise disorder within each block.  We have calculated the
  interference-induced magnetoresistance in a wide interval of
  magnetic fields, in particular, beyond the diffusion
  regime. Remarkably, each Dirac cone taken separately gives a linear contribution
  to the low-field magnetoresistance, which turns out to be asymmetric in magnetic field $B$. We present an interpretation of this result in terms of the Berry phase formalism.
  The contributions of the two blocks are related to each other by replacing $B$ to $-B$,
  so that the total magnetoresistance is symmetric and parabolic in the limit $B\to 0$. However, in some range of parameters field dependence
turns out to be strongly non-monotonous.
    We also demonstrate that block mixing gives rise to additional singular diffusive modes
  which do not show up in the   absence of mixing.
 \end{abstract}
\maketitle
\thispagestyle{plain}

\section{Introduction}
\label{s1}

It is well known that the low-temperature transport in disordered systems is  crucially affected by  quantum interference  effects.
The first-order term in a series expansion of the conductivity in $1/k_F l$ (here $k_F$ is the Fermi wave vector and $l $ is
the mean free path) contains the weak localization (WL) correction \cite{weak1} (for a review, see Ref.~\onlinecite{weak2}).
The underlying physics is the coherent enhancement of the backscattering amplitude which comes from
the constructive interference of the waves propagating in the opposite directions (clockwise and counterclockwise) along
a closed loop formed by scatterers (see Figs.~1 and 2). A remarkable feature of this
correction is the logarithmic divergence at low temperatures in the two-dimensional (2D) case.
Such a divergency is a precursor of the strong localization effects and, therefore, reflects universal symmetry properties
of the system.  An external  magnetic field breaks the time-reversal symmetry and, as a consequence,
leads to a suppression of the WL correction.
On the other hand, spin-orbit coupling  does not violate the time-reversal symmetry, but strongly modifies the quantum
conductivity correction because of the interference of the spin parts of the waves. In particular, strong spin-orbit
coupling leads to a destructive interference\cite{TI12} between clockwise- and counterclockwise-propagating paths,
thus changing the sign of the quantum correction.
The change from the WL to the weak antilocalization (WAL) behavior was considered for the
first time in Ref.~\onlinecite{TI12} and was further addressed, both theoretically and experimentally, in
a number of more recent papers
(see  Refs.~ \onlinecite{TI62, TI62a,TI63, TI64, TI65, TI66, TI67, TI67aa, TI67a, TI67b,TI68, TI68a, TI69, TI69a, TI70,
TI70a, TI70aa, TI70aaa, TI70aaaa, TI70aaaaa, TI70ab, TI70b, TI70c,TI70d,TI70e,TI70g,TI70i,TI71} and references therein).

Recently, the interest in quantum transport in the systems with strong spin-orbit coupling
dramatically increased after the discovery that such systems may exhibit a topological insulator (TI)
phase \cite{Hasan10RMP,Qi11,kane,BernevigHughesZhang,Koenig07,Fu07,hasan}
with preserved time-reversal invariance.
In the 2D case, the TI behavior was predicted by Bernevig, Hughes, and
Zhang (BHZ) \cite{BernevigHughesZhang} for HgTe/HgCdTe quantum wells (QWs). Soon after this prediction,
the existence of a TI phase was experimentally demonstrated in Refs.~\onlinecite{Koenig07,Roth09} in
HgTe/HgCdTe QWs with band inversion.
The latter resulted in emergence of delocalized (topologically protected) helical modes at the edge of the sample.
Another realization of a 2D  TI based on InAs/GaSb structures was proposed in Ref.~\onlinecite{Liu08}
and was experimentally discovered in Ref. \onlinecite{Du}.

When the Fermi energy in a HgTe/HgCdTe QW is shifted away from the band gap,
the system exhibits a 2D metallic phase known as  a 2D spin-orbit
metal (SOM).  The spectrum of the SOM can be well approximated by the Dirac spectrum of massive fermions
(with the mass proportional to the band gap), so that the quantum transport in such a system is strongly affected by the Dirac
nature of carriers \cite{ostrovsky10,Tkachov11,ostrWAL,myPSS,Richter12,Tkachov13} and, owing to the presence of the Berry phase,
bears certain similarity to interference phenomena in graphene,\cite{McCann,Nestoklon,Ostrovsky06,aleiner-efetov}
in 2D semiconducting hole structures, \cite{Richter11,TI71} and in surface layers of 3D TI. \cite{ostrovsky10,Glazman,Koenig13}

The remarkable property of the  2D SOM  is the dependence of the effective spin-orbit coupling
on the particle concentration.  For low concentration, the spectrum is approximately parabolic,
while the coupling is weak and can be neglected in the first approximation. Then, the system
belongs to so-called orthogonal symmetry class and one can expect conventional WL behavior of
the quantum correction. On the other hand, at large  concentrations, when the  Fermi energy
becomes much larger than the gap width, the spectrum is quasi-linear, while the spin is strongly coupled
to the particle momentum, so that the underlying symmetry class is symplectic and the SOM would
demonstrate  WAL similar to graphene-based structures.
Recently, the magnetoresistivity of HgTe/HgCdTe structures was experimentally
studied away from the insulating regime in Refs. \onlinecite{kvon,minkov,minkov13,bruene-unpub},
both for inverted and normal band ordering.
The experiments demonstrate that the system may show both  WL and WAL behavior.
Hence, there is a clear need in a theory which can rigorously describe the crossover
between the two symmetry classes that correspond to WL and WAL regimes.

A theoretical study of the  quantum transport in the 2D SOM in HgTe/HgCdTe QWs was undertaken in
Ref. \onlinecite{Tkachov11} for the case when the chemical potential is located in the almost linear
range of the spectrum.  The conductivity correction was calculated within the diffusive approximation.
The WAL nature of the correction was demonstrated and attributed to the Berry phase mechanism
characteristic of the Dirac systems.
It was also shown  that the weak nonlinearity of the dispersion (due to a finite bandgap) suppresses
the quantum interference on large scales. Another approach to the problem  based on the analysis
of the symmetry properties of the underlying Dirac-type Hamiltonian
and physically important symmetry-breaking mechanisms was proposed in Ref.~\onlinecite{ostrWAL}.
This approach captures the universal properties of the system  and, consequently,
allows one to find without microscopic calculations the singular (logarithmic in 2D) interference corrections
within the diffusion approximation.
This approximation is sufficient near the band bottom and in the regime of almost linear spectrum
for relatively weak magnetic fields.  The ballistic case  was discussed in  Ref.~\onlinecite{myPSS} for zero  magnetic field. Also, the quantum transport in HgTe/HgCdTe QWs was numerically
simulated both in the ballistic and diffusive samples in Ref. \onlinecite{Richter12} where the influence of the
Berry phase (similar to the effect of Berry phase in semiconducting hole structures\cite{Richter11}) on magnetoresistance was emphasized.

In this paper we present  a systematic analysis of the interference
corrections in a  2D SOM biased by gate away from the TI
regime. Having in mind  to describe the crossover between two symmetry classes we  generalize the approach of Ref.~\onlinecite{myPSS} for nonzero magnetic fields and develop a ballistic theory of the quantum transport
valid beyond the diffusion approximation and, therefore, applicable for  a wide range of particle concentrations.

 We start with discussing the basic equations and properties of
such a SOM (see Sec.~\ref{basic}).  As a starting model, we will use
BHZ Hamiltonian,  \cite{BernevigHughesZhang} which allows one to describe TI in a wide range of particle concentration both in  the normal and in the inverted states.  We  discuss the  dependence of the conductivity correction on key parameters of the model. Although our theory is valid for a  wide  range of parameters, in order to simplify intermediate calculations we will use the simplified version of this model which corresponds to two equivalent massive Dirac cones with momentum-independent mass and velocity.

  Next, we derive a kinetic equation
for the Cooperon propagator within a single cone and find its exact solution valid beyond
the diffusion approximation (see Sec.~\ref{Coo}). In Sec.~\ref{inter}
we use these results to evaluate the interference corrections in a
wide interval of electron concentration and magnetic fields, assuming
that two equivalent blocks of the system (BHZ
blocks) are not mixed by any perturbation.  One of the main results of this section is
the demonstration of the strong asymmetry (with respect to the inversion of the sign of the magnetic field)
of the contribution of a single block.  This result can be explained within the Berry-phase formalism and turns out to be in a good agreement with the previous
numerical simulations. \cite{Richter12}  Summing of the contributions of two blocks restores symmetry with respect to the field inversion.
However, field dependence turns out to be strongly non-monotonous in  some range of the parameters.
We also predict a non-monotonous dependence of the quantum correction on the phase-breaking rate. Similar
result was predicted earlier for weak localization of holes in conventional heterostructures.\cite{TI71}

In Sec.~\ref{mix}, we
generalize obtained results assuming that blocks are weakly coupled to
each other.  Most interestingly, the block mixing gives rise to additional singular diffusive modes
which do not show up in the   absence of mixing. One of these modes is a
``purely singular'' mode which is responsible for the WAL at very low temperatures.
Existing of such singular modes leads to an additional mechanism of the non-monotonous magnetoresistance.

In the end of the paper, we present plots demonstrating  dependence of  the interference correction on different parameters
(see Sec.~\ref{num}). We also discuss  asymptotical behavior of the quantum correction in the strong-field
limit (see Sec.~\ref{large}).  In this limit, the main contribution to the correction is given by scattering on
complexes of three impurities separated by untypical  distances much smaller than the mean free path
(for discussion of such an asymptotic in the parabolic spectrum see Refs.~\onlinecite{TI72, nonback}).
Technical details of the calculations are placed
in Appendices \ref{mix1}, \ref{diffusion1}, \ref{kinBblock}, \ref{IM},
and \ref{strongA}.

\section{Basic equations}\label{basic}
\subsection{BHZ Hamiltonian}\label{BHZ}

Effective Hamiltonian for a narrow symmetric HgTe quantum well (QW)
was derived  by  Bernevig, Hughes, and Zhang (BHZ) in
Ref.\ \onlinecite{BernevigHughesZhang} in the framework of the
$\mathbf{k}\cdot\mathbf{p}$ method. The BHZ
Hamiltonian has a $4 \times 4$ matrix
structure in the spin (sign of the $z$-projection of the total momentum $\textbf{J}$,
where the $z$-axis is perpendicular to the QW plane)
and E1 -- H1 ($|J_z|=1/2$ and $|J_z|=3/2$ bands, respectively) space,
\be
 H_\text{BHZ}\label{1}
  =
  \left[\begin{array}{cc}
         H_{\rm I}(\mathbf{k}) & 0 \\
      0 & H_{{\rm II}} (\mathbf{k})
     \end{array}\right],
\ee
where
\be
H_{{\rm I}} (\mathbf{k})
  =
  \left[\begin{array}{cc}
       \epsilon(\mathbf{k}) + m(\mathbf{k}) & v (k_x + i k_y) \\
      v(k_x - i k_y) & \epsilon(\mathbf{k}) - m(\mathbf{k})
        \end{array}\right],
   \ee
and
\be
H_{{\rm II}}(\mathbf{k})=H^*_{{\rm I}}(-\mathbf{k})=\left[
                    \begin{array}{cc}
                      \epsilon(\mathbf{k}) + m(\mathbf{k}) & -v (k_x - i k_y)  \\
      -v (k_x + i k_y) & \epsilon(\mathbf{k}) - m(\mathbf{k}) \\
                    \end{array}
                  \right].
\ee
Here we have used the form given in Refs. \onlinecite{Qi11, Liu08} with the following
arrangement of components in the spinor: $E1+,H1+,E1-,H1-$.

The functions $\epsilon(\mathbf{k})$ and $m(\mathbf{k})$ are effective energy
and mass which are assumed to be isotropic in the momentum space. Within the $\mathbf{k}\cdot\mathbf{p}$ expansion in the vicinity of
the $\Gamma$ point, they are given by
\begin{equation}
 \epsilon(\mathbf{k})
  = C + D \mathbf{k}^2,
 \qquad
 m(\mathbf{k})
  = m v^2 + B \mathbf{k}^2.
  \label{3}
\end{equation}
We note that, in general, $v$ might also depend on $\mathbf{k}.$\cite{rothe}

The two phases of normal and topological insulator correspond to $m>0$ and
$m<0$, respectively. The sign of $m$ changes at the critical thickness $d_c$ of the
QW of about $6.2$ nm. \cite{Koenig07} The quantities $B$ and $D$ are positive with
$B > D$. The parameter $C$ can be eliminated by a shift of the  chemical potential.

As seen from Eq.~\eqref{1}, the Hamiltonian $H_\text{BHZ}$ breaks up into two blocks
acting independently in
the spin-up and spin-down subspaces. The blocks have  the same spectrum
\begin{equation}
 E^\pm_\mathbf{k}
  = \epsilon(\mathbf{k}) \pm \sqrt{v^2 k^2 + m^2(\mathbf{k})}.
  \label{4}
\end{equation}
The eigenfunctions for each block are two-component spinors in E1-H1 space:
\be
\psi_\mathbf{k}^{(\pm)}(\mathbf{r})=\chi_\mathbf{k}^{(\pm)}\, e^{i\mathbf{k r}}.
\ee
The spinors $\chi_\mathbf{k}^{(\pm)}$ are different in different blocks
\bee \chi_\mathbf{k}^{(\text{I},+)}&=&
\frac{1}{\sqrt{1+ \eta}}\
\begin{pmatrix}
1\\
{\sqrt{\eta}e^{-i \phi_\mathbf{k}}} \\0\\0
\end{pmatrix},\\
\chi_\mathbf{k}^{(\text{I},-)}&=&
\frac{1}{\sqrt{1+ \eta}}\
\begin{pmatrix}
{-\sqrt{\eta}e^{i \phi_\mathbf{k}}}\\
1 \\0\\0
\end{pmatrix},\\
 \chi_\mathbf{k}^{(\text{II},+)}&=&
\frac{1}{\sqrt{1+ \eta}}\
\begin{pmatrix}
0\\
0 \\1\\{-\sqrt{\eta}e^{i \phi_\mathbf{k}}}
\end{pmatrix},\\
 \chi_\mathbf{k}^{(\text{II},-)}&=&
\frac{1}{\sqrt{1+ \eta}}\
\begin{pmatrix}
0\\
 0\\{\sqrt{\eta}e^{-i \phi_\mathbf{k}}}\\1
\end{pmatrix},
\eee
with
$\phi_\mathbf{\bf k}$ being the polar angle of the momentum $\mathbf{k}$ and
\begin{equation}
\eta=\eta(k)=\left[\frac{v k}{m(\mathbf{k})+ \sqrt{v^2k^2+m^2(\mathbf{k})}}\right]^2.
\label{eta0}
\end{equation}
Parameter $\eta$ is the key  quantity  of the problem.  At  temperatures smaller than the Fermi energy one may replace $k$
    with $k_F,$ so that we put throughout the paper
    \be \eta =\eta(k_F). \label{etaF}\ee
    %
    This parameter is the function of the particle concentration.
As we will see  below,  $\eta$
governs the crossover from WL
to WAL.
In particular,    in the absence  of dephasing processes and for zero magnetic field, the conductivity correction  depends  on  $\eta$ only:
$\delta\sigma=\delta \sigma (\eta),$ so that all essential information  about the particular dependencies  $m(\mathbf k)$ and $v(\mathbf k)$ is encoded in
$\eta,$  while the specific dependence $\epsilon(\mathbf k) $ drops out from the resulting equations.\cite{comment1}   It is worth stressing that $\eta$  is the
same for two  spectrum branches [see Eq.~\eqref{4}]. This implies that the electron and hole conductivity corrections coincide (for a given value of  $k_F$).
Moreover, as we will see below, the dependence of the correction on $\eta$ obeys an essential property  $\delta \sigma (\eta)=\delta \sigma (1/\eta).$ As follows from Eq.~\eqref{eta0} this property ensures that inversion of sign of the mass, $m(k_F) \to -m(k_F),$ does not change the correction.  In other words,
conductivity corrections  for normal material with the mass $m(k_F)>0$ and inverted material with the mass $-m(k_F)$  are equal. This allows us to limit
our consideration to the normal case and upper branch of the spectrum. In this case,
$$
0\leq \eta \leq 1.
$$
We will see that for $\eta \to 0$ the quantum correction is negative  (weak localization), while   for $\eta \to 1,$ the sign of the quantum correction changes
(weak antilocalization). Hence, with changing the electron (hole) concentration the system undergoes a transition between orthogonal and symplectic classes.

In a more general case, when magnetic field is applied  perpendicular to the quantum well plane and  dephasing processes  are taken into account, the
conductivity correction is expressed in terms of three parameters (here we neglect block mixing):
\be
\delta \sigma=\delta\sigma \left(\eta, \frac{l}{l_B}, \frac{\tau}{\tau_\phi}\right),
\label{dd}
\ee
where $\tau$ and $ l=v_F \tau$     are the mean free time and mean free path at the Fermi level, respectively,   $v_F = |(\partial E_\mathbf k/\partial
k)_{k=k_F}|$ is the Fermi velocity, $l_B$ is the magnetic length, and $\tau_\phi$ is the phase-breaking time. This equation is valid for both spectrum branches
[which, in general, have slightly different $l$ and $\tau/\tau_\phi$ for fixed $k_F$ due to the term $\epsilon(\mathbf k)$] and is invariant under replacement:
$m(k_F)\to -m(k_F)$, $\eta \to 1/\eta$ (one can check  that  $\tau$, $l$ and $v_F$ are invariant under this replacement).
Therefore,  the symmetry with respect to mass inversion, $m(k_F) \to -m(k_F),$   still holds.

Although our results are valid for general Hamiltonian  Eq.~\eqref{1} with arbitrary parameters $B$,$C$, and $D$, in what follows, in order to simplify
intermediate calculations, we will consider the simplified model with $B=C=D=0.$

\subsection{Single massive Dirac cone}
  Having in mind symmetry arguments presented in the previous section, we only consider normal case with the positive mass.
  We  assume that the Fermi level lies outside the gap in the region of the positive energies, thus focusing on
   the upper branches of the spectrum in both blocks.
    We start with the discussion of the properties of a single block (for definiteness, the block $\text{II}$). The contribution of the other block and the
effects caused by the inter-block transitions will be considered in  Secs.~\ref{another} and ~\ref{mix}, respectively.

  In the framework of the simplified model with $B=C=D=0$,  the expressions for the spectrum and the wave functions  simplify:
\bee E_\mathbf k &=&\sqrt{m^2v^4+v^2k^2}, \\ \chi_\mathbf{k}&=&
\frac{1}{\sqrt{1+\eta}}\
\begin{pmatrix}0\\0\\
1\\
-\sqrt \eta e^{i \phi_\mathbf{k}}
\end{pmatrix}, \label{chichi}\\ \eta&=&\left(\frac{ k}{ m v+ \sqrt{m^2v^2+ k^2}}\right)^2\label{chi1}.
\eee
Within this approximation, $\eta \to 0$ for
low concentration in the region of the parabolic spectrum and $\eta \to 1$ for large concentrations when the Fermi energy is
large compared to the band gap and the spectrum is quasi-linear.
The Fermi velocity, density of states (per block), Fermi energy and  Fermi momentum
are  expressed explicitly in terms of $\eta$ as follows:
\begin{eqnarray}
v_F&=&\frac{2\sqrt{\eta}}{1+\eta} ~v, \quad \nu_F=\frac{m}{2\pi\hbar^2}\frac{1+\eta}{1-\eta},
\label{nuF}
\\
E_F&=&mv^2\frac{1+\eta}{1-\eta}, \quad
 k_F=\frac{2mv}{\hbar} \frac{\sqrt{\eta}}{1-\eta}.
  \label{kF}
 \end{eqnarray}
 Importantly, the particular  dependence of these  quantities on $\eta$  enters only the last two arguments in Eq.~\eqref{dd}, $l/l_B$  and $\tau/\tau_\phi.$  At the same time, the explicit dependence of $\delta \sigma$ on $\eta$   arises solely  from the spinor structure of the wave functions and hence is not specific for the model with $B=C=D=0.$

The standard way \cite{Tkachov11} to introduce disorder in the model is to add the fully
diagonal term
\begin{equation}
 H_\text{dis}
  = V(\mathbf{r}) \hat I,
  \label{5}
\end{equation}
with a random potential $V(\mathbf{r})$ to the effective Hamiltonian $H_\text{BHZ}$ (here $\hat I$ is unit $4\times 4$ matrix).
While the diffusive behavior of the quantum interference
correction is universal, the precise from of the correction in the ballistic regime depends
on the particular form of the disorder correlation function.
We will assume the white-noise disorder with the correlation function
\begin{equation}
\langle V(\mathbf{r}) V(\mathbf{r}') \rangle = W \delta(\mathbf{r}-\mathbf{r}').
\label{corrW}
\end{equation}

As usual for the weak localization (antilocalization) regime, we assume
that $k_F l \gg 1$ and   $E_F \tau \gg 1 $
that allows us to neglect mixing of the upper and lower branches.
Then, the  matrix Green's function  can be written as:
\begin{equation}
\hat{G}(E,\mathbf{k})\simeq \hat{P}(\mathbf{k})G(E,\mathbf{k}),
\label{G}
\end{equation}
where
\begin{equation} \label{grin}
{G}(E,\mathbf{k})=
\frac{1}{E-E_\mathbf{k}-\Sigma},
\end{equation}
$ \Sigma$ is the disorder-induced   self-energy
and $\hat{P}$ is the upper-band  projector  given by
\begin{equation}\label{Pkk}
\hat{P}(\mathbf{k})=|\chi_\mathbf{k}\rangle\,\langle\chi_\mathbf{k}|.
\end{equation}

The spinors $|\chi_\mathbf{k}\rangle$  ``dress'' the matrix element of disorder by an
angular-dependent factor $\langle\chi_{\mathbf{k}}|\chi_\mathbf{k}'\rangle$:\cite{myPSS}
$$V_{\mathbf{k},\mathbf{k}'} \to \tilde V_{\mathbf{k},\mathbf{k}'} = V_{\mathbf{k},\mathbf{k}'}\langle\chi_{\mathbf{k}}|\chi_\mathbf{k}'\rangle.$$
 Using Eq.~\eqref{chichi} we find
\begin{equation}
\tilde{V}_{\mathbf{k} \mathbf{k}'}={V}_{\mathbf{k} \mathbf{k}'}\langle\chi_{\mathbf{k}}|\chi_\mathbf{k}'\rangle
={V}_{\mathbf{k} \mathbf{k}'}\frac{ 1+\eta\,\exp[i(\phi_{\mathbf{k}'}-\phi_\mathbf{k})]}{ 1+\eta}.
\label{tildeV}
\end{equation}
In other words, one can use a single-band approximation  with the particles described by the  Green's functions \eqref{grin}
with  self-energy $\Sigma$ determined by the spinor-dressed disorder, Eq.~(\ref{tildeV}).

The  quantum scattering rate  $\gamma=1/\tau$ (the imaginary
part of the self-energy) is related to the disorder correlation function (\ref{corrW}) as follows:
\be
\gamma = \int_0^{2\pi} \frac{d\phi}{2\pi}~\gamma_D(\phi) =\gamma_W\frac{1+\eta^2}{(1+\eta)^2}, \label{gammaq}
\ee
where
\bee
&&\gamma_D(\phi_\mathbf k-\phi_{\mathbf k'})=\frac{2\pi}{\hbar} \int \langle |\tilde{V}_{\mathbf{k} \mathbf{k}'}|^2 \rangle
 \delta(E_\mathbf k-E_{\mathbf k'})\frac{k'd k'}{2\pi}
 \nonumber
\\&&=\gamma_W
 \left|\frac{1+\eta e^{-i(\phi_\mathbf k-\phi_{\mathbf k'})}}{1+\eta}\right|^2
 \nonumber
 \\&&=\gamma_W
 \frac{1+2\eta \cos(\phi_\mathbf k-\phi_{\mathbf k'}) +\eta^2}{(1+\eta)^2}
\label{gammaD}
\eee
($\langle \cdots \rangle$ denotes the disorder averaging), and
\be
\gamma_W=\frac{2\pi \nu_F}{\hbar}  {W}.
\ee

Although we consider the short-range scattering potential,
function $\gamma_D(\phi_\mathbf k-\phi_{\mathbf k'})$  turns out to be
angle-dependent due to the ``dressing'' by the  factor $\left|
  \left\langle \chi_\mathbf k | \chi_{\mathbf k'} \right\rangle
\right|^2.$ As a consequence, the
transport scattering rate, \be \gamma_{tr}= \int_0^{2\pi}
\frac{d\phi}{2\pi}~\gamma_D(\phi)(1-\cos\phi)=\gamma_W\frac{1+\eta^2-\eta}{(1+\eta)^2},\ee
differs from the total (quantum) rate: \be
\frac{\gamma}{\gamma_{tr}}=\frac{1+\eta^2}{1+\eta^2-\eta}.\ee

It is well known (see Ref.~\onlinecite{schmid} for detailed
discussion), that the quantum conductivity corrections can be
expressed in terms of the Cooperon propagator. Such a propagator
obeys kinetic equation which can be derived by means of standard
diagrammatic technique.  The collision integral of this equation
contains both ingoing
and outgoing
terms describing the scattering from the momentum $\mathbf k'$ to the
momentum $\mathbf k.$
The outgoing rate
is determined by the rate $\gamma=\rm{Im} \Sigma$ entering the single-particle Green's
function \eqref{grin}.
To find ingoing rate we notice that the disorder
vertex lines in the Cooperon propagator are also ``dressed'' by the Dirac
spinor factors.
The   two-particle amplitude of the scattering from $\mathbf k'$ to $\mathbf k$ is dressed by $
\left\langle \chi_\mathbf k | \chi_{\mathbf k'} \right\rangle
\left\langle \chi_\mathbf{ -k} | \chi_{\mathbf {-k'}} \right\rangle$ (here we neglected the momentum transferred through disorder
vertex lines in these factors). Hence,
the ingoing rate $\gamma_C(\phi_\mathbf{k}-\phi_{\mathbf{k}'})$
is given by Eq.~\eqref{gammaD} with the replacement $\langle
|\tilde{V}_{\mathbf{k}\mathbf{k}'}|^2 \rangle$ with $\langle
\tilde{V}_{\mathbf{k}\mathbf{k}'}\tilde{V}^*_{-\mathbf{k'},-\mathbf{k}}\rangle.$
Simple calculation yields (here and below, for the sake of brevity we
replace $\phi_\mathbf k \to \phi $ and $\phi_\mathbf k' \to \phi' $):
 \begin{equation}
\gamma_C(\phi-\phi')=\gamma_W \left[\frac{1+\eta~ e^{-i (\phi-\phi')}}{1+\eta} \right]^2
=\sum\limits_M \gamma_M e^{iM(\phi-\phi')} ,
 \label{Wc}
\end{equation}
where
\be
\gamma_0=\frac {\gamma}{1+\eta^2},~\gamma_{-1}=\frac{2\eta \gamma}{1+\eta^2},~~\gamma_{-2}=\frac{\eta^2 \gamma}{1+\eta^2},
\label{gammaM}
\ee
and
$\gamma_M=0$ for $M\neq 0,-1,-2.$
We see that the outgoing rate considered as a function of $\phi$ contains only three harmonics:
$M=0,-1,-2.$
Remarkably,
\be \gamma_ M\neq \gamma, ~~{\rm for~ any}~~  M,\ee
that means that the
  the Cooperon propagator has a finite decay rate even in the absence of the inelastic  scattering. \cite{Tkachov11}
More specifically, the kernel of the collision integral can be written as:
\bee \nonumber
&&\gamma_C(\phi-\phi')-\gamma \delta (\phi-\phi')
\\
&& \nonumber
=-\frac{\gamma}{1+\eta^2}\left[ \eta^2 +(1-\eta)^2 e^{-i(\phi-\phi')}+e^{-2i(\phi -\phi')} \right]
\\ \nonumber && - \gamma \sum\limits_{ \mathsmaller{ M\neq 0,-1,-2}} e^{i M (\phi-\phi')}.
\nonumber
\eee
We see that in the  limit $\eta \to 0$ (orthogonal class) the Cooperon mode with the moment $M=0$ does
not decay, while for $\eta \to 1$ (symplectic class)  there is also a non-decaying mode with the moment $M=-1.$
However, for arbitrary $\eta$ all modes decay. We also note that the  function
$\gamma_C(\phi)$ is asymmetric with respect to inversion of the scattering angle, $\phi \to -\phi:$
 \be \gamma_C(\phi)=\gamma_C^*(-\phi)\neq\gamma_C(-\phi). \label{asimm}\ee

Equation \eqref{Wc} for the ingoing scattering rate in the Cooper channel
was derived for the block ${\rm II}.$ One can easily  show that in the block ${\rm I}$
the ingoing rate is given by the equation which is complex conjugated to  Eq.~\eqref{Wc}:
$\gamma_C(\phi) \to \gamma_C^*(\phi). $

In the end of  this section, we note that  for $\eta=0$ the system is in the orthogonal
symmetry class (the two-particle scattering amplitude has no angular dependence), whereas the limit $\eta=1$ corresponds to the symplectic
symmetry class with the disorder
scattering dressed by the ``Berry phase.''
The intermediate case corresponds to the {unitary} symmetry class,  where the quantum interference is partially killed by the Berry phase fluctuations (see discussion in  Sec.~\ref{berry-sec}).

\section{Kinetic equation for the Cooperon} \label{Coo}

With the projection on the upper band and dressing the disorder
correlation functions  in the Cooperon ladders, the calculation of the correction
to the conductivity reduces to the solution of a kinetic equation for
a Cooperon propagator moving in an effective disorder characterized by
the correlation functions (\ref{Wc}) in the ingoing part of the
collision integral and by (\ref{gammaD}) in the outgoing term.  This
equation is analyzed below separately for the cases of zero and
nonzero magnetic fields. In this section we focus on the contribution
of the block ${\rm II}. $ Contribution of the block ${\rm I}$ will be
discussed in the Sec.~\eqref{another}.

\subsection{Zero magnetic field} \label{CooB=0}

The kinetic  equation for
the Cooperon in $(\mathbf{q},\omega)$ domain at $\omega=0$
has the form:
\bee  \nonumber
&&\left[1/\tau_\phi + i \mathbf q \mathbf v_F   \right]C_{\mathbf{q}}(\phi,\phi_0)
= \int \frac{d\phi'}{2\pi} \left[\gamma_C(\phi-\phi') C_{\mathbf{q}}(\phi',\phi_0) \right.
\\ && \left. -\gamma_D(\phi-\phi') C_{\mathbf{q}}(\phi,\phi_0)\right]
+\gamma \delta(\phi-\phi_0).
\label{Cooperon}
\eee
 Introducing dimensionless variables
\be
\Gamma= 1/\gamma \tau_\phi,\quad \mathbf Q= \mathbf q l,\ee
where
\be l=\frac{v_F}{\gamma}=\frac{2\hbar^3v}{m W_0} \frac{\sqrt{\eta}(1-\eta)}{1+\eta^2}
\ee
is the mean free path, we rewrite Eq.~\eqref{Cooperon} as follows:
\bee \label{Cooperon1}
&&(1 +  \Gamma+ i \mathbf Q \mathbf n  ) C_{\mathbf{Q}}(\phi,\phi_0)
\\&&= \int \frac{d\phi'}{2\pi} \frac{[1+\eta e^{-i(\phi-\phi')}]^2}{1+\eta^2} C_{\mathbf{Q}}(\phi',\phi_0)
+\delta(\phi-\phi_0),
\nonumber
\eee
where $\mathbf n=(\cos \phi, \sin \phi ).$

Before solving this equation, we note that  the Fourier transform of the
Cooperon propagator gives (see Ref.~\onlinecite{schmid})  the quasiprobability
(per unit area)
for electron starting with momentum  direction $\mathbf n_0$ from initial point
$\mathbf r_0$    to arrive at the point $\mathbf r$
with the momentum direction $\mathbf n$:
\be
C(\phi,\phi_0,\mathbf r-\mathbf r_0)=\frac{1}{l^2}\int\frac{d^2\mathbf Q}{(2\pi)^2}e^{i\mathbf Q(\mathbf r- \mathbf r_0)/l}C_\mathbf Q (\phi,\phi_0).
\label{Cooperon-coord}
\ee
In particular, the conductivity can be expressed in terms of
this ``probability'' taken at $\mathbf r-\mathbf r_0=0,$  so-called
``return probability'' \cite{schmid}
 \be
  W(\phi-\phi_0)=C(\phi,\phi_0,0).\label{W-def}
  \ee
It is worth noting that  return ``probability'' which was defined above as a formal
solution of the kinetic equation for the Cooperon can be negative or complex.

Usually, the solution of the Cooperon kinetic equation is searched in
the diffusion approximation.  Within such an approximation one can
obtain a solution for an arbitrary type of the disorder.  It is also known
that in the case of the short-range disorder, when the collision
integral contains only zero harmonic with the moment $0,$ the exact
solution can be found which is valid beyond the diffusion
approximation and, consequently, describes the ballistic case.   As seen from Eq.~\eqref{Cooperon1}, in our
case the incoming term of the
collision integral contains only three angular harmonics with the
moments $0,-1,$ and $-2.$ This allows one to find the exact solution of
Eq.~\eqref{Cooperon1} valid beyond the diffusion approximation. The
details of the calculations are presented in
Appendix~\ref{kinBeq0}. We find that the return probability is given
by
\be \label{W} W(\phi)=\frac{1}{2 \pi l^2}\sum_{n=-\infty}^{\infty}
w_n e^{i(n-1)\phi}, \ee where \be \label{wn} w_n=\int \frac{d^2\mathbf
  Q}{(2\pi)^2} \left[\begin{array}{cc} P_{n-1} \\ P_{n} \\
    P_{n+1} \end{array}\right]^T \left (\hat M^{-1}-\hat
  M^{-1}_{Q=\infty} \right)\left[\begin{array}{cc} P_{n-1} \\ P_{n} \\
    P_{n+1} \end{array}\right], \ee and matrix $\hat M$ reads as
\bee
\hat{M} =\left[\begin{array}{cccc} 1+\eta^2-P_0 & - P_1 & - P_2\\
    -P_1 & \dfrac{1+\eta^2}{2\eta} - P_0 & - P_1 \\
    -P_2 & - P_1 & \dfrac{1+\eta^2}{\eta^2} - P_0
\end{array}\right].
\label{M}
\eee
Here \bee \label{Pn}
&&P_n=\int \frac{d\phi}{2\pi} \frac{e^{-i n \phi}}{1 + \Gamma + i Q \cos\phi }\\ && =(-i)^{|n|}P_0 \left[\frac{1-P_0(1+\Gamma)}{1+P_0(1+\Gamma)} \right]^{|n|/2},
\eee
and
\be
P_0=\frac{1}{\sqrt{(1+\Gamma)^2+Q^2}}.
\ee

On the technical level,  the logarithmic divergency of conductivity, specific for WL (WAL) corrections,
comes from singular behavior of the matrix $\hat M^{-1}$ at $Q\to 0.$
Let us
consider the limiting case
$Q=0$, $\Gamma=0$. In this case $P_0=1$, $P_1=P_2=0$, and we find:
 \begin{equation}
\hat M^{-1}=\begin{pmatrix} \dfrac{1}{\eta^2} & 0 & 0\\
0 & \dfrac{2\eta}{(1-\eta)^2} & 0 \\
0 & 0 & \eta^2
\end{pmatrix}.
\end{equation}
Then in the limit $\eta\to 0$ (orthogonal class) the singular mode corresponds to $M=0$ [see Eq.~\eqref{Cooperon0-1-2}] and $C_{\mathbf Q =0}(\phi,\phi_0) \propto 1/\eta^2$,
while in the limit $\eta\to 1$ (symplectic class) the singular mode corresponds to $M=-1$ and $C_{\mathbf Q=0}(\phi,\phi_0) \propto e^{-i(\phi-\phi_0)}/(1-\eta)^2$.

\subsection{Nonzero magnetic field} \label{CooB}

Now we assume that
the magnetic field $B$ is applied to the system perpendicular to the
plane of the well.  For sufficiently weak fields when the mean free
path is smaller than the cyclotron radius, the bending of the
trajectories by the magnetic field can be neglected, and the only
effect of the field is the phase difference between two trajectories
propagating along a closed loop in the opposite directions. This phase
can be taken into account by rewriting Eq.~\eqref{Cooperon} in the $(\omega,\mathbf r)$
representation and modification of the operator $\mathbf q$ as follows
$\hat{ \mathbf  q } = -i{\partial}/{\partial \mathbf r} +{2e \mathbf A}/{c\hbar},$
where $\mathbf A=(By,0)$ is the vector potential in the Landau gauge. The components of this operator
  \be  \hat q_x = -i\frac{\partial}{\partial  x} + \frac{y}{l_B^2},~~ \hat q_y= -i\frac{\partial}{\partial  y} \label{qxqy}\ee
    obey the following commutation rule:
\be \label{com}
\left[\hat q_x, \hat q_y\right ]=\frac{i}  {l_B^2} {\rm sign}(B).
\ee
Here,
\be
l_B=\sqrt{\frac{\hbar c}{2e|B|}}
\label{magnetic}
\ee
is the magnetic length for the particle with the charge $2e.$

Instead of Eq.~\eqref{Cooperon1} we get
\bee  \label{CooperonB}
&&[1 +  \Gamma+
i l\hat{\mathbf q} \mathbf n
]
C(\mathbf r, \mathbf r_0,\phi,\phi_0)
\\
&&= \int \frac{d\phi'}{2\pi} \frac{[1+\eta e^{-i(\phi-\phi')}]^2}{1+\eta^2}
C(\mathbf r, \mathbf r_0,\phi',\phi_0)
\nonumber \\&&+\delta(\mathbf r-\mathbf r_0)\delta(\phi-\phi_0),
\nonumber
\eee
where $ \hat{\mathbf q} \mathbf n   =\hat{ q_x} \cos \phi+\hat{ q_y} \sin \phi$ and  $\hat{ q_x},\hat{ q_y}$ are given by Eq.~\eqref{qxqy}.
Exact solution of this equation valid beyond the diffusion approximation is obtained in Appendix~\ref{kinBneq0}.
We  demonstrate that return probability is given by  Eq.~\eqref{W} where $w_n$  are now given by
\bee \label{wnB}
&& w_n=\frac{l^2}{2\pi l_B^2}
  \sum_{m=-\infty}^{\infty} \\ && \left[\begin{array}{cc} P_{n+m,m+1}\\ P_{n+m,m}\\ P_{n+m,m-1}\end{array}\right]^T
\left(\hat M_m^{-1}-\hat M_{m=\infty}^{-1}\right) \left[\begin{array}{cc} P_{n+m,m+1} \\P_{n+m,m} \\P_{n+m,m-1} \end{array}\right].
\nonumber
\eee
Here
\begin{widetext}
\be\label{MB}
\hat{M}_m  = \left[\begin{array}{cccc} 1+\eta^2-P_{m+1,m+1} & - P_{m+1,m} & - P_{m+1,m-1}\\
-P_{m+1,m} & \dfrac{1+\eta^2}{2\eta} - P_{m,m} & - P_{m,m-1} \\
-P_{m+1,m-1} & - P_{m,m-1} & \dfrac{1+\eta^2}{\eta^2} - P_{m-1,m-1}
\end{array}\right],
\ee
\end{widetext}
and \bee \label{Pnm} &&P_{nm}= \frac{l_B}{l}
\left[\frac{\min(n,m)!}{\max(n,m)!} \right]^{1/2} \theta(n)\theta(m)
\\ && \times\int_0^\infty \hspace{-4mm}dt e^{-t(1+\Gamma)l_B/l}
e^{-t^2/4} \left(\frac{-it}{\sqrt 2}\right)^{|n-m|}
\hspace{-2mm}L_{\min(n,m)}^{|n-m|}\left(\frac{t^2}{2}\right) ,
\nonumber \eee where $L_N^k$ are generalized Laguerre
polynomials. Although $P_{nm}$ are defined for $n \geqslant 0$ and $m
\geqslant 0,$ it is convenient to introduce theta functions
$\theta(n)$ and $\theta(m)$ [$\theta(n)=1$ for $n\geq 0$ and
$\theta(n)=0$ for $n < 0$] in the definition of $P_{nm},$ assuming
that they are defined for arbitrary integers $n$ and $m.$ From
Eq.~\eqref{Pnm} we see that $P_{nm}=P_{mn}.$ One can check that $w_n$
are real, in spite of the fact that $P_{nm}$ contains factor
$i^{|n-m|}.$

Let us now demonstrate that for $B \to 0$ Eq.~\eqref{wnB} coincides with
Eq.~\eqref{wn}. Using the asymptotical Tricomi expression for generalized
Laguerre polynomials \bee && L_n^\alpha(t^2)
\approx \frac{e^{t^2/2}}{t^\alpha}\left(n+\frac{\alpha+1}{2}
\right)^{\alpha/2} \nonumber \\ && \nonumber \times J_\alpha\left(
  2\sqrt{n+\frac{\alpha+1}{2}}t\right), ~~\text{for}~n\to \infty, \eee
(here $J_\alpha$ is the Bessel function) we find from Eq.~\eqref{Pnm}
\be \label{Passym} P_{m,m+\alpha} \to
P_\alpha(Q_{m,\alpha}),~~\text{for}~~ m\to\infty, \ee where \be
Q_{m,\alpha}=\frac{l}{l_B}\sqrt{2m+{\alpha+1}} \ee and $P_\alpha(Q)$
is given by Eq.~\eqref{Pn}.  Substituting Eq.~\eqref{Passym} into
Eq.~\eqref{MB} and \eqref{wnB}, neglecting at large $m$ dependence of
$Q_{m,\alpha}$ on $\alpha$ and replacing summation over $m$ in
Eq.~\eqref{wnB} with integration over $dQ^2l_B^2/2l^2,$ after simple
calculations we arrive to Eq.~\eqref{wn}.

Finally, we note that Eqs.~\eqref{wnB} and \eqref{MB} were derived for block ${\rm II}.$ Analogous equations valid in
the block ${\rm I}$ are presented in the Sec.~\ref{another}.

\subsection{Diffusion approximation}\label{diffusion}

The diffusion  approximation is valid provided that the typical length of
interfering diffusive trajectory is long compared to the mean free
path $l.$ On the formal level this implies that there is a diffusive
Cooperon mode with the gap much smaller than $\gamma.$ To guarantee
that this condition is fulfilled, we assume that $\Gamma \ll 1, $
dimensionless magnetic field,
\be \label{b} b=\frac{l^2}{2l_B^2} {\rm sign (B)} , \ee
is weak,
$
|b| \ll 1 ,$
and one of the conditions, $\eta \ll 1$ or $1-\eta \ll 1$ is satisfied.
In this case, calculation of $w_n$ essentially simplifies.  Let us
start with discussion of the case of zero field and then generalize
obtained results for the case of finite fields.

\vspace{2mm}
\underline{(a) $b=0.$ } \vspace{2mm}

The formal solution of Eq.~\eqref{Cooperon} is written as
\be
\hat C =\frac{\gamma}{\gamma+\gamma_\varphi-\hat \gamma_C +i\mathbf q\mathbf n},
\label{Coop-formal}
\ee
where $\gamma_\varphi=1/\tau_\varphi$ and the kernel of the operator $\hat \gamma_C$ is given by
Eq.~\eqref{Wc}.

For $ql \ll 1$ we find from Eq.~\eqref{Coop-formal}
\bee
\nonumber &&C_{\mathbf q}(\phi, \phi_0)\approx \frac{\gamma}{2\pi} \\ \nonumber && \times \sum \limits_M
\frac{e^{iM(\phi-\phi_0)}}{\tau_M^{-1} +v_F^2 \left\langle  \mathbf q \mathbf n
 \frac{1}{\gamma+\gamma_\varphi-\hat\gamma_C}\mathbf q \mathbf n
\right\rangle_M }\\ \nonumber
&&= \frac{\gamma}{2\pi}\sum \limits_M \frac{e^{iM(\phi-\phi_0)}}{\tau_M^{-1}+q^2D_M },
\eee
where
\be
 \tau_M=\frac{1}{\gamma_\varphi +\gamma-\gamma_M},
  \label{tauMM}
 \ee
\be
D_M=\frac{v_F^2(\tau_{M+1}+\tau_{M-1})}{4},
\ee
  $\langle \cdots\rangle_M$ stands for projection on the $M$th harmonic and $\gamma_M$ are given by Eq.~\eqref{gammaM}
  We note that for calculation $D_M$ one may set $\gamma_\varphi=0$ in Eq.~\eqref{tauMM}. Within this approximation  we find
  $\tau_0={(1+\eta^2)}/{\gamma \eta^2},
 \tau_{-1}={(1+\eta^2)}/{\gamma(1- \eta)^2},
 \tau_{-2}={(1+\eta^2)}/{\gamma},$ and $
 \tau_M={1}/{\gamma},~~\rm{for}~~M\neq 0,-1,-2.$
  Here, we took into account that $\Gamma \ll 1~ (\gamma_\varphi \ll \gamma).$ For $w_M$ we find
 \be
 w_M=\int\limits_{ql\lesssim 1} \frac{l^2d^2\mathbf q }{(2\pi)^2}\frac{\gamma}{\tau_{M-1}^{-1} +q^2D_{M-1} }
 \label{wn0}
 \ee
Since we uses the diffusion approximation, the integration in Eq.~\eqref{wn0} is limited by small $q:$  $ql\lesssim 1.$

\vspace{2mm}
\underline{(b) $b \neq 0.$ } \vspace{2mm}
In this case one should
take into account that  operators $\hat q_x$ and $\hat q_y$
no longer commute [see Eq.~\eqref{com}]. Simple calculation yields
\bee
&& \label{lin} v_F^2 \left\langle  \hat{ \mathbf  q} \mathbf n
 \frac{1}{\gamma+\gamma_\varphi-\hat\gamma_C}\hat{ \mathbf q} \mathbf n\right\rangle_M \\ &&= ( \hat q_x^2+\hat q_y^2)D_M + i[\hat q_x,\hat q_y]\tilde D_M
 , \nonumber \eee
where
\be \tilde D_M=\frac{v_F^2(\tau_{M-1}
  -\tau_{M+1}) }{4}.
\ee
As a next step one should replace in
Eq.~\eqref{lin} the operator $\hat q_x^2 + \hat q_y^2$ by it's
eigenvalue $\hat q_x^2 + \hat q_y^2 = (4/l^2)|b| (n+1/2). $
Importantly, the operator $\hat q_x^2 +
\hat q_y^2$ is positively defined and depends on the absolute value of
the magnetic field. In contrast, commutator $[\hat q_x,\hat q_y]$
changes sign with inversion of field. Hence, Eq.~\eqref{lin} contains
the sum of two equations of different parity with respect to the field inversion:
\be v_F^2 \left\langle \hat{
    \mathbf q} \mathbf n
  \frac{1}{\gamma+\gamma_\varphi-\hat\gamma_C}\hat{ \mathbf q} \mathbf
  n\right\rangle_M = \frac{2}{l^2}\left[ |b|\left(2n+1\right)D_M - b
  \tilde D_M\right] ,
 \label{lin1}
 \ee
  One should also modify  Eq.~\eqref{wn0}
 by replacing the  integral over $d^2\mathbf q$ with the sum over $n$:
\be
\int\limits_{ql <1} \frac{l^2 d^2 \mathbf q}{(2\pi)^2}  \to \frac{|b|}{\pi} \sum\limits _{n=0}^{N},
\ee
where $N \propto 1/|b| $ limits the region where diffusion approximation is applicable.
Analytical expression for $w_M$  reads as
  \bee \label{wn00}
 && w_M =\frac{|b|}{\pi} \\ && \times \sum\limits_{n=0}^N \frac{\gamma l^2}{l^2 /\tau_{M-1} + {2}|b|(2n+1) D_{M-1}-2b\tilde D_{M-1} }.
 \nonumber
 \eee
 We see that $w_M$ turns out to be asymmetric with respect to
 inversion of $b$: $w_M(b)\neq w_M(-b),$ so that contribution to the
 conductivity correction coming from a single cone is an asymmetric
 function of $b$. Of course, after summing contributions of two cones,
 conductivity becomes even function of magnetic field as it should be.

\section{Interference correction to the conductivity}\label{inter}
\subsection{Contribution of the single Dirac cone}\label{single}
In this subsection, we calculate the contribution to the interference
correction coming from the single cone. We will use expressions
derived in the previous section for the Cooperon propagator in the
block II.

\begin{figure}[ht]\center
\includegraphics[width=0.45\textwidth]{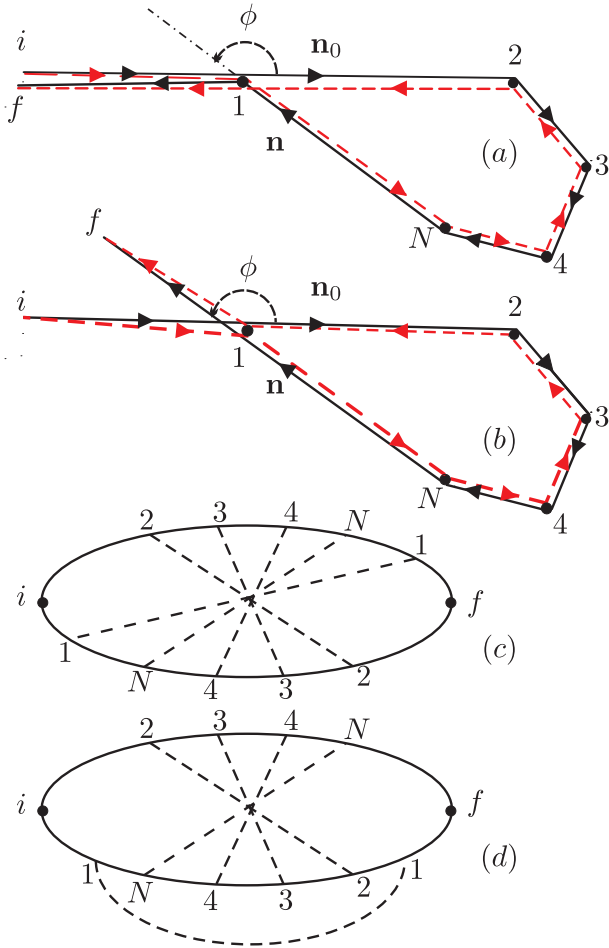}
\caption{Backscattering (a) and  nonbackscattering (b)
scattering processes on a complex of impurities consisting of $1,2, \ldots, N$ impurities and corresponding diagrams
(c) and (d) (For simplicity, we do not show in (a) and (b) diffusive motion of the particle before and after the loop. Such motion leads  to ``transportization''  of  vertexes  $i$ and $f$).
}
\label{F1}
\end{figure}

\begin{figure}[ht] \center
\includegraphics[width=0.45\textwidth]{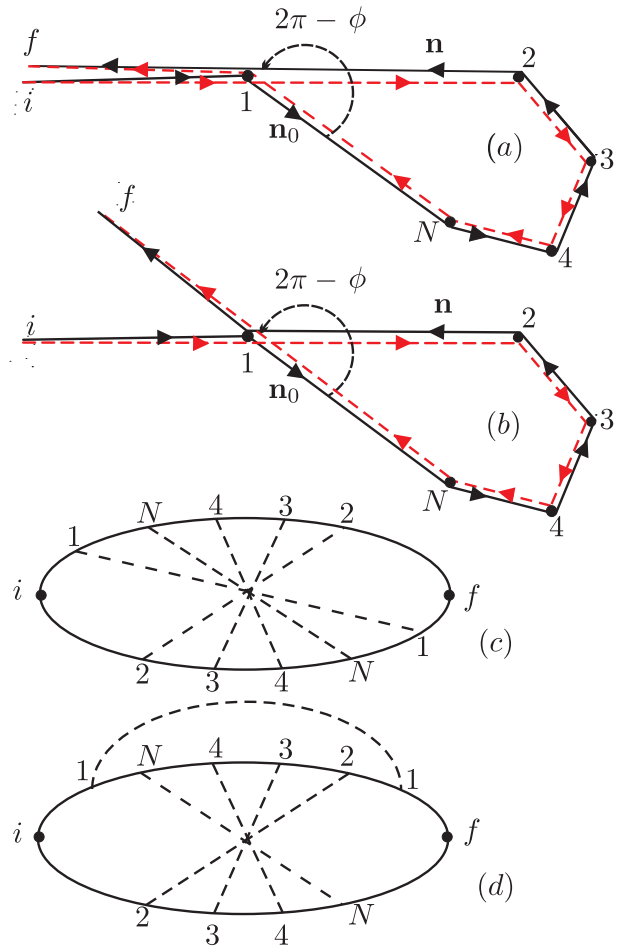}
\caption{ The processes (a), (b) and corresponding diagrams leading to the
contribution which is complex conjugated to the one given by the processes shown in Fig.~1.
}
\label{F2}
\end{figure}

The Cooperon propagator enters as a building block into diagrams describing the quantum
correction to the conductivity. Such diagrams
can be calculated in a standard way.
However, instead of formal summation of the diagrams, one can use a
semiclassical method developed in Ref.~\onlinecite{nonback}.   It was shown in this paper that there are  two contributions to the coherent scattering,
so-called backscattering and nonbackscattering contributions. Both of
them can be expressed in terms of renormalization of the cross section
of the scattering by a single impurity.  The main advantage of this
method compared to the standard diagrammatic machinery is the simplicity
of the calculations and the physical transparency that will allow us
to clarify the physical sense of the obtained results. This method is
based on considering of the trajectories corresponding to the
condition of the phase stationarity. Such trajectories give the
dominant contribution in the weak localization regime, when
$k_Fl\gg1.$
The method was developed in Ref.~\onlinecite{nonback} for the case of
the isotropic scattering of spinless particle by a short-range
potential, that corresponds to $\eta\to 0$ in our
notations. Generalization for the case of finite $\eta$ is
straightforward. The only additional ingredient of the discussed
problem is the spin-projectors, Eq.~\eqref{Pkk}, entering Green's functions,
 Eq.~\eqref{G}. As we already mentioned above such projectors can be taken
into account by dressing of the disorder correlation functions.  Below
we demonstrate it by discussing the simplest coherent scattering processes
at a scattering complex consisting of $N=5$ impurities. Such processes
and corresponding diagrams are shown in Figs.~ \ref{F1} and \ref{F2}.

The coherent backscattering process plotted in Fig.~1(a) corresponds to
diagram shown in Fig.~1(c). In this process, two electron waves with the
amplitude $A$ (solid line) and $B$ (dashed line) start at the point
$i$ and interfere at the point $f$ after traveling in the opposite
directions around a closed loop formed by impurities $1,2, \ldots , N.$
The contribution of this process to the conductivity is proportional
to $AB^*.$ This process leads to correction to the scattering cross
section (delta-peaked at the scattering angle $\phi=\pi$) of the
impurity $1$ (for a more detailed discussion see
Ref.~\onlinecite{nonback}). The scatterings on impurities $2,3, \ldots
, N$ are dressed by a spinor factors and described by the Cooperon
correlation function $\gamma_C$ defined by Eq.~\eqref{Wc}. One can
easily show that the impurity $1$ is also dressed by spinor factor
proportional to $\gamma_C(\pi-\phi),$ where $\phi$ is the angle
between $\mathbf n$ and $\mathbf n_0$ [see Fig.~1(a)].  The process
shown in Fig.~1(b) leads to nonbackscattering contribution described by
diagram shown in Fig 1(d). One can easily check that in this process the
impurity $1$ is dressed by the same spinor factor $\propto
\gamma_C(\pi-\phi).$ The sum of these two processes and the processes
Figs.~2(a) and 2(b) [described by diagrams 2(c), and 2(d), respectively] which are complex
conjugated to Figs.~1(c), and 1(d) leads to the correction to the scattering cross
section:
\bee \nonumber && \frac{\delta \Sigma (\phi)}{\Sigma_{tr}} \approx
\lambda_F l_{tr} \frac{1}{\gamma} \left[ \delta(\phi-\pi) \int
  d\phi' { \gamma_C(\pi-\phi')} W(\phi') \right.\\ &&\left. -
  \gamma_C(\pi-\phi) W(\phi) \right]. \nonumber \eee
where first and
second term represent backscattering and nonbackscattering
contributions, respectively. Here, $\lambda_F l_{tr}$ is the effective
return area and the coefficient $1/\gamma$ appeared due to the
normalization of the Cooperon propagator [see coefficient $\gamma$ in
front of the delta function in Eq.~\eqref{Cooperon}]. The quantum
correction to the cross section  incorporating both types of the
coherent processes is given by
$\int (d\phi/2\pi) \delta \Sigma
(\phi)(1-\cos\phi)$ and corresponding  conductivity correction reads as
\be \label{sigma} \delta
\sigma_{{\rm II}}= -\frac{e^2}{ \hbar} \left(
  \frac{l_{tr}}{l}\right)^2 l^2\int
\frac{d\phi}{2\pi}~\frac{\gamma_C(\pi -\phi)}{\gamma}
W(\phi)(1+\cos\phi).
\ee
Equation~\eqref{sigma} can be also derived within the standard diagrammatic approach, in which factor $l^2$ appears due to the
integration over coordinates of the points $i$ and $f$, whereas factor
$\left( {l_{tr}}/{l}\right)^2 $ accounts for the transportization of
the vertexes.  As a result, we obtain
\bee \label{sigma1} &&\delta
\sigma_{{\rm II}}= -\frac{e^2}{ \hbar}
\frac{(1+\eta^2)l^2}{(1+\eta^2-\eta)^2} \\ && \nonumber \times \int
\frac{d\phi}{2\pi} (1-2\eta e^{i\phi}+\eta^2 e^{2i\phi})
W(\phi)(1+\cos\phi), \eee
 or, finally, with the use of Eq.~\eqref{W}
\bee \nonumber && \delta \sigma_{{\rm II}}=-\frac{e^2}{2\pi \hbar}
\frac{1+\eta^2}{(1+\eta^2-\eta)^2}\hspace{-1mm}\left[(1-\eta)w_1+\frac{1+\eta^2-4\eta}{2}w_0
\right.\\ &&\left.+\frac{w_2}{2}+(\eta^2-\eta)w_{-1}+\frac{\eta^2}{2}
  w_{-2} \right] ,\label{sigma2}
\eee
where $w_n$ are given by
Eqs.~\eqref{wn} and \eqref{wnB} for $B=0$ and $B\neq 0,$
respectively.

As we mentioned above, for $\eta \to 0$ and $ 1$ one of the
modes becomes singular and diffusion approximation is applicable
provided that $\Gamma \ll 1, |b|\ll 1.$ Next, we consider limiting cases
described within this approximation. One can use two alternative
approaches: to explore simplified equations obtained in
Sec.~\ref{diffusion} or obtain results directly from rigorous
equations \eqref{wn} and \eqref{wnB}. In the following two subsections
we use the first approach. In Appendix \ref{diffusion1} we demonstrate
that the second approach yields the same results.

At the end of this Section we note that  the conductivity correction, Eq.~\eqref{sigma2}, is invariant over the replacement $\eta \to 1/\eta.$ As was explained in the Sec.~\ref{BHZ} this means that although  this equation  was derived for normal case it equally applies for inverted semiconductor both for electron and the hole spectrum branches.

\subsubsection{Limiting cases for $B=0$}

\underline{(a)~$\eta \to 0.$} As seen from Eq.~\eqref{wn0}, in this case the singular
contribution comes from $w_1$ because $\tau_{0}^{-1}= \gamma_\varphi +\gamma \eta^2/(1+\eta^2) \to  \gamma_\varphi+ \gamma \eta^2 \ll \gamma. $
The diffusion coefficient entering  Eq.~\eqref{wn0} reads as $D_0\approx v_F^2/2\gamma,$ so that from Eqs.~\eqref{wn0} and \eqref{sigma2}
we find
\be
\label{ds}\delta \sigma_{{\rm II}}\approx-\frac{e^2}{4\pi^2\hbar}\ln\left(\frac{1}{\eta^2+\Gamma}\right).
\ee

\underline{(b)~$\eta \to 1.$} The singular contribution comes from $w_0,$
because $\tau_{-1}^{-1} \to \gamma_\varphi+\gamma(1-\eta)^2/2\ll 1.$ The diffusion coefficient
in this mode reads as $D_{-1}\approx v_F^2/\gamma.$ From Eqs.~\eqref{wn0} and \eqref{sigma2} we obtain
\be \label{ds1}\delta \sigma_{{\rm II}} \approx\frac{e^2}{4\pi^2\hbar}\ln\left[\frac{1}{(1-\eta)^2/2+\Gamma}\right].\ee

A more general approach based on exact equation \eqref{wn}
allows one to find analytically corrections to Eqs.~\eqref{ds} and \eqref{ds1}:\cite{myPSS}
\bee \label{ds-ball}
&& \delta \sigma_{{\rm II}} \approx \frac{e^2}{4\pi^2 \hbar} \\ && \times  \begin{cases}
 -\left( 1-\frac{5\eta^2}{2}\right)\ln\left(\frac{1}{\eta^2}\right) +\ln2 +4\eta \ln 2- \eta \\+O(\eta^2) ,~~ {\rm for}~ \eta\to 0,\\[10pt]
 \left[ 1-\frac{7(1-\eta)^2}{2}\right]\ln\left[\frac{1}{(1-\eta)^2}\right] +3(\ln2-1) + \eta -1 \\+O[(1-\eta)^2],~~  {\rm for}~ \eta\to 1.
 \end{cases}
\nonumber
\eee
For simplicity,  we neglected  here dephasing processes thus  setting $\Gamma=0$. [For comparison of Eq.~\eqref{ds-ball}  in the limit $\eta \to 1$
with the previously obtained results,\cite{Tkachov11}   see Ref.~\onlinecite{myPSS}.]

\subsubsection{Limiting cases for $B\neq0$}
\underline{(a)~$\eta \to 0.$} In this case, $\tilde D_0 \approx \eta v_F^2/2\gamma $
and from Eq.~\eqref{wn00} we find the expression for the singular mode
\be
w_1=\frac{|b|}{\pi} \sum\limits_{n=0}^{N}\frac{1}{\Gamma+\eta^2 +2|b|(n+1/2)-b\eta}.
\label{w11}
\ee
Let us introduce  the function
\bee
\nonumber
&&h(|b|,A)=\sum\limits_{n=0}^{N \sim 1/|b|}\frac{|b|}{ |b|(n+1/2)+A}\\&\approx& \ln\left(\frac{C_*}{|b|}
\right)- \psi(A/|b|+1/2),
\label{h}
\eee
where $$\psi(z)=\sum \limits_{k=1}^{\infty}\left( \frac{1}{k} -\frac{1}{k+z-1}\right) -C ,$$
is the digamma function, $C$ is the Euler constant and the field-independent coefficient  $C_* \sim N |b|\sim 1$ is determined by the  ballistic effects  and can not be found within the diffusion approximation. We will see that this coefficient drops out from  the equation for the  magnetoconductivity [see Eq.~\eqref{WL1}].
Asymptotical  behavior  of $h(|b|,A)$ in the limits of weak and strong fields can be found with the use of the following asymptotic of $\psi(z)$:
\be \label{psi}
\psi(z)
\approx
\ln z-\frac{1}{2z} -\frac{1}{12z^2}  ,\ \  \text{for} ~~ z\gg1.
\ee
In our case, $A=(\Gamma+\eta^2-b\eta)/2 $ and the variable $z=b_*/|b|+1/2-\eta b/2|b|$ changes in the following interval:  $(1-\eta)/2< z <\infty.$
Here,
$$
b_*=\frac{\Gamma+\eta^2}{2}.
$$
From Eqs.~\eqref{h} and \eqref{psi}  we find
\bee
\label{sum}
&&h\left(|b|,\frac{\Gamma+\eta^2-b\eta}{2}\right)
\\ \nonumber
&&\approx   \begin{cases}
\displaystyle  \ln \left( \frac{C_*}{b_*}\right)+\frac{\eta b}{2b_*}-\frac{b^2}{24 b_*^2},~~~\text{for}~~ |b|\ll b_*
,\vspace{5mm}\\
\displaystyle
\ln\left( \frac{C_*}{|b|}\right)-\psi(1/2)-\frac{\pi^2}{2} \frac{b_*-\eta b/2}{|b|} , \\\text{for}~~ b_*\ll |b| \ll 1.
 \end{cases}
\eee
We see that function $h$ is asymmetric with respect to inversion of the magnetic field. In the low-field limit the asymmetry is due to the term $\eta b/2b_*,$ which shifts the position of the maximum of $h$ from point $b=0$ to the point $b=6\eta b_*.$ In the opposite strong-field limit, a small asymmetry comes from  the term $\pi^2 b/4|b|,$ which arises as a result of  expansion of the digamma function near the point $z=1/2$     [in order to avoid cumbersome equations, in the following  we neglect everywhere small correction $-\psi(1/2)-({\pi^2}/{2}) (b_*-\eta b/2)/{|b|}$ to the logarithmic strong-field asymptotic of $h$].

The function $w_1$ is expressed in terms  of function $h$ as follows:
\be \label{w1h} w_1={h\left(|b|,\frac{\Gamma+\eta^2-b\eta}{2}\right)}/{2\pi}.\ee
Using  Eqs.~\eqref{sigma2},  \eqref{sum}, and \eqref{w1h} we find
\bee \label{WL1}
&&\Delta \sigma_{{\rm II}}=\delta \sigma_{{\rm II}}(b)-\delta \sigma_{{\rm II}}(0)
\\\nonumber &&=- \frac{e^2}{4\pi^2\hbar}\left[h\left(|b|,\frac{\Gamma+\eta^2-b\eta}{2}\right)-h\left(0,\frac{\Gamma+\eta^2}{2}\right)\right]
\\\nonumber&&\approx- \frac{e^2}{4\pi^2\hbar}
\left\{\begin{array}{ll} \displaystyle \frac{\eta b}{2b_*}-
\frac{b^2}{24b_*^2},\ \ & \text{for} ~~|b|\ll b_* ~~\vspace{2mm}  \\  \displaystyle
\ln\left(\frac{b_*}{|b|}\right),\ \  & \text{for}~~|b|\gg b_*
\,.  \end{array} \right.
\eee
Remarkably, Eq.~\eqref{WL1} contains linear-in-$b$ term in the limit
of low fields.
It is worth noting that this term can be taken into account on the formal level as renormalization of the Cooperon decay rate
 $\eta^2 \to \eta^2 -\eta b$ in the argument of the function $h.$
 The
existence of linear contribution to the conductivity means that in a single Dirac cone the
minimum of the conductivity correction as a function of $b$ is shifted
from the point $b=0$ to the point $b=6\eta b_*.$  However, as we already mentioned, after summing
contributions of two blocks the linear-in-$b$ terms existing in these
blocks cancel each other.

\underline{(b)~$\eta \to 1.$}
In this case,   $\tilde D_{-1} \approx -(1-\eta) v_F^2/\gamma $ and from Eq.~\eqref{wn00} we find
\be
w_0=\frac{|b|}{\pi} \sum\limits_{n=0}^{N}\frac{1}{\Gamma+(1-\eta)^2/2 +2|b|(2n+1)+2b(1-\eta)}.
\label{w111}
\ee
By using the
asymptotic of the digamma function
we find for the conductivity correction
\bee \label{WL2}
&&\Delta \sigma_{{\rm II}}=\delta \sigma_{{\rm II}}(b)-\delta \sigma_{{\rm II}}(0)\\ \nonumber
&&=\frac{e^2}{4\pi^2\hbar} \left\{ h\left[|b|,\frac{\Gamma+(1-\eta)^2/2+2b(1-\eta)}{4}\right]\right.
\\\nonumber && \left. -h\left[0,\frac{\Gamma+(1-\eta)^2/2}{4}\right]\right\}
\\ \nonumber &&\approx \frac{e^2}{4\pi^2\hbar}
\left\{\begin{array}{ll} \displaystyle  -\frac{(1-\eta) b}{2b_*}-
\frac{b^2}{24b_*^2},\ \ & \text{for} ~~|b|\ll b_* ~~\vspace{2mm}  \\  \displaystyle
\ln\left(\frac{b_*}{|b|}\right),\ \  & \text{for}~~ b_*\ll |b| \ll 1
\,,  \end{array} \right.
\eee
where
$$
b_*=\frac{\Gamma+(1-\eta)^2/2}{4}.
$$
Again, we see that Eq.~\eqref{WL2} contains the linear-in-$b$ term in the limit of low fields.
 We also see that this term  can be formally taken into account by renormalization of the Cooperon decay rate $(1-\eta)^2/2 \to (1-\eta)^2/2 + 2(1-\eta) b$ in the argument of the function $h.$

\subsection{Contribution of the other Dirac cone and inversion of the sign of the magnetic field}
\label{another}

Next we take into account the contribution to the conductivity
correction coming from the block I.  The only difference is that the
scattering rate $\gamma_C(\phi-\phi')$ is given by the equation which
is complex conjugated to Eq.~\eqref{Wc}:
 \begin{equation}
\gamma_C(\phi-\phi')=\gamma_W \left[\frac{1+\eta~ e^{i (\phi-\phi')}}{1+\eta} \right]^2
.
\label{Wc1}
\end{equation}
The calculations are quite analogous to the ones presented above.
  The return probability  reads as
  \be \label{Wphi1}
 W(\phi)=\frac{1}{2 \pi l^2}\sum_{n=-\infty}^{\infty} w_n e^{-i(n-1)\phi},
 \ee
  [we notice sign minus in the exponent compared to Eq.~\eqref{W}], where
  \bee \label{wnB1}
&& w_n=\frac{l^2}{2\pi l_B^2} \\ && \times \sum_{m=-\infty}^{\infty} \hspace{-1mm}
\left[\begin{array}{cc} P_{m-n,m-1}\\ P_{m-n,m}\\ P_{m-n,m+1}\end{array}\right]^T
\hspace{-4mm}\left(\hat { M}_m^{-1}-\hat  M_{m=\infty}^{-1}\right)\hspace{-1mm}
\left[\begin{array}{cc} P_{m-n,m-1} \\P_{m-n,m} \\P_{m-n,m+1} \end{array}\right],
\nonumber
\eee
 and
 \begin{widetext}
 \be
 \label{MB1}
\hat{M}_m  = \left[\begin{array}{cccc} 1+\eta^2-P_{m-1,m-1} & - P_{m,m-1} & - P_{m+1,m-1}\\
-P_{m,m-1} & \dfrac{1+\eta^2}{2\eta} - P_{m,m} & - P_{m+1,m} \\
-P_{m+1,m-1} & - P_{m+1,m} & \dfrac{1+\eta^2}{\eta^2} - P_{m+1,m+1}
\end{array}\right].
\ee
\end{widetext}
 The contribution to the conductivity correction is expressed in terms of $w_n$
according to Eq.~\eqref{sigma2}.
For  zero field,
 Eq.~\eqref{wnB1} transforms to Eq.~\eqref{wn}, so that the account for the second
cone doubles  the conductivity correction:
 \be \delta \sigma_{tot}=2\delta \sigma_{{\rm
    II}}.\label{sumb0} \ee
 However, this is not the case for the nonzero field when the total conductivity correction is given by
\be \label{tot}
\delta \sigma_{tot}= \delta \sigma_{\text I}+\delta \sigma_{\text I \text I},
\ee
where $\delta \sigma_{\text I}$ and  $\delta \sigma_{\text I \text I}$  are found from  Eqs.~\eqref{wnB1}, \eqref{MB1}, \eqref{Pnm}, \eqref{sigma2},
and Eqs.~\eqref{wnB}, \eqref{MB},   \eqref{Pnm}, \eqref{sigma2}, respectively.

Let us now discuss what happens with the inversion of the sign of the magnetic field.
Such inversion changes the sign of the commutator~\eqref{com}.
The calculations fully analogous to those presented in Sec.~\ref{CooB} and Appendix~\ref{kinBneq0}
yield that the  magnetic field inversion is equivalent to  the replacement of
$\delta \sigma_{\text I}$ and $\delta \sigma_{\text I \text I}:$
\be
\delta \sigma_{\text I}(b)=\delta \sigma_{\text I \text I}(-b).
\label{bto-b}\ee
Importantly,
$\delta \sigma_{\text I}(b)\neq \delta \sigma_{\text I \text I}(b)$,
$\delta \sigma_{\text I}(b)\neq \delta \sigma_{\text I}(-b)$
and $\delta \sigma_{\text I \text I}(b)\neq \delta \sigma_{\text I \text I}(-b).$
As we demonstrate in the following (see Sec.~\ref{num}), numerical evaluation of the conductivity correction shows strong
magnetic field asymmetry of the functions $\delta \sigma_{\text I}(b)$ and
$\delta \sigma_{\text I \text I}(b). $   Only the total conductivity correction is an even function of the magnetic field
\be
\delta \sigma_{tot}(b)=\delta \sigma_{tot}(-b).
\label{sigma_tot}
\ee

 \subsection{Interpretation of the obtained results in terms of the Berry phase} \label{berry-sec}
   In this subsection we will  use  the Berry-phase  picture for  a qualitative interpretation of the obtained results.
   We will follow the approach discussed in Ref.
   \onlinecite{Richter11} in connection with the weak localization of holes and later on used  for description of the interference corrections in the  2D  HgTe/CdTe  based TI \cite{Richter12} and topological crystalline insulators with a quadratic surface spectrum \cite{Tkachov13}.

 The Berry ``vector potential''  $\boldsymbol{\mathcal  A}$  is defined as (see Refs.~\onlinecite{Richter12,Richter11} and references therein)
  \be
  \boldsymbol{\mathcal  A}=i\left \langle \chi_\mathbf k \left| \frac{\partial}{\partial \mathbf k}   \right| \chi_\mathbf k \right \rangle.
  \label{berry1}
  \ee
The Berry phase is given by  the integration of the vector potential along a closed loop on the Fermi surface:  $\Gamma_B=\oint \boldsymbol{\mathcal  A}d\mathbf k .$
Substituting into Eq.~\eqref{berry1} spinor functions  $\chi_\mathbf k^{(\text{I},+)}$ and $\chi_\mathbf k^{(\text{II},+)}$  we find
after simple algebra
\be
\Gamma_B^{\text{I}}=-\Gamma_B^{\text{II}}=\frac{2\pi\eta}{1+\eta}.
\label{GammaB}
\ee
We see that the Berry phases in   two  blocks have opposite signs.
For $\eta \ll 1,$ we have  $\Gamma_B^{\text{I}}=-\Gamma_B^{\text{II}}\approx {2\pi\eta}.$ In the opposite limit, $1-\eta \ll 1,$ we get $\Gamma_B^{\text{I}}=-\Gamma_B^{\text{II}}\approx \pi -\pi(1- \eta)/2.$

 Next, we use  Eq.~\eqref{GammaB}  to clarify  the underlying physics of the results obtained in the previous sections.
Since we  have already  presented the rigorous derivations, in this subsection we will limit ourselves to the  qualitative estimates omitting coefficients on the order of unity. The estimates are valid in the diffusion approximation, so that we will only  discuss  the regions $\eta \ll 1 $ and $1-\eta \ll 1.$

Let us start with the case of zero magnetic field. Redrawing  in the momentum space  coherent scattering processes shown in Figs. 1,2  and taking into account that the amplitudes  of the trajectories shown by dashed lines should be complex-conjugated, one easily  finds that the total  geometrical phase factor  entering the return probability $W(\phi)$ is given by
\be
e^{i\theta }= e^{i\Gamma_B (1+2n)},
\label{theta}
\ee
where $-\infty <n<\infty $ is an integer number (winding number) and $\Gamma_B$ equals  $\Gamma_B^{\text{I}}$ or $\Gamma_B^{\text{II}}.$
For given value of $\phi$ the winding number strongly fluctuates due to the diffusive nature of the particle  motion. In the diffusion approximation, the distribution of the directions of the particle momentum obeys the diffusion law  $  P(\phi_\mathbf k) =  \exp(-\phi_\mathbf k^2 /4\gamma t)/\sqrt{4\pi \gamma t}$ (since we present  order-of-magnitude estimates only, we do  not distinguish   between total and transport scattering rates). We  used  here extended Fermi surface,   assuming that  $-\infty <\phi_\mathbf k <\infty.$  Putting $\phi_\mathbf k \approx 2\pi n$ and averaging Eq.~\eqref{theta} over $n$ with the function $P(\phi_\mathbf k)=P(2\pi n)$  we find
 \be
\left \langle e^{i\theta } \right \rangle  \propto  \left \{ \begin{array}{c}
                                                               \exp{(-\eta^2\gamma t)},~~\text{for} ~~\eta \to 0, \\
                                                               -\exp{[-(1-\eta)^2 \gamma t]},~~\text{for} ~~\eta \to 1.
                                                             \end{array}
 \right.
\label{theta-av}
\ee
 Equation~\eqref{theta-av} allows one to interpret the  finite gap in the Cooperon propagator   in terms of decay of the averaged Berry phase factor
 due  to the  diffusive fluctuations of the winding number. Indeed, multiplying Eq.~\eqref{theta-av} with the product of the diffusive return probability $1/Dt$ (here $D \sim v_F^2/ \gamma$) and  the decoherence factor  $\exp(-t/\tau_\varphi),$  and integrating over $t$ from $t \sim 1/\gamma$ to $t=\infty,$ one easily reproduces logarithms entering    Eqs.~\eqref{ds} and \eqref{ds1}.    It is worth noting that the negative sign in front of the exponent in the second line of Eq.~\eqref{theta-av} is responsible for the change of WL to WAL and also arises due to the geometrical reasons (because $\Gamma_B=\pi$ for $\eta=1$).

The approach based on the Berry-phase picture also allows one to give a transparent physical interpretation of the low-field  linear magnetoresistance. For $B\neq 0,$ the return probability  acquires the  phase factor\cite{schmid} $\exp(2ieBS/\hbar c  )=\exp(iS/l_B^2) $  in addition to the Berry phase factor. Here, $S$ is the algebraic area covered by the particle while propagating along the closed loop. Let us assume that $D\tau_\varphi \ll l_B^2.$ We also  assume that $1/\tau_\varphi \gg  \eta^2\gamma $ ($\Gamma \gg \eta^2$) in the region $\eta\ll 1$ and $1/\tau_\varphi \gg  (1-\eta)^2\gamma $ [$\Gamma \gg (1-\eta)^2$] in the region $1-\eta\ll 1.$ In this case, both phase factors can be expanded in the Taylor series. As a next step, one should average over diffusion motion of the particle. Evidently,
\bee && \langle S \rangle =0,~~ \langle S^2 \rangle \sim (D\tau_\varphi)^2,  \label{Save}\\
&& \langle n \rangle =0,~~ \langle n^2 \rangle \sim \tau_\varphi\gamma.
   \label{nave}\eee
 It is also  clear physically,  that some correlations between $S$  and $n$ should  exist. To calculate the correlation function $\langle  n S   \rangle $  at a given time  $t$ ($t\sim \tau_\varphi \gg \gamma^{-1}$), we write
 \be n S \approx \frac{\phi(t)}{2\pi} \int\limits_0^t dt_2 \int\limits_0^{t_2} dt_1 v_F^2\sin[\phi(t_2)- \phi(t_2)]/2, \ee
  and average by functional integration  over  $\{D\phi \}$ with the weight $\exp\left( - \int_0^t \dot{\phi}^2 dt \tau \right).$ Standard calculation yields
  \be  \left \langle  n S\right \rangle \sim D\tau_\varphi. \label{nS} \ee
 Magnetoconductivity is estimated as \be   {\delta \sigma(b) - \delta \sigma(0) } \propto -e^{i\Gamma_B}\left \langle -S^2/2l_B^4 - 2\Gamma_B   n S/l_B^2 \right \rangle       \ee
Using Eqs.~\eqref{Save}, and \eqref{nS} we find
\be
\delta \sigma^{\text{I,II}}(b)- \delta \sigma^{\text{I,II}}(0) \propto \left\{ \begin{array}{c}
                                                                                 b^2/b_*^2 \pm \eta b/b_*,~~~\text{for}~~\eta\to 0, \\
                                                                                  -b^2/b_*^2 \pm (1-\eta) b/b_*,~~~\text{for}~~\eta\to 1,
                                                                               \end{array}
 \right.
\ee
where signs $+$ and $-$ stand for $\text{I}$  and $\text{II}$ blocks, respectively, and  $b_* \sim \Gamma$ under approximations used in this subsection. Hence, we reproduce  by the order of magnitude the low-field magnetoconductivity given by Eqs.~\eqref{WL1} and \eqref{WL2}.
We see, that within the Berry-phase formalism    linear-in-$b$ terms appear due to   correlation between $S$ and $n.$

\section{Block mixing}\label{mix}

\subsection{Zero magnetic field, \underline{$B=0$}.} \label{zero}

Let us now assume that blocks are mixed by weak perturbation.  We
consider only positive energies $E>M,$ so that the Hilbert space is
limited to the states
$\chi_\mathbf{k}^{(\text{I},+)},
\chi_\mathbf{k}^{(\text{II},+)}.$ It is convenient to redefine these
states multiplying
$\chi_\mathbf{k}^{(\text{I},+)}$ by a phase factor
$e^{i\phi_\mathbf k}.$
Hence, the spinors corresponding to the states
in two blocks with positive energies are chosen as follows:
\be |{\rm  I}_\mathbf{k}\rangle= \frac{1}{\sqrt{1+\eta}}\
\begin{pmatrix}
e^{i \phi}\\\sqrt \eta\\0\\0
\end{pmatrix},
~~|{ \rm II}_\mathbf{k}\rangle=
\frac{1}{\sqrt{1+\eta}}\
\begin{pmatrix}
0\\0\\1\\
-\sqrt \eta e^{i \phi}
\end{pmatrix}.
\label{chi12}
\ee
The general form of a perturbation which mixes blocks and conserves the time-reversal
symmetry is presented in Ref.~\onlinecite{ostrWAL}. In the following, we assume the simplest form of
the mixing potential:
\begin{equation}\label{Vmix}
\hat{V}=V(\mathbf r)\left[\begin{array}{cccc} 1 & 0 & 0& -\Delta\\
0 & 1& \Delta & 0\\
0 & \Delta^* & 1 &0\\ -\Delta^* &0&0&1
\end{array}\right],
\end{equation}
where $V(\mathbf r)$ is a short-range potential with the correlation function
given by Eq.~\eqref{corrW} and $\Delta$ is a parameter responsible for the block mixing.
We assume that $\Delta $ is small  and real: $$\Delta \ll 1,~\Delta=\Delta^*.$$
The potential~\eqref{Vmix} obeys the symmetry: $\hat V=\hat U \hat V^* \hat U^{-1},$ where
\begin{equation}\label{U}
\hat{U}=\left[\begin{array}{cccc} 0 & 0 & -i& 0\\
0 & 0& 0 & -i\\
i & 0& 0 &0\\ 0 &i&0&0
\end{array}\right],~~\hat U \hat U^*=-1.
\end{equation}

The matrix elements of $\hat V$ read:
\bee
&&\langle {\rm I}_\mathbf k| \hat V | {\rm I}_{\mathbf k'}\rangle
= V_{\mathbf k \mathbf k'} \frac{\eta +e ^{i(\phi'-\phi)}}{1+\eta},
\nonumber \\
&&\langle {\rm II}_\mathbf k| \hat V | {\rm II}_{\mathbf k'}\rangle
= V_{\mathbf k \mathbf k'} \frac{1+\eta e ^{i(\phi'-\phi)}}{1+\eta},
 \nonumber \\
&&\langle {\rm I}_\mathbf k| \hat V | {\rm II}_{\mathbf k'}\rangle
= \langle {\rm II}_\mathbf k| \hat V | {\rm I}_{\mathbf k'}\rangle
\nonumber \\
&& =V_{\mathbf k \mathbf k'} \frac{1+ e ^{i(\phi'-\phi)}}{1+\eta} \Delta \sqrt{\eta}.
\label{mat}
\eee
Using these equations one can derive the equation for the Cooperon using standard diagrammatic rules.
However, one can see that this representation is inconvenient because the single-particle Green's
functions (calculated in the self-consistent Born approximation) turns out to be matrices
$2\times 2.$ It is more convenient to make a unitary transformation which diagonalizes
the Green's functions. Such a transformation looks like
\be
|{\rm 1}_\mathbf{k}\rangle = \frac{|{\rm I}_\mathbf{k}\rangle
+ |{\rm II}_\mathbf{k}\rangle}{\sqrt 2 },~~|{\rm 2}_\mathbf{k}\rangle
= \frac{|{\rm I}_\mathbf{k}\rangle - |{\rm II}_\mathbf{k}\rangle}{\sqrt 2 }.
\label{new}
\ee
Derivation of kinetic equation for the Cooperon in the new basis is presented in Appendix~\ref{block1}.
This equation has a matrix form:
\bee \label{Cooperonnm} &&
\left[1/\tau_\phi + i \mathbf q \mathbf v_F +\hat \gamma_D   \right]\hat C_{\mathbf{q}}(\phi,\phi_0)
\\ &&= \int \frac{d\phi'}{2\pi} \hat \gamma_C(\phi-\phi') \hat C_{\mathbf{q}}(\phi',\phi_0)
+\gamma \hat I \delta(\phi-\phi_0),
\nonumber
\eee
where $\hat I$ is the unit matrix $4\times 4,$ and the matrix  $\hat\gamma_C(\phi) $ contains three angular  harmonics
\be
\hat\gamma_C(\phi)=\hat \gamma_{0}+\hat \gamma_{-1}e^{-i\phi} +\hat \gamma_{-2}e^{-2i\phi}.
\label{gammac}
\ee
The expressions for matrices $\hat \gamma_{0}, ~\hat \gamma_{-1},~\hat \gamma_{-2}$ and $\hat \gamma_D$ are presented in the Appendix~\ref{block1}.

The conductivity correction  is expressed in terms of the matrix return probability as follows
\be
\label{sigma-matrix}
\delta \sigma_{tot}=-\frac{e^2}{\hbar} \frac{l_{tr}^2}{\gamma}\int\frac{d\phi}{2\pi} {\rm Tr} \left [ \hat \gamma(\pi -\phi)\hat
\xi
\hat W(\phi)   \right ] (1+\cos \phi),
\ee
where
\be
\hat W(\phi-\phi_0)=\int \frac{d^2\mathbf q}{(2\pi)^2}\hat C_{\mathbf{q}}(\phi,\phi_0),
\label{Wphi11}
\ee
and
\be
\label{xi}
\hat \xi=\left[\begin{array}{cccc}
1 & 0 & 0& 0\\
0 & 1& 0 & 0 \\
0 & 0& -1 &0\\
0 & 0 &0& 1
\end{array}\right].
\ee

Expanding $\hat W(\phi)$ in the Fourier series
\be \label{Wnm}
 \hat W(\phi)=\frac{1}{2 \pi l^2}\sum_{M=-\infty}^{\infty} \hat w_M e^{i(M-1)\phi},
 \ee
we can write the conductivity correction in a form similar to Eq.~\eqref{sigma2}:
\bee \label{sigmanm}
&& \delta\sigma_{tot}=-\frac{e^2}{2\pi\hbar}
\left(\frac{l_{tr}}{l}\right)^2\frac{1}{\gamma}{\rm Tr}\left[\frac{\hat\gamma_{-2} }{2}\hat\xi \hat w_{-2} \right.
\\
\nonumber &&
+\left(\hat\gamma_{-2}-\frac{\hat\gamma_{-1}}{2}\right)\hat \xi \hat w_{-1}
 -\left(\hat\gamma_{-1}-\frac{\hat\gamma_{0}+\hat\gamma_{-2}}{2}\right)\hat \xi \hat w_{0} \\ &&\left.
 +\left(\hat\gamma_{0}-\frac{\hat\gamma_{-1}}{2}\right)\hat \xi \hat w_{1}
 +\frac{\hat\gamma_{0} }{2}\hat\xi \hat w_{2}
 \right].
 \nonumber \eee
Since $\Delta \ll 1,$  the matrices standing in fronts of $\hat w_n$ can be calculated for $\Delta=0.$
They are written down in Appendix~\ref{block2}.

For $q=0,$ Eq.~\eqref{Cooperonnm} is easily solved by expansion
$\hat C_{\mathbf{q}}(\phi,\phi_0)$ in the Fourier series over $\exp(i
M\phi).$ Doing so, one can find ``masses'' of the diffusive modes which
are given by the eigenvalues of the matrices $\hat\gamma_D-\hat
\gamma_M$ ($M=0,-1,-2$).  These matrices are presented in
Appendix~\ref{block3}. For $q\neq 0,$ different harmonics couple with
each other. In the diffusion approximation, when $ql \ll 1,$ the
coupling is weak and the mode $e^{i M \phi}$ is only effectively
coupled with the nearest modes $e^{i (M\pm 1) \phi}.$ The main
formulas describing diffusion approximation in the presence of the
block mixing are quite similar to those obtained in
Sec.~\ref{diffusion}. We present them in Appendix~\ref{diffusion2}.

Next, we demonstrate that the interblock mixing leads to appearance of two
additional singular modes which do not show up in the absence of
mixing.  To this end let us analyze eigenvalues of the matrix
$\hat\gamma_D-\hat \gamma_{-1}.$ This matrix is diagonal, so that its
eigenvalues are given by the diagonal elements. The matrix element
$(\hat\gamma_D-\hat \gamma_{-1})_{44}$ is exactly equal to zero, while
the element $(\hat\gamma_D-\hat \gamma_{-1})_{33}=4\eta \Delta^2$
turns to zero for $\Delta=0$ when inter-block transitions are absent.
Hence, there are two modes which are singular (gapless for any $\eta$)
in the limit $\Delta \to 0.$ We will call these modes
inter-block singular modes (ISM).  In Appendix~\ref{IM} we present
calculation of the contribution of ISM to the conductivity and also
find equations describing limiting cases $\eta \to 0$ or $\eta\to 1.$
Below, we summarize the results of the calculations.
For $b=0,$ conductivity correction coming from  ISM reads
\be
\label{sigma(1)1}
\delta \sigma_{\rm{ISM}}= \frac{e^2}{4\pi^2\hbar} \left[ \ln \left( \frac{1}{\Gamma}\right)- \ln \left( \frac{1}{\Gamma +4\Delta^2\eta/(1+\eta^2)}\right)\right].
\ee
 We see that  contributions  of two modes exactly cancel each other in the limit $\Delta \to 0.$
This explains why these modes do not show up in the model of independent blocks.
In the limit $\Gamma \to 0,$ first logarithm dominates. It represents a standard singlet contribution
responsible for weak antilocalization.

 Importantly, Eq.~\eqref{sigma(1)1} is valid in the whole interval $0<\eta<1,$ because the
applicability of the diffusion approximation which was used in our derivation is
guaranteed by the smallness of  dimensionless gaps of the diffusive modes: $\Gamma \ll 1 $
and $\Gamma +4\Delta^2\eta/(1+\eta^2) \ll 1.$  Next, we notice  that other modes,
discussed in previous sections are not strongly affected by block mixing provided
that $\eta$ is not too close to $0$ or $1.$
This allows one to find  analytical expression for total conductivity correction valid
in the whole interval $0<\eta<1,$ except narrow regions near points $\eta=0$ and $1$:
 \be
 \delta \sigma_{tot} = \delta \sigma_{\rm{ISM}}+2\delta\sigma_{\text I \text I},
 \label{totalb=0}
 \ee
 where $\delta \sigma_{\text I \text I}$ is given by Eq.~\eqref{sigma2}
with $w_n$ determined by Eq.~\eqref{wn} and coefficient $2$ in front of $\delta\sigma_{\text I \text I}$ accounts
for the contribution of two blocks. To avoid confusion we stress again that adding
expression for  $ \delta \sigma_{\rm{ISM}}$ obtained in the diffusion approximation to the ballistic
contribution $2\delta\sigma_{\text I \text I}$ is well-controlled because
for small $\Delta $ and $\Gamma$ gaps of ISM are much smaller than $\gamma.$

\subsection{Nonzero magnetic field, $B\neq 0.$ }\label{nonzero}

In a finite magnetic field, the contribution of ISM becomes
\bee \label{sigma111}
&&\delta \sigma_{\rm{ISM}}=-\frac{e^2}{4\pi^2\hbar}
\nonumber
\\
&&\nonumber
\times \sum\limits_{n=0}^N\left[ \frac{|b|}{|b|(n+1/2) +A_1} -\frac{|b|}{|b|(n+1/2) +A_2}\right]\\
&=&-\frac{e^2}{4\pi^2\hbar}\left[h\left(|b|,{A_1}\right)-h\left(|b|,{A_2}\right)\right]
\label{psipsi}
\eee
where
\be
A_1=\left(\Gamma+ \frac{4\eta\Delta^2}{1+\eta^2}\right) \frac{1+\eta^2-\eta}{2(1+\eta^2)},~~A_2=\Gamma \frac{1+\eta^2-\eta}{2(1+\eta^2)}.
\ee
Next, we write
\bee
\label{dism}
&&\Delta\sigma_{\rm{ISM}}=\delta \sigma_{\rm{ISM}}(b)-\delta \sigma_{\rm{ISM}}(0)
\\\nonumber &&=\frac{e^2}{4\pi^2\hbar}\left[h\left(|b|,{A_2}\right)-h\left(|b|,{A_1}\right) -h\left(0,{A_2}\right)+h\left(0,{A_1}\right)\right].
\eee
and use Eq.~\eqref{sum} to find asymptotes of Eq.~\eqref{dism} for the case $\eta \Delta^2 \lesssim \Gamma$:
\bee
\label{ism1}
&& \Delta \sigma_{ism}\approx-\frac{e^2}{4\pi^2\hbar}  \\ \nonumber &&\times
\left\{\begin{array}{ll} \displaystyle
\frac{b^2}{6 }\frac{(1+\eta^2)^2}{(1+\eta^2-\eta)^2}\left[   \frac{1}{\Gamma^2}  -  \frac{1}{ \left(\Gamma+\frac{4\eta\Delta^2}{1+\eta^2}\right)^2} \right]
,\\  \text{for} ~~|b|\ll \Gamma, ~~\vspace{6mm}  \\  \displaystyle
-\frac{\pi^2\eta\Delta^2(1+\eta^2-\eta)}{(1+\eta^2)^2|b|},~~ \text{for}~~
|b| \gg \Gamma.  \end{array} \right.
\eee
For  stronger  interblock coupling  $\eta \Delta^2 \gg \Gamma,$ we get
\bee \label{ism2}
&& \Delta \sigma_{\rm{ISM}}\approx-\frac{e^2}{4\pi^2\hbar}  \\ \nonumber &&\times
\left\{\begin{array}{ll} \displaystyle
\frac{b^2}{6 }\frac{(1+\eta^2)^2}{(1+\eta^2-\eta)^2\Gamma^2}
,~~  \text{for} ~~|b|\ll \Gamma, ~~\vspace{6mm}  \\  \displaystyle  \ln\left(\frac{|b|}{\Gamma}\right)- \frac{b^2}{96 }\frac{(1+\eta^2)^4}{(1+\eta^2-\eta)^2\eta^2\Delta^4}
,\\ \text{for}~~
\Gamma \ll |b| \ll \eta\Delta^2.
~~\vspace{6mm}  \\  \displaystyle
-\frac{\pi^2\eta\Delta^2(1+\eta^2-\eta)}{(1+\eta^2)^2|b|}-\ln\left(\frac{\Gamma}{\eta\Delta^2}\right),\\ \text{for}~~
|b| \gg \eta\Delta^2.  \end{array} \right.
\eee

The total conductivity correction reads as
\be
\delta\sigma_{tot}=\delta \sigma_{\rm{ISM}}+\delta\sigma_{\text I}+\delta \sigma_{\text I \text I},
\label{totalbneq0}
\ee
where $\delta \sigma_{\text I,\text I \text I}$ are given by Eq.~\ref{sigma2} with $w_n$
determined by Eqs.~\eqref{wnB} and \eqref{wnB1}, respectively (see Sec.~\ref{another}).
Equation \eqref{totalbneq0} yields the most general  expression for conductivity correction in magnetic field.
Just in the absence of field, expression $\delta \sigma_{\rm{ISM}}(b)$ was found in the diffusion approximation,
while $\delta \sigma_{\text I}$,  and $\delta\sigma_ {\text I \text I}$ are calculated
 by using exact ballistic formulas. Such an approach is well controlled
for small $\Delta$ and $\Gamma$ provided that $\eta$ is not too close to $0$ or $1$.

Let us now consider the behavior of the conductivity correction near points $\eta=0$ and $1.$
At zero field, in the absence of block mixing, the  total conductivity correction  is given by
Eqs.~\eqref{ds} and \eqref{ds1}, respectively, multiplied by a factor $2$ which accounts for
the contribution of the block ${\rm I}$.  Block mixing  slightly modifies these equations
because neglect of $\Delta$ in $\delta \sigma_{ \text I}$  and $\delta \sigma_{\text I \text I}$ is no longer justified.
On the other hand, calculation of conductivity near these points can be essentially
simplified because these are the points where the diffusion approximation works well.

\underline{(a)~$\eta \to 0.$} As shown in Appendix \ref{IM},
in the presence of the block mixing instead of Eq.~\eqref{ds}  (multiplied by the factor $2$) we get the  following equation
\be
\delta \sigma_{\eta} \approx -\frac{e^2}{2\pi^2\hbar} \ln\left ( \frac{1}{\Gamma+\eta^2 +2\eta\Delta^2}\right).
\label{sigmaeta}
\ee
which accounts for two weakly mixed blocks at $b=0$. The total conductivity is given by the sum of
$\delta \sigma_{\rm{ISM}}$  and $\delta\sigma _\eta $
\bee
\nonumber
&&\delta \sigma_{tot}=\delta\sigma_\eta+\delta \sigma_{\rm{ISM}}\approx
\frac{e^2}{4\pi^2\hbar}\left[ \ln \left(\frac{1}{\Gamma}\right)-\ln\left(\frac{1}{\Gamma+4\Delta^2\eta}\right)\right.
\\
&&\left. -2 \ln\left(\frac{1}{\Gamma+\eta^2+2\eta\Delta^2}\right)\right],
\label{orth}
\eee
This equation is valid, provided that $\eta \ll 1.$ In the interval
$ \Delta^2 \ll \eta \ll 1 $ it  matches  with Eq.~\eqref{totalb=0}.

The variation of the total conductivity correction with the magnetic field
$\Delta \sigma_{tot}(b)=\delta \sigma_{tot}(b) -\delta \sigma_{tot}(0)$ is presented as a sum of two terms:
\be
\Delta \sigma_{tot}=  \Delta \sigma_{\rm{ISM}}+\Delta \sigma_{\eta}
\label{delta-sigma-tot}
\ee
where $\Delta \sigma_{\rm{ISM}}$ is given by Eqs.~\eqref{dism}, \eqref{ism1}, and \eqref{ism2}, whereas
\bee \label{eta}
&&\Delta \sigma_{\eta}=\delta \sigma_{\eta}(b)-\delta \sigma_{\eta}(0)= -\frac{e^2}{4\pi^2\hbar}
\\&& \nonumber \times \left[ h\left( |b|, \frac{\Gamma+\eta^2-\eta b+2\eta\Delta^2}{2}\right) \right.
\\&& \nonumber
+h\left( |b|, \frac{\Gamma+\eta^2+\eta b+2\eta\Delta^2}{2}\right)
\\&& \nonumber
\left. -2 h\left( 0, \frac{\Gamma+\eta^2+2\eta\Delta^2}{2}\right)\right]
\\&& \nonumber
\approx -\frac{e^2}{2\pi^2\hbar}
\\  &&\times
\left\{\begin{array}{ll} \displaystyle
-\frac{b^2}{6 }   \frac{1}{ \left(\Gamma+\eta^2 +2\eta\Delta^2\right)^2}
,\ \ & \text{for} ~~|b|\ll \Gamma+\eta^2 +2\eta\Delta^2,  ~~\vspace{2mm}  \\  \displaystyle
\ln\left( \frac{\Gamma+\eta^2 +2\eta\Delta^2}{|b|}\right),\ \  & \text{for}~~
|b| \gg  \Gamma+\eta^2 +2\eta\Delta^2 .  \end{array} \right..
\nonumber
\eee

The low-field asymptotic of $\Delta \sigma_{tot}$ deserves special attention.
From Eqs.~\eqref{ism1} and \eqref{eta} we find  for $|b|\to 0,~\eta\ll 1:$
\bee
&& \label{low} \Delta \sigma_{tot}\approx \frac{e^2 b^2}{12\pi^2\hbar} \\ \nonumber
&&\times \left[ \frac{1}{( \Gamma+\eta^2+2\eta\Delta^2)^2}-\frac{4\eta\Delta^2(\Gamma+2\eta\Delta^2)}{\Gamma^2(\Gamma+4\eta\Delta^2)^2} \right].
\eee
 We see that coefficient in the square bracket changes sign with increasing of $\eta$ or $\Delta.$
Consequently, the magnetoconductivity also changes sign and becomes negative.

\underline{(b)~$\eta \to 1.$} Instead of Eq.~\eqref{ds1} (multiplied by the factor $2$) we get the following equation:
\bee \label{sigma(2)}
&&\delta \sigma_{1-\eta}=
 \frac{e^2}{4\pi^2\hbar} \left[ \ln \left( \frac{1}{\Gamma+(1-\eta)^2/2}\right) \right.
\\
&& \left.+ \ln \left( \frac{1}{\Gamma + (1-\eta)^2/2+2\Delta^2}\right)\right].
\nonumber
\eee
which should be added to  $\delta\sigma_{\rm{ISM}}$:
\bee \label{sym}
&& \delta \sigma_{tot}=\delta\sigma_{1-\eta}+\delta \sigma_{\rm{ISM}}\\
\nonumber
&&\approx \frac{e^2}{4\pi^2\hbar}\left[ \ln \left(\frac {1}{\Gamma}\right)-\ln\left(\frac{1}{\Gamma+2\Delta^2}\right) \right.
\\
&& \left.+ \ln\left(\frac{1}{\Gamma+(1-\eta)^2/2+2\Delta^2} \right) + \ln \left(\frac{1}{\Gamma+(1-\eta)^2/2}\right)\right].
\nonumber
\eee
This equation for zero-field correction is valid, provided that $1-\eta \ll 1.$ In the interval  $ \Delta^2 \ll 1-\eta \ll 1 $ it  matches  with Eq.~\eqref{totalb=0}.

The variation of the total conductivity correction with the magnetic field is written as
\be
\Delta \sigma_{tot}=  \Delta \sigma_{\rm{ISM}}+\Delta \sigma_{1-\eta},
\label{delta-sigma-tot1}
\ee
where $\Delta \sigma_{\rm{ISM}}$ is given by Eqs.~\eqref{dism}, \eqref{ism1}, and \eqref{ism2}, whereas
\bee \label{1-eta}
&&\Delta\sigma_{1-\eta}=\delta \sigma_{1-\eta}(b)-\delta \sigma_{1-\eta}(0)=\frac{e^2}{8\pi^2\hbar}
\\\nonumber &&\times \left\{ h\left[ |b|, \frac{\Gamma+(1-\eta)^2/2 + 2 b (1-\eta)}{4}\right]\right.\\&& \nonumber\left.+h\left[ |b|, \frac{\Gamma+(1-\eta)^2/2+2 b (1-\eta) +2\Delta^2}{4}\right]\right.
\\\nonumber &&
+ h\left[ |b|, \frac{\Gamma+(1-\eta)^2/2 - 2 b (1-\eta)}{4}\right]
\\&& \nonumber+h\left[ |b|, \frac{\Gamma+(1-\eta)^2/2-2 b (1-\eta) +2\Delta^2}{4}\right]
\\\nonumber && - 2 h\left[ 0, \frac{\Gamma+(1-\eta)^2/2}{4}\right]
\\\nonumber &&
\left.
- 2 h\left[ 0, \frac{\Gamma+(1-\eta)^2/2 +2\Delta^2}{4}\right]\right\}
\\&&\approx \frac{e^2}{4\pi^2\hbar} \nonumber\\
\nonumber && \times
\left\{\begin{array}{ll} \displaystyle
-\frac{2b^2}{3 }\left\{   \frac{1}{\left[\Gamma+\frac{(1-\eta)^2}{2}\right]^2}  +  \frac{1}{ \left[\Gamma+\frac{(1-\eta)^2 }{2}+2\Delta^2\right]^2} \right\}
,\\  \text{for} ~~|b|\to 0 ~~\vspace{3mm}  \\  \displaystyle
\ln\left\{ \frac{\left[\Gamma+\frac{(1-\eta)^2}{2}\right]\left[\Gamma+\frac{(1-\eta)^2}{2}+2\Delta^2\right] }{b^2}\right\}  , \\  \text{for}~~
|b| \gg \Gamma+(1-\eta)^2/2+2\Delta^2  .  \end{array} \right.
\nonumber
\eee

\subsection{Crossover between ensembles}
Let us discuss results obtained in Secs. \ref{zero} and \ref{nonzero}
in the context of crossover between orthogonal and symplectic ensembles.

For an arbitrary value of $\eta$ between $0$ and $1$  and $\Delta \neq 0,$ all
symmetries are broken except time-reversal symmetry (see Ref.~\onlinecite{ostrWAL}
for detailed discussion of symmetries existing in the system), so that in the absence of dephasing ($\Gamma=0$)
there exists  only   one gapless mode.  This mode gives rise to antilocalizing conductivity
correction  described by the  first logarithm in Eq.~\eqref{sigma(1)1}.

In the limit $\eta \to 0,$ the spin degree of freedom is irrelevant, so that
we have two copies  of a
system with orthogonal symmetry. Indeed, as seen from Eq.~\eqref{orth} for $\eta=0$ we have two (due to the spin degeneracy) gapless
modes  yielding the WL correction to the conductivity [described by the last logarithm in Eq.~\eqref{orth}].

The case $\eta \to 1$ turns out to be more subtle. Indeed, one may
expect that in this case all symmetries are broken because of the
inter-block transition, and one may conclude that there exists a
single copy of a system with symplectic symmetry corresponding to a
divergent (in the limit $\Gamma\to 0$) logarithm. However, such a
conclusion is not supported by our calculations. Indeed, as seen from
Eq.~\eqref{sym}, in addition to always singular mode [one of the ISM
described by the first term in Eq.~\eqref{sym}] there also exists a
mode which becomes singular when $\eta$ becomes exactly equal to $1.$
This mode corresponds to the last term in Eq.~\eqref{sym}. In other
words, for $\eta =1$ there exist two copies of the symplectic
ensembles instead of one expected. The physical explanation of this
fact follows from analysis of the inter-block matrix
element. Specifically, from the last line of Eq.~\eqref{mat1} we see
that in the limit $\eta \to 1,$ the inter-block matrix element exactly
equals to zero, which means that perturbation \eqref{Vmix} does not
effectively mix blocks and, therefore, is not sufficient to break all
symmetries except the time-reversal one.  One may expect that
this symmetry, however, is broken  by the Rashba term arising in asymmetric quantum wells (see discussion and symmetry arguments
 in Refs.~\onlinecite{ostrWAL,myPSS}), so that  only single singular
antilocalizing mode should  survive. The detailed analysis of the interference correction in the presence of
the Rashba coupling is out of the scope of this work.

\section{
Plots of  the conductivity correction
}\label{num}

As seen from equations derived in the previous sections interference
correction at zero field as well as magnetoconductivity can be
negative or positive depending on $\eta,\Gamma$ and $\Delta.$ In this
section we present corresponding pictures for different values of
parameters.

\subsection{Interference correction at zero field}
\begin{figure}[ht] \center
\includegraphics[width=0.38\textwidth]{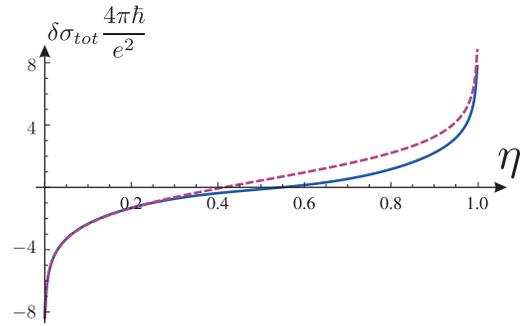}
\caption{Conductivity correction as a function of $\eta$ for infinite dephasing time. Dashed line: sum of two logarithms.
}
\label{F3}
\end{figure}

 In the absence of  magnetic field and block mixing ($b=0,\Delta=0$),
interference correction depends on two parameters, $\eta$ and
$\Gamma:$ $\delta\sigma_{tot}=\delta \sigma_{tot}(\eta,\Gamma).$
In  Fig.~\ref{F3}, the conductivity calculated
by using Eqs.~\eqref{wn}, \eqref{sigma2} and \eqref{sumb0} is plotted as a
function of $\eta$ for $\Gamma=0.$

As seen, $\delta \sigma_{tot}$ diverges at $\eta \to 0$ and $ 1,$ but remains finite for
intermediate values of $\eta.$ As a rough approximation one can describe this dependence as a sum of
two logarithmic terms, Eqs.~\eqref{ds} and \eqref{ds1} (multiplied by a factor 2 accounting for the contribution
of the block $\text I  \text I$). This  sum  is shown by a dashed line.

With increasing $\Gamma$ correction is suppressed.
This is illustrated in  detail in  Fig.~4a, where the dependence of $\delta \sigma_{tot}$ on $\Gamma$
is plotted for different $\eta$ increasing from bottom to top.
Although at  very large  $\Gamma $ correction  always decays (by absolute value), the
dependence of $\delta \sigma_{tot}$ on $\Gamma$ can be non-monotonous for some intermediate
values of $\eta$ as shown in  Fig.~4b for $\eta=0.54.$ Such a counterintuitive
behavior  aries due to the competition between localizing and antilocalizing contributions into interference correction
(an analogous result was obtained for  two-dimensional holes in semiconductor heterostructures\cite{TI71}).

\begin{figure}[h] \center
\centerline{\includegraphics[width=0.38\textwidth]{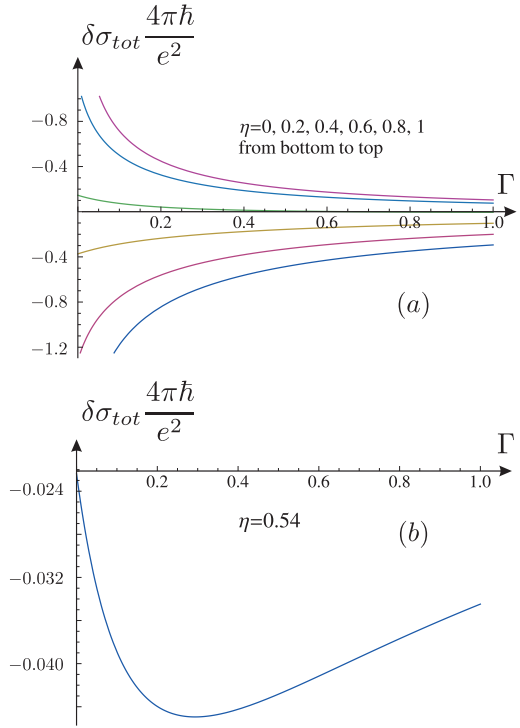} }
\caption{
 Conductivity correction as a function of dephasing rate for different values of $\eta$ $(a)$.
Conductivity correction as a function of dephasing rate for $\eta=0.54$ $(b)$.}
\label{F4}
\end{figure}
\subsection{Magnetoconductivity in a single cone}

\begin{figure}[h]\center
\centerline{\includegraphics[width=0.38\textwidth]{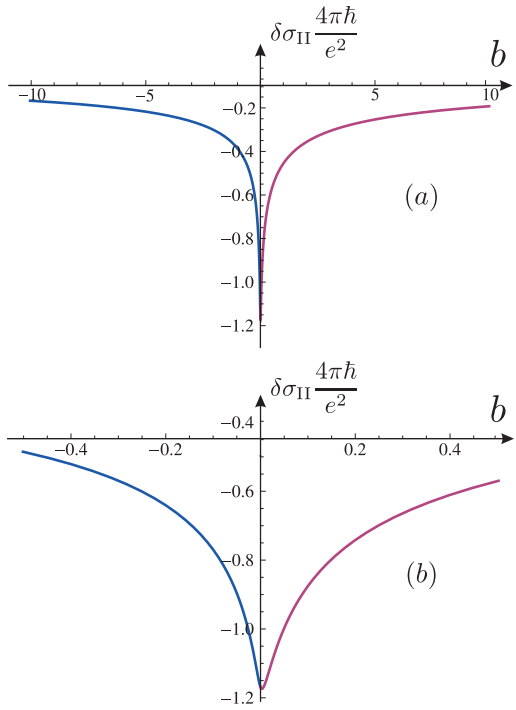} }
\caption{
 Magnetoconductivity within the block II in the intervals   $-10<b<10$ (a) and  $-0.5<b<0.5$ (b) for $\eta=0.1,~\Gamma=0.01.$}
\label{F5}
\end{figure}

\begin{figure}[H]\center
\centerline{\includegraphics[width=0.38\textwidth]{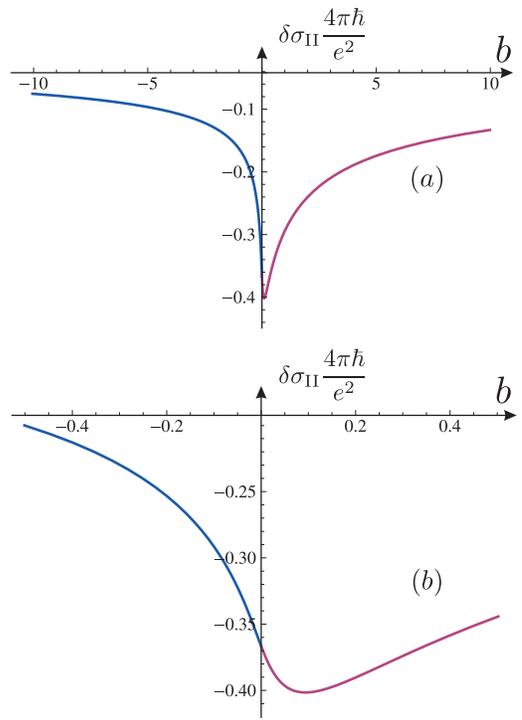} }
\caption{
Magnetoconductivity  within the block II  in the intervals    $-10<b<10$ (a) and  $-0.5<b<0.5$ (b) for $\eta=0.3,~\Gamma=0.01.$}
\label{F6}
\end{figure}

\begin{figure}[H]\center
\centerline{\includegraphics[width=0.38\textwidth]{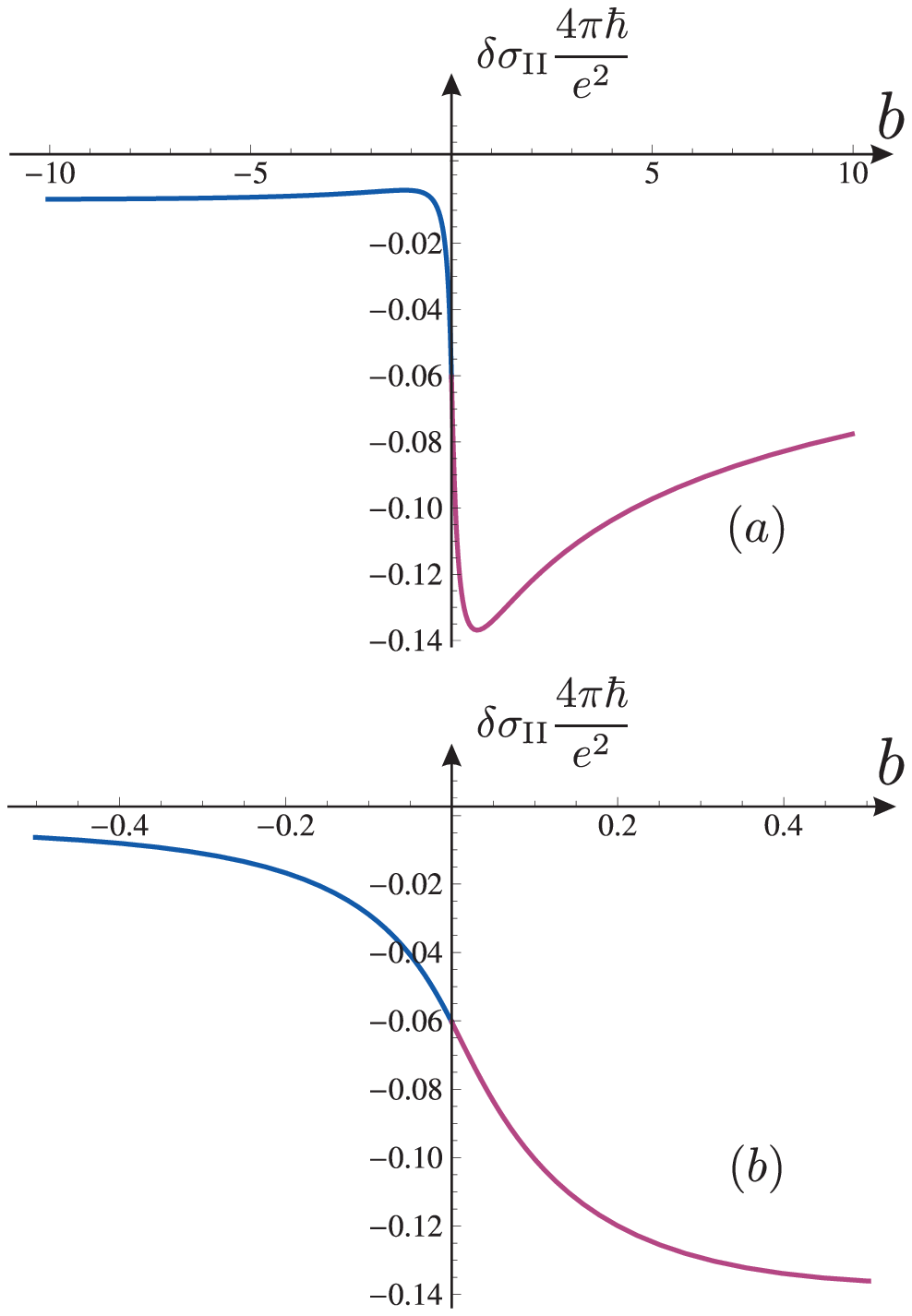} }
\caption{
Magnetoconductivity  within the block II  in the intervals    $-10<b<10$ (a) and  $-0.5<b<0.5$ (b) for $\eta=0.5,~\Gamma=0.01.$}
\label{F7}
\end{figure}

\begin{figure}[H]\center
\centerline{\includegraphics[width=0.38\textwidth]{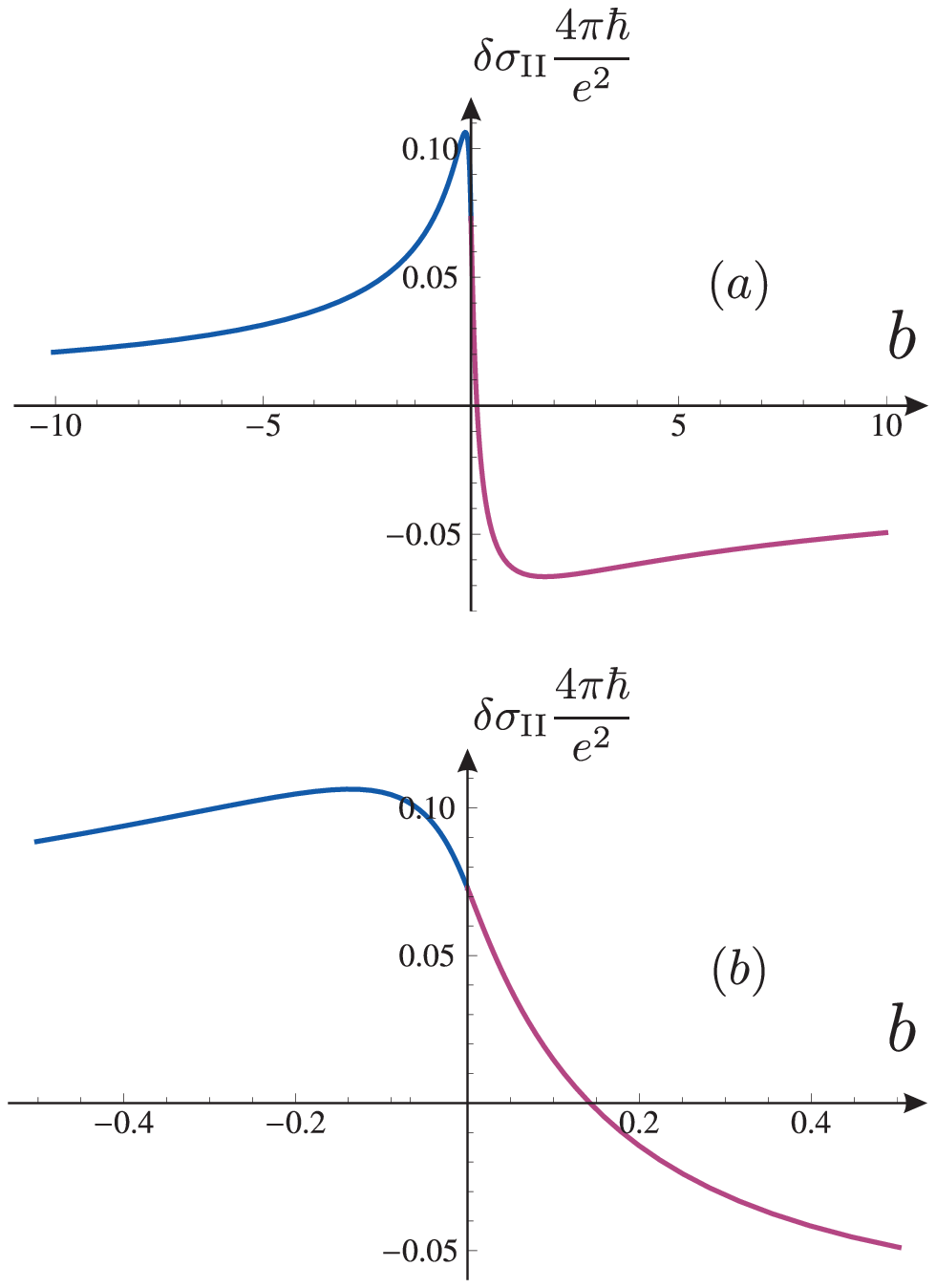} }
\caption{
Magnetoconductivity  within the block II  in the intervals    $-10<b<10$ (a) and  $-0.5<b<0.5$ (b) for $\eta=0.6,~\Gamma=0.01.$}
\label{F8}
\end{figure}

\begin{figure}[H]\center
\centerline{\includegraphics[width=0.38\textwidth]{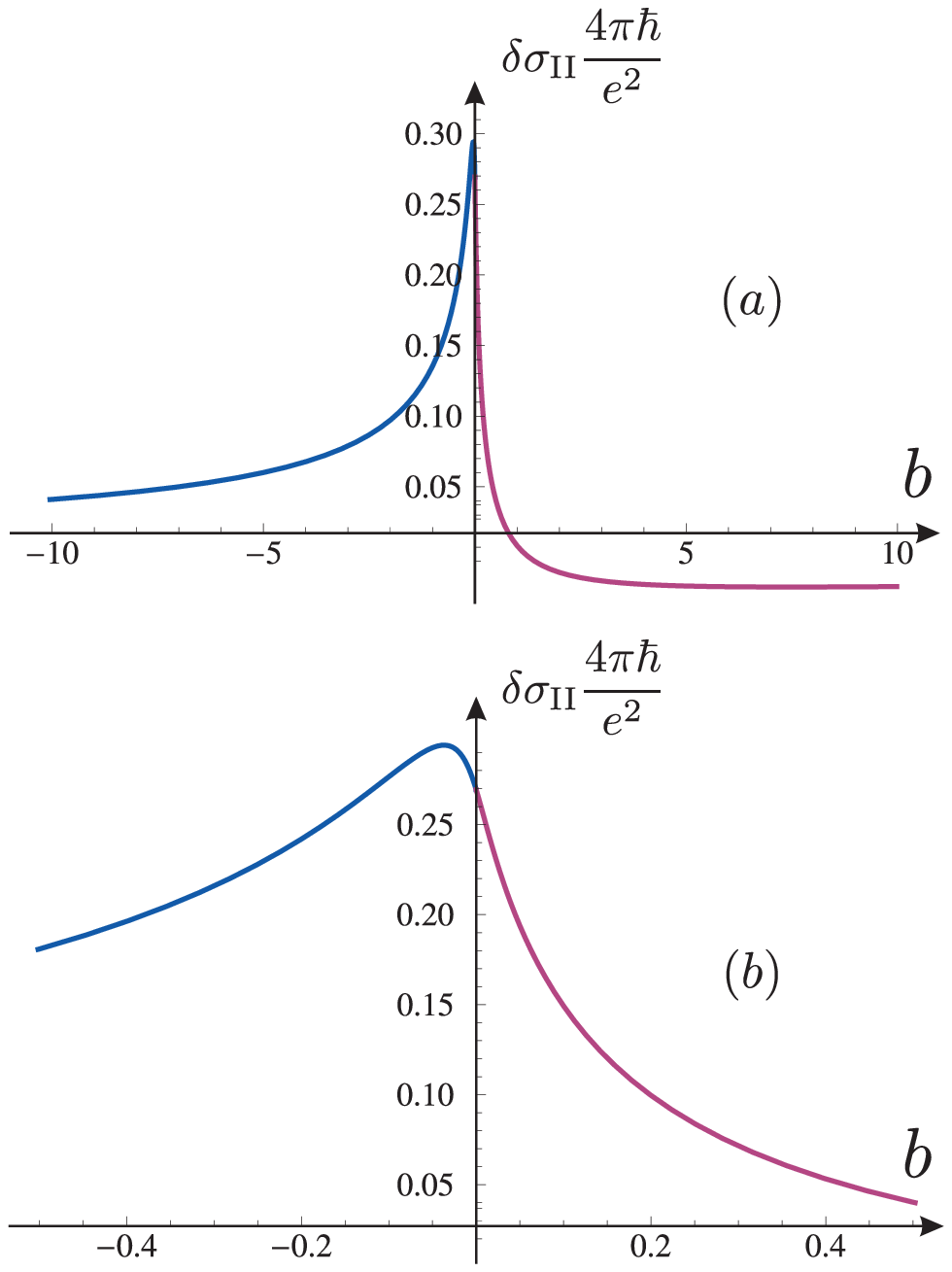} }
\caption{
Magnetoconductivity  within the block II in the intervals    $-10<b<10$ (a) and  $-0.5<b<0.5$ (b) for $\eta=0.7,~\Gamma=0.01.$}
\label{F9}
\end{figure}

\begin{figure}[H]\center
\centerline{\includegraphics[width=0.38\textwidth]{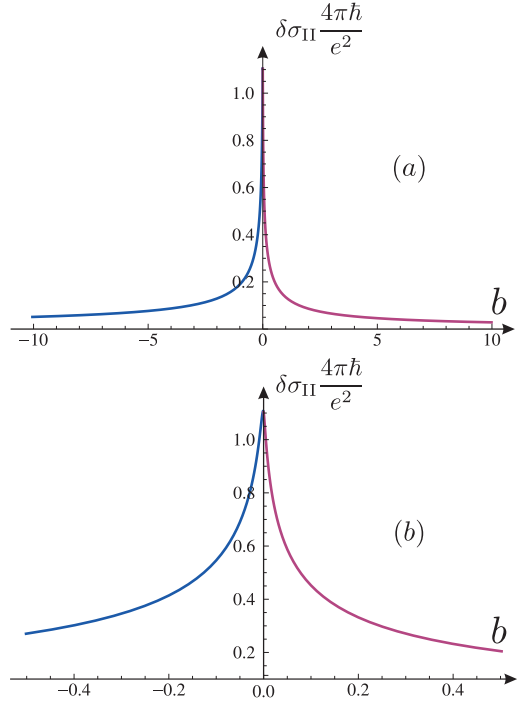} }
\caption{
Magnetoconductivity  within the block II  in the intervals   $-10<b<10$ (a) and  $-0.5<b<0.5$ (b) for $\eta=0.9,~\Gamma=0.01.$}
\label{F10}
\end{figure}

In this section we present results of numerical simulations of
magnetoconductivity in a single Dirac cone (block II)
for $\Delta=0$. In Figs.~\ref{F5}-\ref{F10} we
plotted dependence $\delta \sigma_{\text I \text I} (b)$ for fixed
 low dephasing rate, $\Gamma=0.01,$ and different $\eta.$
In the upper panels of these pictures $\delta\sigma_{\text I \text I} (b)$ is plotted within the interval $-10<b<10,$ while
in the lower panels we plot  in more detail the
low-field behavior, $-0.5<b<0.5.$  The most important information presented in these pictures is the
asymmetry of the function  $\delta \sigma_{\text I \text I} (b).$
(Such an  asymmetry was previously found numerically in Ref.~\onlinecite{Richter12}.)
For $\eta$ close to $0$ and $1,$ the asymmetry is not that strong, however, even in these cases the peak of the
magnetoconductivity is shifted away  from the point $b=0.$
The most asymmetric curves are obtained for intermediate values of $\eta$ [see, for example,
plots of magnetoconductivity  for $\eta=0.5$ (Fig.~\ref{F7}) and $\eta=0.6$ (Fig.~\ref{F8})].


\subsection{Total magnetoconductivity in the absence of the block mixing}

\begin{figure}[H]\center
\centerline{\includegraphics[width=0.38\textwidth]{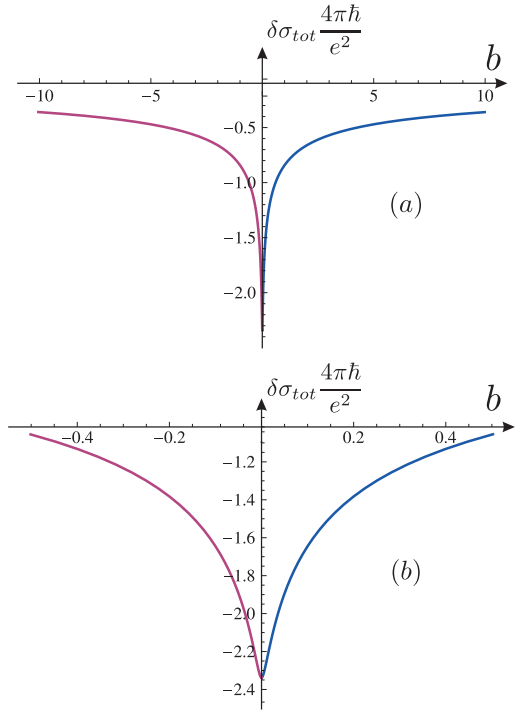} }
\caption{
 Total magnetoconductivity  in the intervals   $-10<b<10$ (a) and  $-0.5<b<0.5$ (b) for $\eta=0.1,~\Gamma=0.01.$}
\label{F11}
\end{figure}

\begin{figure}[H]\center
\centerline{\includegraphics[width=0.38\textwidth]{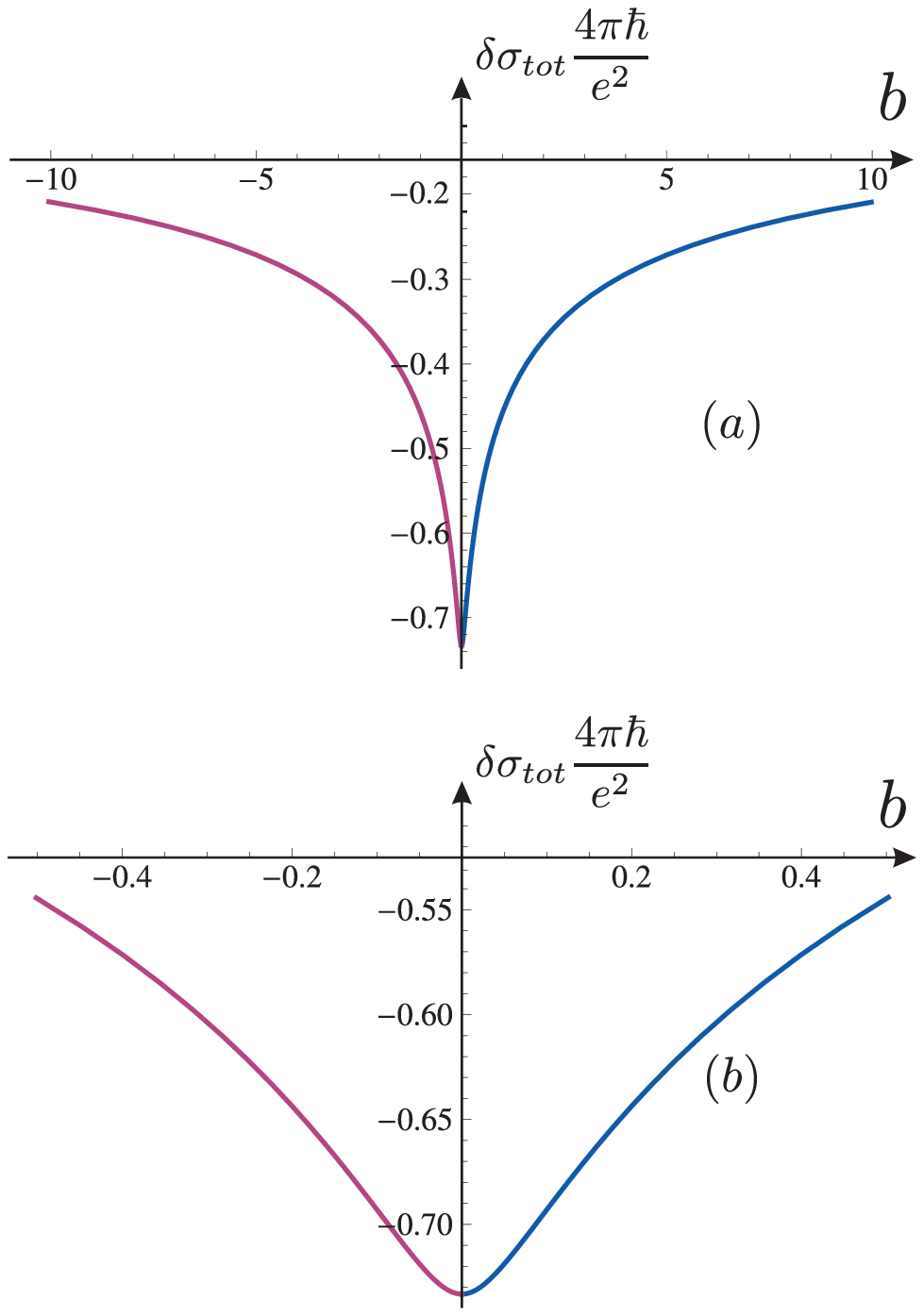} }
\caption{
Total magnetoconductivity  in the intervals    $-10<b<10$ (a) and  $-0.5<b<0.5$ (b) for $\eta=0.3,~\Gamma=0.01.$}
\label{F12}
\end{figure}

\begin{figure}[H]\center
\centerline{\includegraphics[width=0.38\textwidth]{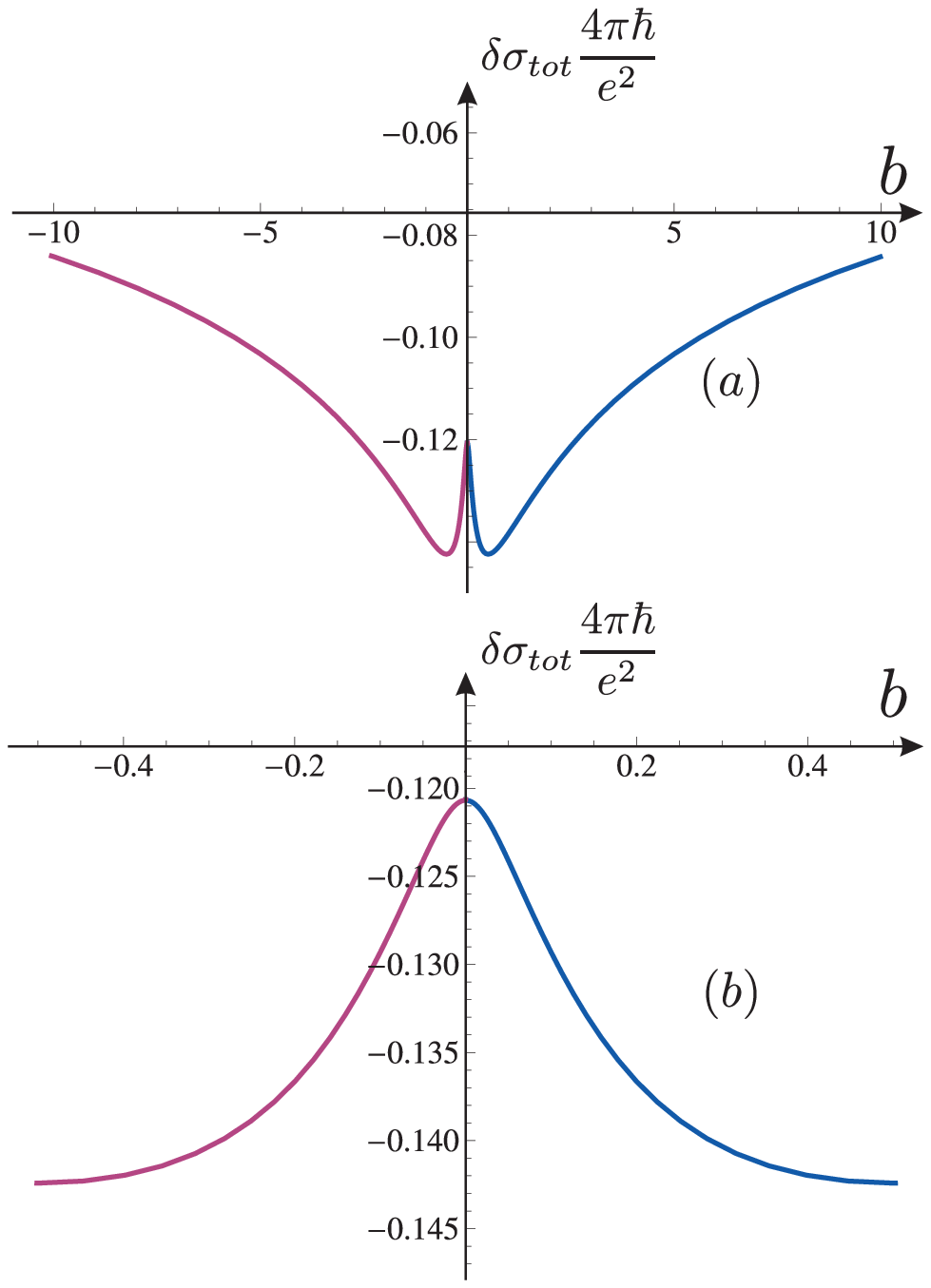} }
\caption{
Total magnetoconductivity  in the intervals    $-10<b<10$ (a) and  $-0.5<b<0.5$ (b) for $\eta=0.5,~\Gamma=0.01.$}
\label{F13}
\end{figure}

\begin{figure}[H]\center
\centerline{\includegraphics[width=0.38\textwidth]{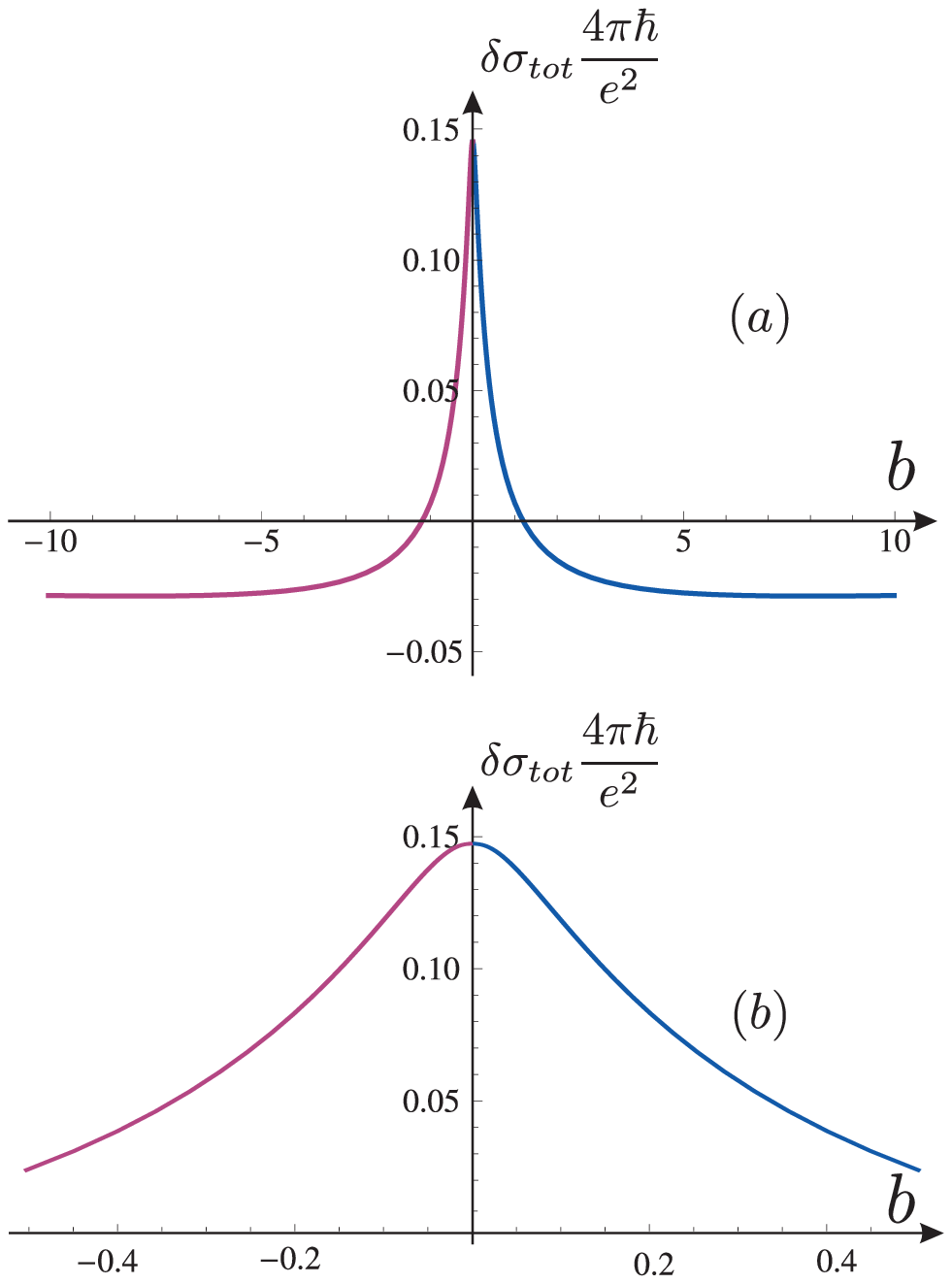} }
\caption{
Total magnetoconductivity  in the intervals    $-10<b<10$ (a) and  $-0.5<b<0.5$ (b) for $\eta=0.6,~\Gamma=0.01.$}
\label{F14}
\end{figure}

\begin{figure}[H]\center
\centerline{\includegraphics[width=0.38\textwidth]{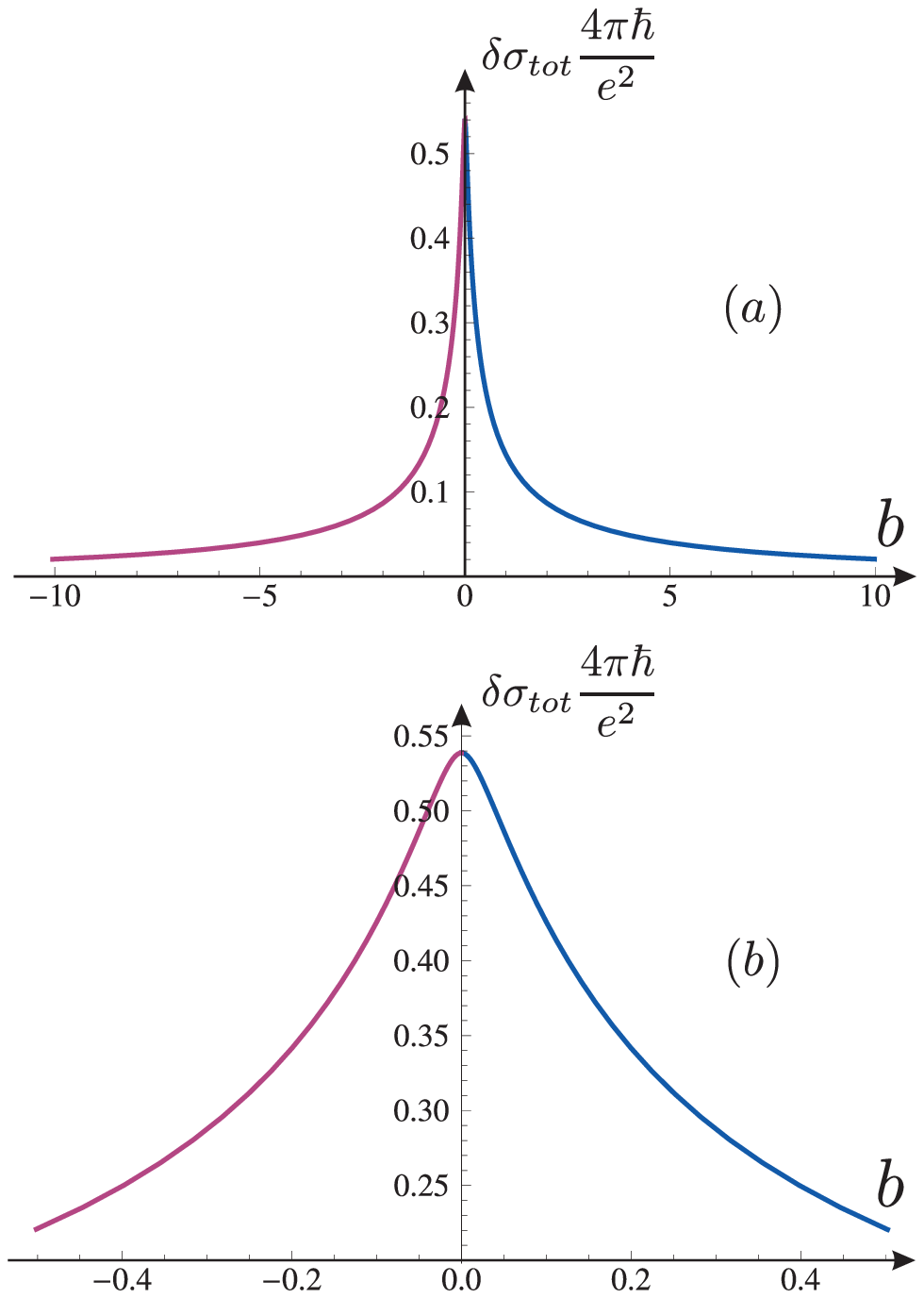} }
\caption{
Total magnetoconductivity in the intervals    $-10<b<10$ (a) and  $-0.5<b<0.5$ (b) for $\eta=0.7,~\Gamma=0.01.$}
\label{F15}
\end{figure}

\begin{figure}[H]\center
\centerline{\includegraphics[width=0.38\textwidth]{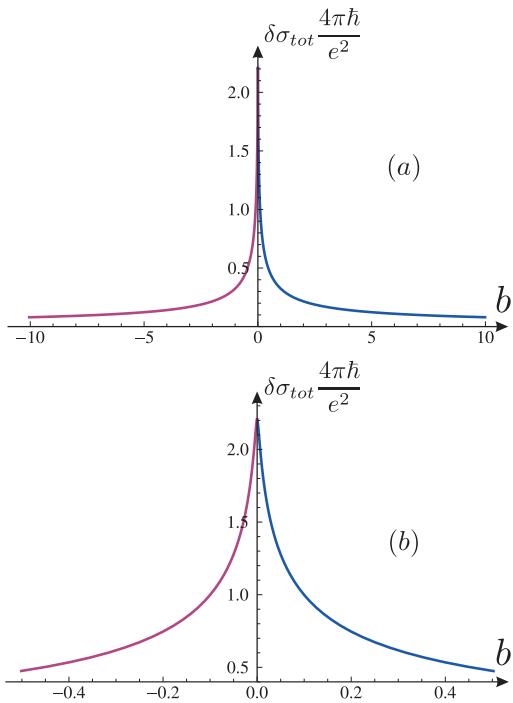} }
\caption{
Total magnetoconductivity  in the intervals   $-10<b<10$ (a) and  $-0.5<b<0.5$ (b) for $\eta=0.9,~\Gamma=0.01.$}
\label{F16}
\end{figure}

In the previous section we demonstrated that in a single Dirac cone the magnetoconducitivty
is a strongly asymmetric function of $b.$  Taking into account contribution of the second block restores
symmetry with respect to field inversion, since as mentioned above
$\delta \sigma_{\text I}(b) = \delta \sigma_{\text I \text I}(-b),$ and, consequently, the total correction,
$\delta \sigma_{tot}=\delta \sigma_{\text I}(b) + \delta \sigma_{\text I \text I}(b),$  is an even function of $b.$
  As an example, let us consider the dependence of $\delta \sigma_{tot} $ on
$b$ in the absence of block mixing ($\Delta=0$)  for $\Gamma=0.01$  and different values of
$\eta$ (see Figs.~\ref{F11}-\ref{F16}).
Comparing this curve with Figs.~\ref{F5}-\ref{F10} plotted for the same parameters  but for a single block, we see that
adding of the contribution of the second block  shifts the minimum of the conductivity back to the point $b=0.$

  The most interesting result is obtained for intermediate values of $\eta$. In particular,
for $\eta=0.5,$ we see two   minima in
the low-field region symmetrical with respect to point $b=0.$

\subsection{Total magnetoconductivity in the presence of the block mixing}

\begin{figure} [H] \center
\centerline{\includegraphics[width=0.38\textwidth]{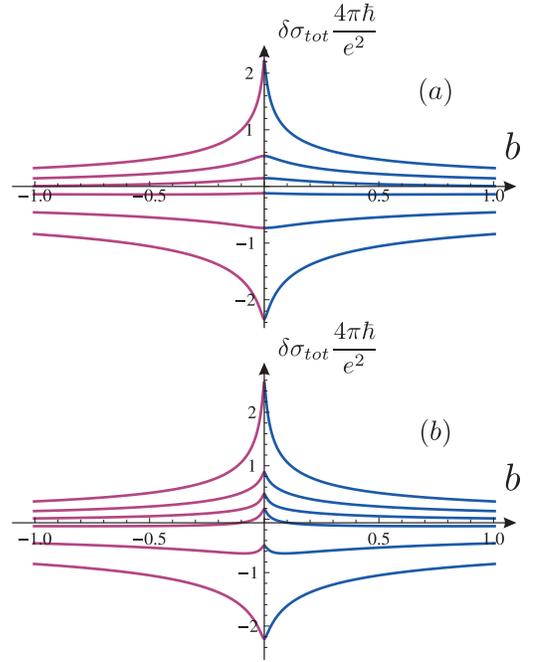} }
\caption{
 Conductivity correction  as a function of $b$  at  $\Gamma=0.01$  and  different $\eta$ ($\eta=0.1,~ 0.3,~ 0.5,~ 0.6,~ 0.7,~ 0.9$)
in the absence of the block mixing, $\Delta=0,$ (a) and for weak  mixing,      $\Delta=0.1,$  (b). }
\label{F17}
\end{figure}

\begin{figure} [H] \center
\centerline{\includegraphics[width=0.38\textwidth]{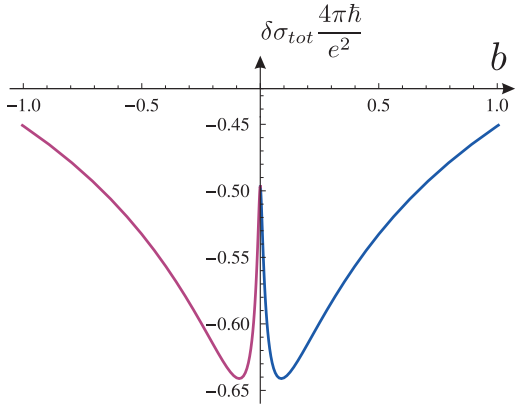} }
\caption{
 Conductivity correction as a function of $b$ at $\Gamma=0.001, \eta=0.3, \Delta=0.1$.
}
\label{F18}
\end{figure}

  In this section we presents   the plots  for magnetoconductivity in the presence of the  block mixing.
  As we mentioned in  Sec.~\ref{mix} for weak coupling between blocks ($\Delta \ll 1$) one can use the diffusion approximation
for calculation of the contribution
  of the ISM. This   contribution should be added to  the term $\delta \sigma_{{\rm I}}+\delta \sigma_{{\rm II}}, $
where $\delta \sigma_{{\rm I}}$ and $\delta \sigma_{{\rm II}} $
  can be calculated by using exact ballistic equations with $\Delta=0.$ Such an approach can be applied
  within the whole interval $ 0< \eta <1$ except vicinities of the points $\eta=0$ and $1.$
  At these special points, one can  also obtain analytical results by using the
   diffusion approximation for all terms contributing to the conductivity correction $\delta \sigma_{tot}$.
Using the
approach described above  we  plotted   in Fig.~\ref{F17} total correction  $\delta \sigma_{tot}$
for different $\eta$  both for the absence of the block mixing [$\Delta=0,$ see Fig.~\ref{F17}(a)]
and for weak inter-block coupling [$\Delta=0.1,$ see Fig.~\ref{F17}(b)].
Comparing these plots we see that the main effect of the mixing is the appearance of the positive peak
in the region of low $b$ for $ \eta \gtrsim 0.1.$ This peak is most pronounced for $\eta =0.3.$ as shown
in Fig.~\ref{F18}.  Physically,  low-field negative magnetoconductivity arises due to the contribution of ISM.
The peak  at $b=0$ is well described by Eq.~\eqref{psipsi} [see  also Eq.~\eqref{low} and discussion after this equation].

\subsection{Strong-field asymptotic of the conductivity.
}\label{large}
 Finally, we present the results for the  asymptotical behavior of
the conductivity correction at sufficiently strong $B,$ such that the magnetic
length becomes much smaller than the mean free path ($|b| \gg 1 $). In this case, the main contribution
to the quantum correction comes from the short electron trajectories involving scattering on untypical impurity
configurations, namely on the  complexes of three impurities separated by untypical  distance $l_B\ll l.$  \cite{nonback,TI72}
\begin{figure} [H] \center
\centerline{\includegraphics[width=0.38\textwidth]{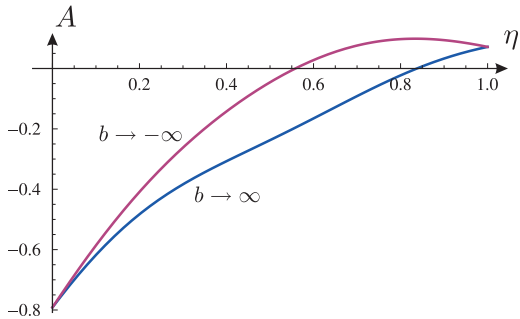} }
\caption{
 Coefficient $A$ as a function of $\eta$ for  $b\to \infty$ (lower curve) and for $b\to -\infty$ (upper curve).
}
\label{F19}
\end{figure}

For simplicity, we neglect here block mixing and consider $\delta \sigma_{{\rm II}}.$
As shown in the Appendix \ref{strongA}  in the  strong-field  limit conductivity correction can be presented as
\be
\delta \sigma_{{\rm II}}=\frac{e^2}{2\pi\hbar}\frac{l_B}{l} A_{\rm II}(\eta) \propto \frac{1}{\sqrt b}, \label{sigmastrong}
\ee
so that conductivity decays as a square root of the field: $\delta \sigma_{{\rm II}} \propto 1/\sqrt |b|.$
The coefficient $A_{\rm II}(\eta)$ is different for the positive and negative $b.$ Its analytical dependence is found in  Appendix \ref{strongA}.
The plots of $A_{\rm II}(\eta)$ for positive and negative fields are shown in the Fig.~\ref{F19}. Due to the  property $\delta \sigma_{{\rm I}}(b)=\delta \sigma_{{\rm II}}(-b),$    Fig.~\ref{F19} presents at the same time dependence of  $A_{\rm I}(\eta)$ for negative and positive fields, respectively.   The sum $A_{\rm I}(\eta)+A_{\rm II}(\eta)$ which determines asymptotical behavior of the total correction, turns to zero at $\eta \approx 0.715.$

\section{Summary}

We  have developed  a microscopic theory of the quantum transport in spin-orbit metals realized in HgTe quantum wells away from
the topological insulator phase.
Our theory is applicable to a wide range of particle concentrations and describes the crossover between WL and WAL regimes.
We demonstrated that this crossover is governed by the single parameter  $\eta$ ($0\leq \eta \leq 1$). All essential information about details of the
spectrum and eigenfunctions at the Fermi energy is encoded in this parameter. Hence, our results are applicable not only to the
HgTe quantum wells, but also to other systems governed by the generic Hamiltonian Eq.~(\ref{1}), in particular, to surfaces of 3D topological
insulators,~\cite{Glazman} to massive Dirac fermions in graphene on the BN substrate,~\cite{GeimBN} and to
semiconductors with strong Rashba splitting of the spectrum. \cite{TI67}

We have found analytically an exact expression for the Cooperon propagator in magnetic field valid beyond the diffusion approximation.
Using this equation, we have calculated the   interference-induced magnetoresistance in a wide interval of
  magnetic fields.  We found that the contributions of the two massive Dirac cones do not coincide,
  $\delta \sigma_{\text I}(b)\neq \delta \sigma_{\text I \text I}(b),$ and both are asymmetric functions of the magnetic field:
  $\delta \sigma_{\text I}(b)\neq \delta \sigma_{\text I}(-b)$
and $\delta \sigma_{\text I \text I}(b)\neq \delta \sigma_{\text I \text I}(-b).$
Only the total conductivity correction $\delta \sigma_{\text I}+ \delta \sigma_{\text I \text I}$ is an even function of
the magnetic field.
Special attention was given to the low- and strong-field limits.
In particular, we have found that each Dirac cone taken separately gives a linear contribution
  to the low-field magnetoresistance, whereas the total correction is parabolic in the limit $B\to 0$.
    In the opposite limit of large $B,$ the magnetoresistance decays as $1/\sqrt{B}$
    with the prefactor being a function of the electron concentration.
  We have also demonstrated that the block mixing gives rise to additional singular diffusive modes
  which do not show up in the absence of the mixing. One of these modes
  remains singular at $B=0$ and $T=0$ for arbitrary electron concentration and yields the WAL correction.
  This implies that any small but finite coupling between blocks turns the system into WAL regime at sufficiently low temperatures.
  We have shown that the quantum correction might change in non-monotonous way both with the phase breaking rate
  and with the magnetic field  and that the  block mixing might lead to an additional
  mechanism of the non-monotonous magnetoresistance.

  Finally, we note that the crossover from WL to WAL with increasing carrier concentration in HgTe-based  quantum wells was already observed in a number of
experiments. \cite{minkov, minkov13, kvon, bruene-unpub}
Detailed analysis of experimental data  is out of scope of this paper. It is worth, however, stressing that on the qualitative level our
theoretical predictions  for  magnetoconducitivity  (see Fig.~\ref{F17}) are in a good agreement with experimental results (see Fig.~3 of
Ref.~\onlinecite{minkov}).

\section{Acknowledgments}

We are grateful to N. Averkiev, C. Br\"une, A. Germanenko, L. Golub, E. Hankiewicz, E. K\"onig, D. Kvon, G. Minkov, A. Mirlin, M. Titov,
and G. Tkachov for useful discussions.
The work was supported by DFG within DFG SPP ``Semiconductor
spintronics'' and  DFG SPP  ``Topological insulators'', by DFG CFN,
by grant FP7-PEOPLE-2013 of the EU IRSES network InterNoM, and by BMBF.

\appendix

\section{Solution of the kinetic equation for the Cooperon in the absence of the block mixing  }
\label{mix1}
\subsection{Zero magnetic field}  \label{kinBeq0}

Here we find  analytically rigorous solution of Eq.~\eqref{Cooperon1} valid beyond the diffusion approximation.
As seen from this equation, the incoming term  of the collision integral contains only three angular
harmonics: $0,-1,-2.$  This allows us to present the solution of Eq.~\eqref{Cooperon1} in the following form:
\bee \label{Cooperon0-1-2}
 &&
C_{\mathbf{Q}}(\phi,\phi_0)
\\
&& =  \frac{C_0+e^{i(\phi_\mathbf{Q}-\phi)}C_{-1}+e^{2i(\phi_\mathbf{Q}-\phi)}C_{-2}+\delta(\phi-\phi_0)}{1 +  \Gamma+ i \mathbf Q \mathbf n}
,
\nonumber
\eee
where the coefficients
\bee
C_{0}&=& \frac{1}{1+\eta^2}\int \frac{d\phi}{2\pi} C_{\mathbf{Q}}(\phi,\phi_0) ,\label{C0}\\
C_{-1}&=& \frac{2\eta}{1+\eta^2}\int \frac{d\phi}{2\pi} C_{\mathbf{Q}}(\phi,\phi_0)e^{i(\phi-\phi_\mathbf{Q})},\label{C-1}\\
C_{-2}&=& \frac{\eta^2}{1+\eta^2}\int \frac{d\phi}{2\pi} C_{\mathbf{Q}}(\phi,\phi_0)e^{2i(\phi-\phi_\mathbf{Q})},\label{C-2}
\eee
do not depend on $\phi$ being the functions of $\phi_0$    and $\phi_\mathbf{Q}$  only.  Here  $\phi_\mathbf{Q}$ is the  angle of the vector $\mathbf{Q}.$ From Eqs.~\eqref{Cooperon0-1-2}, \eqref{C0}, \eqref{C-1}, and \eqref{C-2},
we find a system  of  equations  for  $C_0,C_{-1},$ and $C_{-2}:$
\be \label{MC}
\hat M  \left[\begin{array}{cc} C_0 \\C_{-1} \\ C_{-2}\end{array}\right]=\frac{1}{2\pi(1 +  \Gamma+ i \mathbf Q \mathbf n_0)}\left[\begin{array}{cc} 1
\\
e^{i(\phi_0-\phi_\mathbf{Q})} \\ e^{2i(\phi_0-\phi_\mathbf{Q})} \end{array}\right].
\ee
Here matrix $\hat M$ and its elements are given, respectively, by Eqs.~\eqref{M} and~\eqref{Pn} of the main text.
From Eqs.~\eqref{Cooperon0-1-2}, \eqref{MC}, and \eqref{M} we find
\bee \label{Cooperon2}
 &&C_{\mathbf{Q}}(\phi,\phi_0)
=  \frac{\delta(\phi-\phi_0)}{1 +  \Gamma+ i \mathbf Q \mathbf n} \\ \nonumber &&+\frac{1}{2\pi(1 +  \Gamma+ i \mathbf Q \mathbf n)
 (1 +  \Gamma+ i \mathbf Q \mathbf n_0)} \\  \nonumber && \times \left[\begin{array}{cc} 1 \\e^{i(\phi_\mathbf{Q}-\phi)}
\\ e^{2i(\phi_\mathbf{Q}-\phi)} \end{array}\right]^T \hat M^{-1} \left[\begin{array}{cc} 1 \\e^{i(\phi_0-\phi_\mathbf{Q})}
\\ e^{2i(\phi_0-\phi_\mathbf{Q})} \end{array}\right]
\eee
where $\mathbf n=(\cos \phi_0, \sin \phi_0 ).$

Different terms entering  the r.h.s. of Eq.~\eqref{Cooperon2} have  transparent physical sense.
The first term corresponds to
 ballistic propagation. The second one can be presented as a series   over functions $P_n$
(by expanding of
the matrix $\hat M^{-1}$) which, in fact, is an expansion  over number $N$
of collisions (the zero term in this expansion corresponds to
$N=1$).\cite{schmid} Having in mind to calculate interference-induced
magnetoresistance, we can exclude the term $N=1$ from the summation.\cite{comment}
Physically, this term
describes return to the initial point after a single scattering, so
that corresponding trajectory does not cover any area and,
consequently, is not affected by the magnetic field.  We neglect both
ballistic ($N=0$) and $N=1$ terms in the Cooperon propagator, and  find
\bee \label{Cooperon3}
&&C_{\mathbf{Q}}(\phi,\phi_0) =\frac{1}{2\pi(1
  + \Gamma+ i \mathbf Q \mathbf n) (1 + \Gamma+ i \mathbf Q \mathbf
  n_0)} \\ \nonumber && \times\left[\begin{array}{cc} 1
    \\e^{i(\phi_\mathbf{Q}-\phi)} \\
    e^{2i(\phi_\mathbf{Q}-\phi)} \end{array}\right]^T \left (\hat
  M^{-1}-\hat M^{-1}_{Q=\infty} \right)\left[\begin{array}{cc} 1
    \\e^{i(\phi_0-\phi_\mathbf{Q})} \\
    e^{2i(\phi_0-\phi_\mathbf{Q})} \end{array}\right].  \eee
Here we
took into account that $P_n \to 0$ for $Q\to\infty.$ Let us now find
the return probability. To this end, we make expansions \bee \nonumber
&& \frac{1}{1 + \Gamma+ i \mathbf Q \mathbf
  n}=\sum_{n=-\infty}^{\infty}P_n e^{in(\phi-\phi_\mathbf Q)}, \\ &&
\frac{1}{1 + \Gamma+ i \mathbf Q \mathbf
  n_0}=\sum_{m=-\infty}^{\infty}P_m e^{-im(\phi_0-\phi_\mathbf Q)}
\nonumber \eee in Eq.~\eqref{Cooperon3}, substitute the obtained
equation into Eq.~\eqref{Cooperon-coord}, take $\mathbf r=\mathbf
r_0,$ and average over $\phi_\mathbf Q,$ thus arriving to
Eqs.~\eqref{W} and \eqref{wn} of the main text.

\subsection{Nonzero magnetic field } \label{kinBneq0}

In this appendix, we find the exact solution of Eq.~\eqref{CooperonB} valid beyond the diffusion approximation.
Below, for the sake of brevity, we omit arguments $\mathbf r_0$ and $\phi_0$ in the Cooperon propagator.
First, we make a Fourier transform with respect to $x$ coordinate
\be \label{Fourier}
C(\mathbf r,\phi)=\int\frac{dk}{2\pi} e^{ikx}C(k,y,\phi)
\ee
and rewrite Eq.~\eqref{CooperonB} in a form similar to Eq.~\eqref{Cooperon1}
\bee \label{CooperonB1}
&& C(k,y,\phi)=
\hat R
\left[C_0(k,y)+e^{-i\phi} C_{-1}(k,y) \right.\\ && \left.+ e^{-2i\phi} C_{-2}(k,y)+e^{-ikx_0}\delta(y-y_0)\delta(\phi-\phi_0) \right],
\nonumber
\eee
where
\bee
C_{0}(k,y)&=& \frac{1}{1+\eta^2}\int \frac{d\phi}{2\pi} C(k,y,\phi), \label{CB0}\\
C_{-1}(k,y)&=& \frac{2\eta}{1+\eta^2}\int \frac{d\phi}{2\pi} C(k,y,\phi)e^{i\phi},\label{CB-1}\\
C_{-2}(k,y)&=& \frac{\eta^2}{1+\eta^2}\int \frac{d\phi}{2\pi} C(k,y,\phi)e^{2i\phi},\label{CB-2}
\eee
and
\bee
&& \hat R=\frac{1}{1+\Gamma +i \hat{ \mathbf q} \mathbf n l}
\\&&=\frac{1}{1+\Gamma +il[ \cos \phi(k + {y}/{l_B^2})+  \sin \phi (-i{\partial}/{\partial  y})]}.
\nonumber
\eee
Next, we introduce the canonically conjugated  variables

\be \xi_x = { l_B}\left(k + \frac{y}{l_B^2}\right),~~ \xi_y=-i{ l_B}\frac{\partial}{\partial  y}, ~~\left[\hat \xi_x, \hat \xi_y\right ]=i,\ee
and use the property
\be \label{com1}
\boldsymbol{\xi}\mathbf n=e^{i\phi a^\dagger a}\xi_x e^{-i\phi a^\dagger a},
\ee
where \be
a^\dagger =\frac{\xi_x-i\xi_y}{\sqrt{2}},~~a =\frac{\xi_x+i\xi_y}{\sqrt{2}}, ~[a,a^\dagger]=1.
\ee
Using this property we can transform the operator $\hat R$  entering the r.h.s. of Eq.~\eqref{CooperonB1} as follows:
\be
\hat R=
e^{i\phi a^\dagger a}
\frac{1}{1+\Gamma +i\xi_xl/l_B}
 e^{-i\phi a^\dagger a}\ee
that allows us to present the kernel of this operator in the $\xi_x$ representation in a simple form
\be
\langle \xi_x|\hat R|\xi_x'\rangle=\sum\limits_{n=0}^{n=\infty}\sum\limits_{m=0}^{m=\infty} e^{i\phi (n-m)} P_{nm}\Psi_n^*(\xi_x)\Psi_m(\xi_x'),
\ee
where
$\Psi_n(\xi)=\pi^{-1/4}(2^n n!)^{-1/2} \exp(-\xi^2/2)H_n(\xi)$ are the eigenfunction
of the harmonic oscillator with the Hamiltonian $a^\dagger a +1/2$ [here, $H_n(\xi)$ are Hermitian polynomials]  and
\be \label{p0nm}
P_{nm}=\int_{-\infty}^{\infty} d\xi \frac{\Psi_n^*(\xi)\Psi_m(\xi)}{1+\Gamma +i\xi l/l_B}.
\ee
By writing $[1+\Gamma +i\xi l/l_B]^{-1}=(l_B/l)\int_0^\infty dt \exp[-t(1+\Gamma)l_B/l-it\xi],$ after simple calculations we
find Eqs.~\eqref{Pnm} of the main text.

As a next step, we change in Eq.~\eqref{CooperonB1} variable $y$  to $\xi_x$ and expand
functions $C(k,y,\phi), C_0(k,y), C_{-1}(k,y,\phi), $ and $ C_{-2}(k,y,\phi)$ over full set of
functions $\Psi_n(\xi_x)=\Psi_n\left({ k l_B} + {y}/{l_B}\right)$
\bee
\label{expansion}
&&C(k,y,\phi)=\sum_{n=0}^{\infty} C^{(n)}(k,\phi)\Psi_n(\xi_x),
\\&&C_l(k,y)=\sum_{n=0}^{\infty}C^{(n)}_l(k)\Psi_n(\xi_x),~~l=0,-1,-2.
\nonumber
\eee
Doing so, we find
\bee  \nonumber
&& C^{(n)}(k,\phi)=\sum_{m=0}^{\infty}e^{i\phi(n-m)}P_{nm}\left[C_0^{(m)}(k)+e^{-i\phi}C_{-1}^{(m)}(k)\right.
\\
&& \left. +e^{-2i\phi}C_{-2}^{(m)}(k)+{l_B^{-1}}e^{-ik x_0}\delta(\phi-\phi_0)\Psi_m\left(\xi_x^0 \right)\right],
\label{CooperonB2}
\eee
where $\xi_x^0={ k l_B} + {y_0}/{l_B}.$  Now, we multiply Eq.~\eqref{CooperonB2}
consequently by $1,e^{i\phi},e^{2i\phi}$ and average over $\phi$ having in mind  Eqs.~\eqref{CB0}, \eqref{CB-1}, and \eqref{CB-2}.
Next, we make a replacement $n \to n-1$ and  $n \to n-2$  in the second and third
of the obtained equations, respectively. As a result, we obtain a system of closed equations
for $C_0^{(m)}, C_{-1}^{(m-1)},$ and $C_{-2}^{(m-2)},$ the solution of which can be written in a matrix form
\bee \label{MnB}
 &&\left[\begin{array}{cc} C_0^{(m)} \\C_{-1}^{(m-1)} \\ C_{-2}^{(m-2)}\end{array}\right]=e^{-ikx_0}
\\
&& \times \sum_{s=0}^{\infty}\frac{e^{i\phi_0(m-s)}}{{2}\pi \ l_B} \hat M_{m-1}^{-1}
\left[\begin{array}{cc} P_{m,s} \\P_{m-1,s} \\P_{m-2,s} \end{array}\right]\Psi_s\left(\xi_x^0 \right),
\nonumber
\eee
where matrix $\hat M_m$ is given by Eq.~\eqref{MB} of the main text
and we took into account that $P_{nm}=P_{mn}.$
To find the Cooperon propagator we need to substitute Eq.~\eqref{MnB} into \eqref{CooperonB2} and
then into \eqref{expansion}. Before doing so, we notice that one can extend summation
in Eq.~\eqref{CooperonB2} over negative $m$ because by definition  $P_{nm}=0$ for $m<0.$ Neglecting also ballistic
contribution described by the term with delta function, we can rewrite  Eq.~\eqref{CooperonB2} as
\bee \label{CooperonB3}
&&C^{(n)}(k,\phi)=\sum_{m=-\infty}^{\infty}e^{i\phi(n-m)}\left[ P_{nm} C_0^{(m)}(k) \right.
\\
&&\left.+ P_{n,m-1} C_{-1}^{(m-1)}(k) + P_{n,m-2}C_{-2}^{(m-2)}(k)\right].
\nonumber
\eee
Substituting now Eq.~\eqref{MnB} into~\eqref{CooperonB3} and using
Eqs.~ \eqref{Fourier}, \eqref{expansion}, we finally obtain the equation for the Cooperon propagator in the magnetic field
\bee \label{CooperonB4}
 &&C(\mathbf r, \mathbf r_0,\phi,\phi_0)=\sum_{m=-\infty}^{\infty}\sum_{s=-\infty}^{\infty}\sum_{n=-\infty}^{\infty}
\\
&& \nonumber \int \frac{dk}{2\pi} e^{ik(x-x_0)}\frac{e^{i\phi(n-m)}e^{-i\phi_0(s-m)}}{{2}\pi \ l_B}
\\
\nonumber
 &&\times\left[\begin{array}{cc} P_{nm}\\ P_{n,m-1}\\ P_{n,m-1}\end{array}\right]^T
\left(\hat M_{m-1}^{-1}-\hat M_{m=\infty}^{-1}\right)
\left[\begin{array}{cc} P_{m,s} \\P_{m-1,s} \\P_{m-2,s} \end{array}\right]
\\
&&\times  \Psi_s\left(\xi_x^0 \right)\Psi_n(\xi_x).
\eee
Similar to Eq.~\eqref{Cooperon3}, we excluded contribution coming from the processes with a single scattering act.
The expression for return probability turns out to be less complicated because for $\mathbf r=\mathbf r_0$
the integration over $k$ yields
\bee
&& \int dk \Psi_s\left(\xi_x \right)\Psi_n(\xi_x)
\\
&&=\int dk\Psi_s\left({ k l_B} + {y}/{l_B}\right)\Psi_n\left({ k l_B} + {y}/{l_B}\right)= \delta_{n,s} l_B^{-1},
\nonumber
\eee
so that we obtain Eq.~\eqref{W} where $w_n$  are now given by Eq.~\eqref{wnB} of the main text.

\section{Limiting  cases} \label{diffusion1}
In this appendix we derive Eqs.~\eqref{ds},\eqref{ds1},\eqref{w11} and \eqref{w111} directly from Eqs.~\eqref{wn} and \eqref{wnB}.

 \subsubsection{Limiting cases for $B=0$}
As we mentioned at the end of Sec.~\eqref{CooB=0} for $\eta=0$ and $1$ one of the modes
becomes singular.
Keeping the singular modes only, one can easily obtain  the return
probability and the  conductivity  in vicinities  of the points
$\eta=0$ and $1.$   For $\eta\to 0$
we find from Eq.~\eqref{wn}
\be W(\phi)\approx \frac{w_1}{2\pi l^2}, ~~  w_1\approx\int\frac{d^2\mathbf Q}{(2\pi)^2}\frac{P_0^3}{1+\eta^2-P_0}.
\ee
Substituting this equation to Eq.~\eqref{sigma2} we restore  with   logarithmic precision Eq.~\eqref{ds} of the main text.

For $\eta\to 1,$
we find from Eq.~\eqref{wn}
\bee \label{Wlim1}&& W(\phi)\approx \frac{w_0 e^{-i\phi}}{2\pi l^2}, \\
&& w_0\approx \int\frac{d^2\mathbf Q}{(2\pi)^2}\frac{P_0^3}{1+(1-\eta)^2/2\eta-P_0-2P_1^2},
\nonumber
\eee
Again, with logarithmic precision we restore Eq.~\eqref{ds1}.

\subsubsection{Limiting cases for $B \neq 0.$}

Let us calculate magnetoresistance    assuming that  $\Gamma \ll 1, l_B \gg l,$  and $\eta \ll 1$ or $1-\eta \ll 1,$ respectively.
We will start from exact equation \eqref{wnB} valid for $b>0$ [for $b<0$ calculations are analogous but one should start from Eq.~\eqref{wnB1}].
First, we notice that expression for $P_{nm}$ simplifies for $ l_B \gg l.$ Expanding  Eq.~\eqref{p0nm} in series over $ l/l_B$ and $\Gamma$ we easily find
\bee \label{pb0}
&& P_{mm}\approx 1-\Gamma -b(2m+1), \\&& P_{m+1,m}\approx -i\sqrt{b(m+1)}, ~~\text{for}~~m\geqslant0,
\nonumber
\eee
and $P_{mm}=P_{m+1,m}=0,$ for $m<0.$
Having in mind to invert matrix $M_m$ [see Eq.~\eqref{MB}] we kept linear in $b$ terms in $P_{mm}$ which enters  diagonal elements of $\hat M_m$
and  neglected terms  higher than $\sqrt{b}$ in the $P_{m+1,m}$ which enters  off-diagonal elements.  In this approximation, one can neglect $P_{m+2,m}$
 which are proportional to $b.$  The ballistic contribution  coming from $\hat M_{m=\infty}^{-1}$  can be also disregarded.

 Next, we find eigenvalues and eigenvectors of $\hat M_m$ that allows
 us diagonalize matrix $\hat M_m^{-1}.$ Just as in the case $B=0$ we
 only keep contributions of the singular modes.  One should keep terms
 of the linear order with respect to $b$ in the eigenvalues, while the
 eigenvectors can be taken at $b=\Gamma=0$ and $\eta=0$ (or
 $1-\eta=0$).  Results of calculations are presented below separately
 for $\eta\to 0$ and $1-\eta\to0$.
\vspace{2mm}
\\
\noindent
\underline{(a)~~$\eta \to 0.$} \vspace{2mm}
\\ \noindent
For this case,  the singular contribution comes from $w_1$ that corresponds to zero moment: $M=0.$
Singular eigenvalue of the matrix $\hat M_m$ is given by
$ \lambda_m \approx M_m^{11} -{(M_m^{12})^2}/{M_m^{22}} -{(M_m^{13})^2}/{M_m^{33}}.$
The last two terms in this equation describe mixing of the regular modes with $M=-1$ and $-2,$ respectively,
to the singular mode. With the needed precision we get
\be \lambda_m \approx \Gamma+\eta^2+2b[m(1+\eta) +3/2+\eta] .\ee
In the diffusion approximation we only keep  the singular contribution to the  $\hat M_m^{-1}:$
\be
\hat M_m^{-1}\approx \frac{1}{\lambda_m} \left[\begin{array}{ccc}
                                           1 & 0 & 0 \\
                                           0 & 0 & 0 \\
                                           0 & 0 & 0
                                         \end{array}\right].
\ee

 From Eq.~\eqref{sigma2}  we find that for $\eta \to 0$ the conductivity correction reads as
$\delta \sigma = -(e^2/\pi \hbar)[w_1+w_0/2+w_2/2].$ The main contribution comes from $w_1$ which is given by
$w_1\approx(b/\pi) \sum \limits_{m=-\infty}^{\infty}P^2_{m+1,m+1}/\lambda_m \approx(b/\pi)
\sum \limits_{n=0}^{\infty}1/\lambda_{n-1}=(b/\pi)
\sum \limits_{n=0}^{\infty}1/\{ \Gamma+\eta^2+2b[n(1+\eta) +1/2]\} $ (here we took into account
that $P_{m+1,m+1}=0$ for $m<-1$ and put $P_{m+1,m+1}\approx 1$ for $m \geq 1 $).
Multiplying  both numerator and denominator of the latter equation by $(1+\eta)^{-1}\approx 1-\eta$,
and neglecting terms on the order of $\eta^3,$ $b \eta^2 $ and $\Gamma \eta$ in the denominator and
terms on the order of $b\eta$ in the numerator we obtain Eq.~\eqref{w11} of the main text
\vspace{2mm}
\\ \noindent
\underline{(b)~~$\eta \to 1.$}\vspace{2mm}
\\ \noindent
For this case,  the singular contribution comes from $w_0$ that corresponds to $M=-1.$
Singular eigenvalue of the matrix $\hat M_m$ is given by
$ \lambda_m \approx M_m^{22}-{(M_m^{21})^2}/{M_m^{11}} -{(M_m^{23})^2}/{M_m^{33}} .$
The last two terms in this equation describe mixing of the regular modes with $M=0$ and $-2,$ respectively, to the singular mode.
With the needed precision we get
\be \lambda_m \approx {\Gamma+(1-\eta)^2/2+4b[m +1/2+(1-\eta)/2]}.\ee
In the diffusion approximation we only keep  the singular contribution to the  $\hat M_m^{-1}$:
\be
\hat M_m^{-1}\approx \frac{1}{\lambda_m} \left[\begin{array}{ccc}
                                           0 & 0 & 0 \\
                                           0 & -1 & 0 \\
                                           0 & 0 & 0
                                         \end{array}\right].
\ee
From these equations we find
 \bee \nonumber
&& w_0=\frac{b}{\pi}\hspace{-1mm}\sum\limits_{m=-\infty}^{\infty} \frac{P^2_{m,m}}{\Gamma+(1-\eta)^2/2+4b[m +1/2+(1-\eta)/2]}
\\
&& \approx \frac{b}{\pi}\hspace{-1mm}\sum\limits_{m=0}^{N}\frac{1}{ \Gamma+(1-\eta)^2/2+4b[m +1/2+(1-\eta)/2]}.
\label{hz}
\eee
One can see that for $b>0$ Eq.~\eqref{hz} coincides with Eq.~\eqref{w111} of the main text.

\section{Kinetic equation for the Cooperon in the presence of the block mixing  } \label{kinBblock}
\subsection{Derivation of the kinetic equation} \label{block1}
In the  basis \eqref{new} matrix elements  of the random potential are given by
\bee
\langle {\rm 1}_\mathbf k| \hat V | {\rm 1}_{\mathbf k'}\rangle&=&
V_{\mathbf k \mathbf k'} \frac{1 +e ^{i(\phi'-\phi)}}{1+\eta}\left(\frac{1+\eta}{2}+\Delta \sqrt \eta \right), \nonumber \\
\langle {\rm 2}_\mathbf k| \hat V | {\rm 2}_{\mathbf k'}\rangle&=&
V_{\mathbf k \mathbf k'} \frac{1 +e ^{i(\phi'-\phi)}}{1+\eta}\left(\frac{1+\eta}{2}-\Delta \sqrt \eta \right), \nonumber\\
\langle {\rm 1}_\mathbf k| \hat V | {\rm 2}_{\mathbf k'}\rangle&=&
\langle {\rm 2}_\mathbf k| \hat V | {\rm 1}_{\mathbf k'}\rangle \nonumber \\
&=&
V_{\mathbf k \mathbf k'} \frac{1- e ^{i(\phi'-\phi)}}{1+\eta}\left (\frac{1-\eta}{2} \right) .
\label{mat1}
\eee
The  single-particle  Green's functions are diagonal in this representation and the diagonal elements $G^1$ and $G^2$ read as
\bee
&& G_{R,A}^1(E,\mathbf{k})=\frac{1}{E-E_\mathbf{k}\pm i\gamma_1/2},  \nonumber
\\
&&G_{R,A}^2(E,\mathbf{k})=\frac{1}{E-E_\mathbf{k}\pm i\gamma_2/2},
\label{G12}
\eee
where
\bee
\nonumber
&&\gamma_1=\frac{2\pi}{\hbar}\sum \limits _{\alpha=1,2} \int
|\langle {\rm 1}_\mathbf k| \hat V | {\rm \alpha}_{\mathbf k'}\rangle|^2
\delta(E_\mathbf k-E_{\mathbf k'})\frac{d^2 \mathbf k'}{(2\pi)^2}
\\ \label{gamma1} &&=\gamma_0
 \frac{1+ \eta^2  + 2\eta \Delta^2 + 2\Delta \sqrt \eta (1+\eta)}{(1+\eta)^2},
\eee
\bee
\nonumber
&&\gamma_2=\frac{2\pi}{\hbar} \sum \limits _{\alpha=1,2} \int
|\langle {\rm 2}_\mathbf k| \hat V | {\rm \alpha}_{\mathbf k'}\rangle|^2
\delta(E_\mathbf k-E_{\mathbf k'})\frac{d^2 \mathbf k'}{(2\pi)^2}\\&&=\gamma_0
 \frac{1+ \eta^2  + 2\eta \Delta^2 - 2\Delta \sqrt \eta (1+\eta)}{(1+\eta)^2}.
\label{gamma2}
\eee

\begin{figure} [h] \center
\centerline{\includegraphics[width=0.3\textwidth]{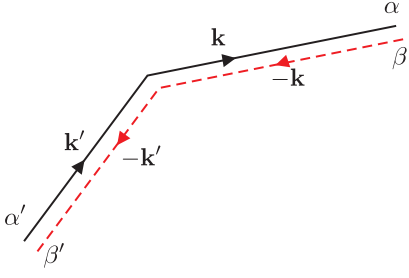} }
\caption{Process corresponding to ingoing term in the collision integral.
}
\label{F20}
\end{figure}

The  Cooperon propagator obeys now the equation which is similar to Eq.~\eqref{Cooperon} but has the matrix form
\bee \label{Cooperonabcd}
&& \left[1/\tau_\phi + i \mathbf q \mathbf v_F  ) \right]C_{\mathbf{q}}^{\alpha\beta,\alpha_0\beta_0}(\phi,\phi_0)
\\ \nonumber &&
= \int \frac{d\phi'}{2\pi} \gamma_C^{\alpha\beta,\alpha'\beta'} (\phi-\phi')
C_{\mathbf{q}}^{\alpha'\beta',\alpha_0\beta_0}(\phi',\phi_0)
\\
&&-\frac{\gamma_\alpha+\gamma_\beta}{2} C_{\mathbf{q}}^{\alpha\beta,\alpha_0\beta_0} (\phi,\phi_0)
+\gamma \delta_{\alpha \alpha_0} \delta_{\beta \beta_0}\delta(\phi-\phi_0).
\nonumber
\eee
Here, we took into account in the outgoing term that the elements of the Cooperon ladder with $\alpha \neq \beta$
decays with the averaged rate $(\gamma_1+\gamma_2)/2.$
The ingoing scattering term describes the process shown in Fig.~\ref{F20}.
The scattering rate  $\gamma_C^{\alpha\beta,\alpha'\beta'} (\phi-\phi')$
is given by  Eq.~\eqref{gammaD} with the replacement $\langle |\tilde{V}_{\mathbf{k}\mathbf{k}'}|^2 \rangle$
with $\langle {\alpha}_\mathbf k| \hat V | { \alpha'}_{\mathbf k'}\rangle\langle {\beta}_{-\mathbf k}| \hat V | { \beta'}_{-\mathbf k'}\rangle$
where matrix elements are given by Eq.~\eqref{mat1}.
The expression for conductivity is given by  Eq.~\eqref{sigma} with the replacement of  $W(\phi)\gamma_C(\pi -\phi)$ with
$\sum\limits_{\alpha\beta\alpha'\beta'}W^{\alpha\beta,\alpha'\beta'}(\phi)\gamma_C^{\alpha' \beta', \beta\alpha}(\pi -\phi),$ where
$W^{\alpha\beta,\alpha'\beta'}(\phi)$ is found from Eqs.~\eqref{Cooperon-coord} and \eqref{W-def} with the replacement of $C$
with $C^{\alpha\beta,\alpha'\beta'}$(we note that indices $\alpha$ and $\beta$
enter in different order in $W^{\alpha\beta,\alpha'\beta'}$ and   $\gamma_C^{\alpha' \beta', \beta\alpha}$).

Equation~\eqref{Cooperonabcd}
can be rewritten in a more compact way by expanding both Cooperon and scattering rate matrices
over the Pauli matrices $\hat \sigma_{(n)},$ ($n=0,1,2,3$ and $\hat\sigma_0$ is the unit $2\times 2$ matrix):
 \be \label{gamma_abcd}
 \gamma_C^{\alpha\beta,\alpha'\beta'} (\phi-\phi')= \sigma_{(n)}^{\alpha\beta}  \gamma_C^{nm} (\phi-\phi') \sigma_{(m)}^{\beta'\alpha'}/2
\ee
\be
C_{\mathbf{q}}^{\alpha\beta,\alpha'\beta'}(\phi,\phi_0)=\sigma_{(n)}^{\alpha\beta}  C_{\mathbf{q}}^{nm}(\phi,\phi_0)  \sigma_{(m)}^{\beta'\alpha'}/2
\ee
Here $C_{\mathbf{q}}^{nm},\gamma_C^{nm} $ are elements of $4\times 4$ matrices.  After simple calculations we arrive to Eqs.~\eqref{Cooperonnm} and \eqref{gammac} of the main text where
\begin{widetext}
\be
\label{gamma-0}
\hat \gamma_{0}=\frac{\gamma_0}{(1+\eta)^2}  \left[\begin{array}{cccc} (1+\eta^2 +2\eta \Delta^2)/2 & (1-\eta^2)/2 & 0& \Delta \sqrt \eta (1+\eta)\\
(1-\eta^2)/2 & (1+\eta^2 -2\eta \Delta^2)/2& 0 & \Delta \sqrt \eta (1-\eta)\\
0 & 0& \eta(1-\Delta^2) &0\\ \Delta \sqrt \eta (1+\eta) & \Delta \sqrt \eta (1-\eta)&0& \eta(1+\Delta^2)
\end{array}\right].
\ee
\be
\label{gamma-1}
\hat \gamma_{-1}=\frac{\gamma_0}{(1+\eta)^2}\left[\begin{array}{cccc} 2\eta(1+\Delta^2) & 0 & 0& 2\Delta \sqrt \eta (1+\eta)\\
0 & 2\eta(1-\Delta^2)& 0 & 0\\
0 & 0& 1+\eta^2 -2\eta \Delta^2 &0\\ 2\Delta \sqrt \eta (1+\eta) & 0&0&1+\eta^2 +2\eta \Delta^2
\end{array}\right].
\ee
\be
\label{gamma-2}
\hat \gamma_{-2}=\frac{\gamma_0}{(1+\eta)^2}\left[\begin{array}{cccc} (1+\eta^2 +2\eta \Delta^2)/2 & -(1-\eta^2)/2 & 0& \Delta \sqrt \eta (1+\eta)\\
-(1-\eta^2)/2 & (1+\eta^2 -2\eta \Delta^2)/2& 0 & -\Delta \sqrt \eta (1-\eta)\\
0 & 0& \eta(1-\Delta^2) &0\\ \Delta \sqrt \eta (1+\eta) & -\Delta \sqrt \eta (1-\eta)&0& \eta(1+\Delta^2)
\end{array}\right].
\ee
\bee
\label{gammaDmatrix}
\hat \gamma_D&=&\left[\begin{array}{cccc} (\gamma_1+\gamma_2)/2 & 0 & 0& (\gamma_1-\gamma_2)/2\\
0 & (\gamma_1+\gamma_2)/2& 0 & 0\\
0 & 0& (\gamma_1+\gamma_2)/2 &0\\ (\gamma_1-\gamma_2)/2 &0&0&(\gamma_1+\gamma_2)/2
\end{array}\right] \\
&=&\frac{\gamma_0}{(1+\eta)^2}\left[\begin{array}{cccc} 1+\eta^2 +2\eta \Delta^2 & 0 & 0& 2\Delta \sqrt \eta (1+\eta)\\
0 & 1+\eta^2 +2\eta \Delta^2& 0 & 0\\
0 & 0& 1+\eta^2 +2\eta \Delta^2 &0\\ \Delta \sqrt \eta (1+\eta) & 0&0& 1+\eta^2 +2\eta \Delta^2
\end{array}\right]
\eee
\end{widetext}

\subsection{Matrices entering expression for the conductivity corrections } \label{block2}

From Eqs.~\eqref{gamma-0}, \eqref{gamma-2}, \eqref{gamma-1}, and \eqref{xi}, we find matrices entering Eq.~\eqref{sigmanm}.
For our purposes, it is sufficient to know these matrices for $\Delta=0$:
\begin{widetext}
\be \label{A1}
\frac{\hat \gamma_{-2}\hat \xi}{2}= \frac{\gamma_0}{(1+\eta)^2}\left[\begin{array}{cccc} \displaystyle
\frac{1+\eta^2}{4} &\displaystyle -\frac{1-\eta^2}{4} & 0& 0\\
\displaystyle -\frac{1-\eta^2}{4} & \displaystyle \frac{1+\eta^2}{4}& 0 & 0 \\
0 & 0& \displaystyle -\frac{\eta}{2} &0\\
0 & 0 &0& \displaystyle \frac{\eta}{2}
\end{array}\right].
\ee

\be \label{A2}
\frac{\hat \gamma_{0}\hat \xi}{2}= \frac{\gamma_0}{(1+\eta)^2}\left[\begin{array}{cccc} \displaystyle
\frac{1+\eta^2}{4} &\displaystyle \frac{1-\eta^2}{4} & 0& 0\\
\displaystyle \frac{1-\eta^2}{4} & \displaystyle \frac{1+\eta^2}{4}& 0 & 0 \\
0 & 0& \displaystyle -\frac{\eta}{2} &0\\
0 & 0 &0& \displaystyle \frac{\eta}{2}
\end{array}\right].
\ee

\be \label{A3}
 \left(\hat \gamma_{-2}-\frac{\hat \gamma_{-1}}{2}\right)\hat \xi= \frac{\gamma_0}{(1+\eta)^2}\left[\begin{array}{cccc}
\displaystyle \frac{(1-\eta)^2}{2}     &\displaystyle  -\frac{1-\eta^2}{2}     & 0        & 0\\

\displaystyle -\frac{1-\eta^2}{2}    & \displaystyle \frac{(1-\eta)^2}{2}   &0         & 0 \\

0     & 0     & \displaystyle \frac{(1-\eta)^2}{2}       & 0\\

0     & 0     & 0        & \displaystyle -\frac{(1-\eta)^2}{2}

\end{array}\right].
\ee

\be \label{A4}
 \left(\hat \gamma_{0}-\frac{\hat \gamma_{-1}}{2}\right)\hat \xi= \frac{\gamma_0}{(1+\eta)^2}\left[\begin{array}{cccc}
\displaystyle \frac{(1-\eta)^2}{2}    &\displaystyle  \frac{1-\eta^2}{2}     & 0        & 0\\

\displaystyle \frac{1-\eta^2}{2}    & \displaystyle \frac{(1-\eta)^2}{2}   &0         & 0 \\

0     & 0     & \displaystyle \frac{(1-\eta)^2}{2}       & 0\\

0     & 0     & 0        & \displaystyle -\frac{(1-\eta)^2}{2}

\end{array}\right].
\ee

\be \label{A5}
 \left(\hat \gamma_{-1}-\frac{\hat \gamma_{0}+\hat \gamma_{-2}}{2}\right)\hat \xi= \frac{\gamma_0}{(1+\eta)^2}\left[\begin{array}{cccc}
\displaystyle 2\eta -\frac{1+\eta^2}{2}    &0     & 0        & 0\\

0    & \displaystyle 2\eta -\frac{1+\eta^2}{2}   &0         & 0 \\

0     & 0     & \displaystyle \eta-1-\eta^2       & 0\\

0     & 0     & 0        & \displaystyle 1+\eta^2-\eta

\end{array}\right].
\ee

\end{widetext}

\subsection{Matrices determining gaps of the diffusive modes } \label{block3}
From Eqs.~\eqref{gamma-0}, \eqref{gamma-2}, \eqref{gamma-1}, and \eqref{gammaDmatrix}
we found matrices whose eigenvalues yield gaps of the diffusive Cooperon modes:
\begin{widetext}
\be \label{gammaD0}
\hat \gamma_{D}-\hat\gamma_0= \frac{\gamma_0}{(1+\eta)^2}\left[\begin{array}{cccc}
\displaystyle \frac{1+\eta^2+2\eta\Delta^2}{2}    & \displaystyle -\frac{1-\eta^2}{2}     & 0        &\displaystyle  \Delta \sqrt \eta (1+\eta)\\

\displaystyle -\frac{1-\eta^2}{2}      & \displaystyle \frac{1+\eta^2+6\eta\Delta^2}{2}      &0         & -\displaystyle  \Delta \sqrt \eta (1-\eta) \\

0     & 0     & \displaystyle 1+\eta^2-\eta +3\eta \Delta^2       & 0\\

\displaystyle  \Delta \sqrt \eta (1+\eta)    & -\displaystyle  \Delta \sqrt \eta (1-\eta)     & 0        & \displaystyle 1+\eta^2-\eta +\eta \Delta^2

\end{array}\right].
\ee

\be \label{gammaD-2}
\hat \gamma_{D}-\hat\gamma_{-2}= \frac{\gamma_0}{(1+\eta)^2}\left[\begin{array}{cccc}
\displaystyle \frac{1+\eta^2+2\eta\Delta^2}{2}    & \displaystyle \frac{1-\eta^2}{2}     & 0        &\displaystyle  \Delta \sqrt \eta (1+\eta)\\

\displaystyle \frac{1-\eta^2}{2}      & \displaystyle \frac{1+\eta^2+6\eta\Delta^2}{2}      &0         & \displaystyle  \Delta \sqrt \eta (1-\eta) \\

0     & 0     & \displaystyle 1+\eta^2-\eta +3\eta \Delta^2       & 0\\

\displaystyle  \Delta \sqrt \eta (1+\eta)    & \displaystyle  \Delta \sqrt \eta (1-\eta)     & 0        & \displaystyle 1+\eta^2-\eta +\eta \Delta^2

\end{array}\right].
\ee

\be \label{gammaD-1}
\hat \gamma_{D}-\hat\gamma_{-1}= \frac{\gamma_0}{(1+\eta)^2}\left[\begin{array}{cccc}
\displaystyle (1-\eta)^2    & 0    & 0        & 0\\

0      & (1-\eta)^2  +4\eta\Delta^2      &0         & 0 \\

0     & 0     & 4\eta \Delta^2       & 0\\

  0    & 0     & 0        & 0

\end{array}\right].
\ee
\end{widetext}
We see that matrix $\hat \gamma_{D}-\hat\gamma_{-1}$ has two eigenvalues which turns to zero at $\Delta=0.$
As shown in the main  text, two corresponding diffusive modes cancel each other in the limit    $\Delta \to 0.$

\subsection{Generalization of the diffusion approximation for the case of block mixing} \label{diffusion2}

In this appendix, we generalize the diffusion approximation
for the case of the block mixing.
This  approximation is valid for $ql \ll 1,$ when the
coupling of different harmonics  is weak and the mode $e^{i M \phi}$  is only effectively
coupled with the nearest modes $e^{i (M\pm 1) \phi}.$  In  a full analogy with the Sec.~\ref{diffusion} we
find from Eqs.~\eqref{Cooperonnm}, \eqref{gammac}, \eqref{Wphi11}, and \eqref{Wnm}
\be
\hat w_{M+1}=\int \frac{\gamma}{\gamma_\varphi +\hat \gamma_D-\hat \gamma_M+q^2 \hat D_M}~\frac{l^2d^2\mathbf q }{(2\pi)^2}.
\label{WM}
\ee
Here, operator $q^2 \hat D_M$ in the denominator of the integrand appeared due to the coupling with $M\pm1$ modes:
\bee
q^2\hat D_M&=&v_F^2\left \langle  \mathbf q\mathbf n \frac{1}{\gamma_\varphi-\hat \gamma_D -
{\hat\gamma}_C } \mathbf q\mathbf n\right\rangle_M
\nonumber
\\
&=& \frac{q^2v_F^2}{4}(\hat \tau_{M+1}+\hat \tau_{M-1}),
 \label{Dq2}
\eee
where $\hat\gamma_C$ is the operator with the matrix kernel given by Eq.~\eqref{gammac},
$\langle \cdots\rangle_M$ stands for projection on the mode $\exp(iM\phi),$
\be
\hat \tau_M=\frac{1}{\gamma_\varphi+\hat \gamma_D-\hat \gamma_M} \approx \frac{1}{\hat \gamma_D-\hat \gamma_M},
\label{tauM}
\ee
and $\hat \gamma_M$ is given by Eqs.~\eqref{gamma-0}, \eqref{gamma-1}, \eqref{gamma-2}
for $M=0,-1,-2,$ respectively ($\gamma_M=0$ for other values of $M$).
Hence,
\be
\hat D_M= \frac{v_F^2(\hat \tau_{M+1}+\hat \tau_{M-1})}{4}.
\ee
For convenience, the matrices $\hat w_n$ entering Eq.~\eqref{sigmanm} are presented below:

\be \label{w-2}
\hat w_{-2}=\int \frac{\gamma}{\gamma_\varphi +\hat \gamma_D+q^2 \hat D_{-3}}~\frac{l^2d^2\mathbf q }{(2\pi)^2},
\ee
\be\label{w-1}
\hat w_{-1}=\int \frac{\gamma}{\gamma_\varphi +\hat \gamma_D-\hat \gamma_{-2}+q^2 \hat D_{-2}}~\frac{l^2d^2\mathbf q }{(2\pi)^2},
\ee
\be\label{w-0}
\hat w_{0}=\int \frac{\gamma}{\gamma_\varphi +\hat \gamma_D-\hat \gamma_{-1}+q^2 \hat D_{-1}}~\frac{l^2d^2\mathbf q }{(2\pi)^2},
\ee
\be\label{w1}
\hat w_{1}=\int \frac{\gamma}{\gamma_\varphi +\hat \gamma_D-\hat \gamma_0+q^2 \hat D_0}~\frac{l^2d^2\mathbf q }{(2\pi)^2},
\ee
\be \label{w2}
\hat w_{2}=\int \frac{\gamma}{\gamma_\varphi +\hat \gamma_D+q^2 \hat D_1}~\frac{l^2d^2\mathbf q }{(2\pi)^2},
\ee
where
\be \label{D-3}
\hat D_{-3}=\frac{v_F^2 (\tau_{-2}+\tau_{-4})}{4}= \frac{v_F^2 }{4} \left( \frac{1}{ \hat \gamma_D-\hat \gamma_{-2}} + \frac{1}{ \hat \gamma_D}\right),
\ee

\be  \label{D-2}
\hat D_{-2}=\frac{v_F^2 (\tau_{-1}+\tau_{-3})}{4}= \frac{v_F^2 }{4} \left( \frac{1}{\hat \gamma_D-\hat \gamma_{-1}} + \frac{1}{\hat \gamma_D}\right),
\ee

\be  \label{D-1}
\hat D_{-1}=\frac{v_F^2 (\tau_{0}+\tau_{-2})}{4}= \frac{v_F^2 }{4} \left( \frac{1}{\hat \gamma_D-\hat \gamma_{0}} + \frac{1}{\hat \gamma_D -\hat \gamma_{-2}}\right),
\ee
\be  \label{D0}
\hat D_{0}=\frac{v_F^2 (\tau_{1}+\tau_{-1})}{4}= \frac{v_F^2 }{4} \left( \frac{1}{\hat \gamma_D} + \frac{1}{\hat \gamma_D -\hat \gamma_{-1}}\right),
\ee
\be  \label{D1}
\hat D_{1}=\frac{v_F^2 (\tau_{2}+\tau_{0})}{4}= \frac{v_F^2 }{4} \left( \frac{1}{\hat \gamma_D} + \frac{1}{\hat \gamma_D-\hat \gamma_0}\right).
\ee

The  results obtained  above in this subsection can be easily generalized for the case of
weak magnetic field such that dimensionless field $b$ is much smaller that unity.
To this end one should slightly modify Eq.~\eqref{Dq2} taking into account that
for $b\neq 0$ operators $\hat q_x$ and $\hat q_y$ are no longer commute. Simple calculation yields
\bee \label{Dq2b}
&& v_F^2\left(  \mathbf q\mathbf n \frac{1}{\gamma_\varphi-\hat \gamma_D -  {\hat\gamma}_C } \mathbf q\mathbf n\right)_M
\\
\nonumber
&& = \frac{(\hat q_x^2 + \hat q_y^2)v_F^2}{4}(\hat \tau_{M+1}+\hat \tau_{M-1}) +\frac{i[\hat q_x,\hat q_y]v_F^2}{4} (\hat \tau_{M-1} -\hat \tau_{M+1})
\eee
Replacing operator  $\hat q_x^2 + \hat q_y^2$ by its  eigenvalue $(4/l^2)|b| (n+1/2) $ we find
that one should make the following  replacement   in Eq.~\eqref{WM}
  \be  q^2\hat D_M  \to   \gamma^2 \left [|b|\left(n+\frac12\right)(\hat \tau_{M+1}+\hat \tau_{M-1}) - b \frac{\hat \tau_{M-1} -\hat \tau_{M+1}}{2}\right]\ee
 and also replace
 integral over ${l^2 d^2 \mathbf q}/{(2\pi)^2} $    with the $({|b|}/{\pi}) \sum\limits _{n=0}^{N\sim 1/|b|}.$
While substituting  $\hat \tau_{M\pm 1}$ into this equation one can set $\Delta=0.$ Corresponding sums can be calculated with the use of Eq.~\eqref{sum}.

\section{Calculation of  conductivity correction  in the presence of the block mixing  }\label{IM}

In this Appendix, we present calculation of conductivity correction in the presence of the block mixing.
\subsection{$B=0$}
   Let us start with calculation of the contributions of ISM. These modes  correspond to  two eigenvalues (elements
${33}$ and $44$ of  the diagonal matrix $\hat\gamma_D-\hat \gamma_{-1}$), one of which equals to zero at
any $\Delta$ and another one turns to zero at $\Delta \to 0.$ These modes give  contribution to conductivity
$\delta \sigma_{\rm{ISM}},$   which comes from $w_0:$
\bee
\nonumber
\delta \sigma_{ism} &\approx& \frac{e^2}{2\pi\hbar} \left( \frac{l_{tr}}{l}\right)^2 \frac{1}{\gamma} ~{\rm Tr}
\left[\left(\hat \gamma_{-1}-\frac{\hat \gamma_{0}+\hat \gamma_{-2}}{2}\right)\hat \xi \hat w_0 \right] \\ \nonumber
&=&\frac{e^2}{2\pi\hbar}  \int \frac{l_{tr}^2d^2\mathbf q }{(2\pi)^2}~{\rm Tr}
\left[\left(\hat \gamma_{-1}-\frac{\hat \gamma_{0}+\hat \gamma_{-2}}{2}\right) \right.
\\
& \times& \left.\hat \xi \frac{1}{\gamma_\varphi +\hat \gamma_D-\hat \gamma_{-1}+q^2 \hat D_{-1}}~\right],
\label{sigma(1)}
\eee
 Further calculations will be performed within the diffusion approximation, which was generalized for the case of
the block mixing in Appendix \ref{diffusion2}. First, we project matrices
$\hat\gamma_D-\hat \gamma_{-1},$ $[\hat \gamma_{-1}-(\hat \gamma_{0}+\hat \gamma_{-2})/{2}]\hat \xi$  and $\hat D_{-1}$ on the space formed
by eigenvectors corresponding to ISM (this means that we take in these $4 \times 4$ matrices their bottom-right $2 \times 2$ blocks).
From Eqs.~\eqref{A5}, \eqref{D-1}, \eqref{gammaD0} and \eqref{gammaD-2} we find projected
matrices (in all these matrices except $\hat\gamma_D-\hat \gamma_{-1}$ one can approximately set $\Delta=0$):
\be
\hat\gamma_D-\hat \gamma_{-1} \to\frac{\gamma_0}{(1+\eta)^2 } \left[\begin{array}{cc} 4\eta\Delta^2&0
\\
0&0 \end{array}\right].
\ee
\be
\left(\hat \gamma_{-1}-\frac{\hat \gamma_{0}+\hat \gamma_{-2}}{2}\right)\hat \xi \to \gamma_0 \frac{1+\eta^2-\eta}{(1+\eta)^2}
\left[\begin{array}{cc}  -1 & 0 \\
 0 &1
\end{array}\right].
\ee
\be
\hat D_{-1} \to \frac{v_F^2}{2\gamma_0}\frac{(1+\eta)^2}{1+\eta^2-\eta}\left[\begin{array}{cc} 1&0
\\
0&1 \end{array}\right].
\ee
Substituting these projected matrices into Eq.~\eqref{sigma(1)}, neglecting $\Delta$
everywhere except gap of one of the singular modes, after simple calculations we find for contribution of ISM Eq.~\eqref{sigma(1)1} of the main text.

Next we discuss special  points $\eta=0$ and $1.$
We start from the case $\eta \approx 1.$ Above we projected
all matrices entering Eq.~\eqref{sigma(1)} on the basis formed by eigenvectors corresponding to
smallest eigenvalues of the matrix  $\hat\gamma_D-\hat \gamma_{-1}.$
Such an approximation works well only far from the point $\eta =1.$
Indeed, as seen from Eq.~\eqref{gammaD-1} for $\eta $ close to $1$ two other eigenvalues of the
matrix  $\hat\gamma_D-\hat \gamma_{-1}$ (matrix  elements $11$ and $22$ of this diagonal matrix)
become small and contribution of two corresponding modes come into play.
Let us now project the matrices
$\hat\gamma_D-\hat \gamma_{-1},$ $[\hat \gamma_{-1}-(\hat \gamma_{0}+\hat \gamma_{-2})/{2}]\hat \xi$  and $\hat D_{-1}$
on the space formed
by these eigenvectors. From Eqs.~\eqref{A5}, \eqref{D-1}, \eqref{gammaD0} and \eqref{gammaD-2} we find
\bee\nonumber
&& \hat\gamma_D-\hat \gamma_{-1} \to\frac{\gamma_0}{(1+\eta)^2 } \left[\begin{array}{cc} (1-\eta)^2&0
\\
0&(1-\eta)^2+4\Delta^2\eta \end{array}\right]\\
&& \approx \frac{\gamma_0}{4} \left[\begin{array}{cc} (1-\eta)^2&0
\\
0&(1-\eta)^2+4\Delta^2\end{array}\right]
\eee
\bee\nonumber
&& \left(\hat \gamma_{-1}-\frac{\hat \gamma_{0}+\hat \gamma_{-2}}{2}\right)\hat \xi \to
\frac{\gamma_0}{(1+\eta)^2}\left (2\eta  -\frac{1+\eta^2}{2} \right)
\left[\begin{array}{cc}  1 & 0 \\
 0 &1
\end{array}\right] \\&&\approx \frac{\gamma_0}{4}
\left[\begin{array}{cc}  1 & 0 \\
 0 &1
\end{array}\right]
\eee
\be
\hat D_{-1} \to \frac{v_F^2(1+\eta)^2(1+\eta^2)}{4\gamma_0\eta^2}\left[\begin{array}{cc} 1&0
\\
0&1 \end{array}\right]\approx \frac{2v_F^2}{\gamma_0}\left[\begin{array}{cc} 1&0
\\
0&1 \end{array}\right].
\ee
Substituting these projected matrices into Eq.~\eqref{sigma(1)}, neglecting $\Delta$ everywhere except gap of one of the diffusive
modes, after simple calculations we find contribution $\delta \sigma_{1-\eta}$ given by Eq.~\eqref{sigma(2)} of the main text.

Now, we turn to the case $\eta \to 0.$
Analyzing Eqs.~\eqref{w-2}-\eqref{w2},   one can see that for $\eta \ll 1$
contributions to conductivity  coming from matrices $ \hat w_{-1} $ and  $ \hat w_{1} $ should be taken into account because
matrices $\hat\gamma_D-\hat \gamma_{0}$ and  $\hat\gamma_D-\hat \gamma_{-2}$ have zero eigenvalues in the
limit $\eta \to 0$ and $\Delta \to 0.$ Let us consider the corresponding contribution to the conductivity:
\bee \label{sigma31}
&&\delta \sigma_{\eta} \approx \frac{e^2}{2\pi\hbar} \left( \frac{l_{tr}}{l}\right)^2 \frac{1}{\gamma}
\\
\nonumber&&
\times {\rm Tr} \left[\left(\hat \gamma_{-2}-\frac{\hat \gamma_{-1}}{2}\right)\hat \xi \hat w_{-1}+
\left(\hat \gamma_{0}-\frac{\hat \gamma_{-1}}{2}\right)\hat \xi \hat w_{1}\right]
\\
\nonumber
&&=\frac{e^2}{2\pi\hbar}  \int\frac{l_{tr}^2 d^2 \mathbf q}{(2\pi)^2}
\\
&& \nonumber\times{\rm Tr} \left[\left(\hat \gamma_{-2}-\frac{\hat \gamma_{-1}}{2}\right)\hat \xi
\frac{1}{\gamma_\varphi +\hat \gamma_D-\hat \gamma_{-2}+q^2 \hat D_{-2}} \right.
\\
&& \nonumber \left.+ \left(\hat \gamma_{0}-\frac{\hat \gamma_{-1}}{2}\right)\hat \xi
\frac{1}{\gamma_\varphi +\hat \gamma_D-\hat \gamma_0+q^2 \hat D_0}\right].
\eee
The  two terms in the square brackets represent contributions of two blocks, respectively.

 From Eqs.~\eqref{gammaD0} and \eqref{gammaD-2}, we see that after neglecting terms proportional to
$\Delta$ in the off-diagonal elements of matrices $\hat\gamma_D-\hat \gamma_{0}$ and  $\hat\gamma_D-\hat \gamma_{-2}$
these matrices  becomes block matrices consisting of   left-upper  and right-bottom $2\times 2$ blocks.
For both matrices, small eigenvalues (for small $\eta$ and $\Delta$) correspond to the left-upper block.
Projecting all matrices entering Eq.~\eqref{sigma31} on this block,
we then transform  projected matrices by unitary transformation
 \be \hat U=\frac{1}{\sqrt 2}\left[\begin{array} {cc}
 1&-1\\
 1&1
 \end{array} \right].\ee
  After this transformation off-diagonal elements of all projected matrices become small and can be neglected.
One can also neglect $\eta$ and $\Delta$ in diagonal elements which remain finite at $\eta \to 0$ and $\Delta \to 0$.
Then, we obtain
\be
\hat U(\hat \gamma_D- \hat \gamma_0)\hat U^{-1} \to \gamma_0 \left[\begin{array}{cc} 1&0\\
0&\eta^2+2\eta \Delta^2\end{array}\right],
\ee
\be
\hat U(\hat \gamma_D- \hat \gamma_{-2})\hat U^{-1} \to \gamma_0 \left[\begin{array}{cc} \eta^2+2\eta \Delta^2&0\\
0&1\end{array}\right],
\ee
\be \hat U \left(\hat \gamma_{-2}-\frac{\hat \gamma_{-1}}{2}\right)\hat \xi \hat U^{-1} \to
\gamma_0 \left[\begin{array}{cc} 1&0\\
0&0\end{array}\right],
\ee
\be
\hat U \left(\hat \gamma_{0}-\frac{\hat \gamma_{-1}}{2}\right)\hat \xi \hat U^{-1} \to
\gamma_0 \left[\begin{array}{cc} 0&0\\
0&1\end{array}\right],
\ee
\be
\hat D_{-2} =\hat D_0 \to \frac{v_F^2}{2\gamma_0} \left[\begin{array}{cc}1&0 \\ 0&1 \end{array}  \right].
\ee
After some simple algebra  we arrive to Eq.~\eqref{sigmaeta} of the main text.

\subsection{ $B \neq 0$.}\label{secb}
The  results obtained in Appendix \ref{diffusion2} can be also used for calculation  of the conductivity correction in weak magnetic field ($b \ll 1$).

  First we calculate   $\delta \sigma_{\rm{ISM}}(b)-\delta \sigma_{\rm{ISM}}(0). $ In this case,
$M=-1$  and projection of the matrix $\hat \tau_{-2} -\hat \tau_{0}$ at the block corresponding to ISM
goes to zero for $\Delta \to 0.$ Hence, the only effect of the magnetic field is the replacement everywhere of the integration over $d^2\mathbf q$   with the summation over $n.$
Simple calculations yield Eq.~\eqref{sigma111} of the main text.

Calculation of   $\Delta \sigma_{1-\eta}= \delta \sigma_{1-\eta}(b)-\delta \sigma_{1-\eta}(0)$ is more tricky.
  In this case, projection  of $\hat \tau_{-2} -\hat \tau_{0}$ to the upper-left block is approximately  given by
$$\hat \tau_{-2} -\hat \tau_{0} \to -\frac{8(1-\eta)}{\gamma_0}\left[ \begin{array}{cc}0&1\\ 1&0 \end{array}\right]$$
and does not commute with the  projection of the matrix
$\hat \gamma _D - \hat \gamma_{-1} $ on the upper-left block which is approximately given by
$$\hat \gamma _D - \hat \gamma_{-1} \to \frac{\gamma_0}{4}\left[ \begin{array}{cc}(1-\eta)^2&0\\ 0& (1-\eta)^2+4\eta\Delta^2\end{array}\right],$$
 (this matrix determines the  diffusive gaps in this channel). Calculations can be performed by using simple generalization of
Eq.~\eqref{sum} to the matrix case:
\bee \label{sum1}
&&{\rm Tr}\left[\sum\limits_{n=0}^{n=N} \frac{b}{ b(n+1/2)+\hat A}\right] \\&&
\nonumber
=2\ln N- \psi(A_1/b+1/2) - \psi(A_2/b+1/2)\\ &&\approx   \begin{cases}
\displaystyle  2 \ln\left( \frac{1}{b}\right),&\qquad  b \gg A_1,A_2,\\[0.5pt]
\displaystyle \ln \left( \frac{1}{{\rm det}\hat A}\right)-\frac{b^2}{24} \left(\frac{1}{A_1^2}+\frac{1}{A_2^2}\right) ,&\qquad  b\ll A_1,A_2
 \end{cases}
\eee
where $A_{1,2}$ are eigenvalues of matrix $\hat A.$
Using Eq.~\eqref{sum1} after some algebra  we find Eq.~\eqref{1-eta} of the main text.

Finally, we calculate $\Delta \sigma_\eta=\delta \sigma_{\eta}(b)-\delta \sigma_{\eta}(0). $ In this case,
additional matrices arising in the denominators of Eqs.~\eqref{w-1} and \eqref{w1} due to non-commutativity
of $\hat q_x$ and $\hat q_y$ are proportional to unit matrix, so that calculations are quite analogous to the
calculations of $\Delta \sigma_{\rm{ISM}}. $ The result is given by Eq.~\eqref{eta} of the main text.

\section {Strong-field asymptotic}
 \label{strongA}
For strong $B,$  such that  $l_B/l \ll 1,$ the expression for the conductivity correction simplifies.
First of all,  from Eq.~\eqref{p0nm} we find  in this limit
\bee  
&& P_{nm}=\frac{l_B}{l}\int_{-\infty}^{\infty} d\xi \frac{\Psi_n^*(\xi)\Psi_m(\xi)}{(1+\Gamma)(l_B/l) +i\xi }\\&& \nonumber \approx
-\frac{il_B}{l}\int_{-\infty}^{\infty} d\xi \frac{\Psi_n^*(\xi)\Psi_m(\xi)}{\xi-i0 }.
\label{p0nm1}
\eee
Hence, $P_{nm}$ decreases as a square root of the field, $P_{nm} \propto 1/\sqrt{b},$ and becomes small, $P_{nm} \ll 1,$ at sufficiently large $b.$
Therefore, one can expand  matrices $\hat M_m^{-1}-\hat M_{m=\infty}^{-1}$ entering
Eqs.~\eqref{wnB} and \eqref{wnB1}  in the series over $P_{nm}$  and keep the terms
of the lowest order only. Doing so, we find
\bee
&&w_n(b\to\infty)=\frac{l^2}{2\pi l_B^2} \frac{1}{(1+\eta^2)^2} \nonumber\\
&& \nonumber\times \sum \limits_{m=-\infty}^{m=\infty} \left[P_{n+m,m+1}^2 P_{m+1,m+1} \right.
\\&&+ \nonumber 4\eta P_{n+m,m+1}P_{m+1,m}P_{n+m,m}\\
&&+ \nonumber 2\eta^2(P_{n+m,m-1}P_{m+1,m-1}P_{n+m,m+1}+2P_{n+m,m}^2 P_{m,m})\\
&&+ \nonumber 4\eta^3P_{n+m,m}P_{m,m-1}P_{n+m,m-1}\\
&&+\eta^4P_{n+m,m-1}^2 P_{m-1,m-1} \left. \right], \label{w_large+}\\
&&w_n(b\to-\infty)=\frac{l^2}{2\pi l_B^2} \frac{1}{(1+\eta^2)^2}\nonumber \\
&&
\times \sum \limits_{m=-\infty}^{m=\infty} \left[P_{m-n,m-1}^2 P_{m-1,m-1} \right.  \nonumber\\
&&+ \nonumber 4\eta P_{m-n,m-1}P_{m-1,m}P_{m-n,m}\\
&&+ \nonumber 2\eta^2(P_{m-n,m+1}P_{m+1,m-1}P_{m-n,m-1}+2P_{m-n,m}^2 P_{m,m})\\
&&+ \nonumber 4\eta^3P_{m-n,m}P_{m,m+1}P_{m-n,m+1}\\
&&+\eta^4P_{m-n,m+1}^2 P_{m+1,m+1} \left. \right] \label{w_large-}.
\eee
For calculation of the conductivity, we need to know $w_n$ for $n=-2,-1,0,1,2.$
Therefore, for our purposes it is sufficient to find $P_{n,m}$ with $|n-m|\leqslant 3.$
From Eq.~\ref{p0nm1} one can find
\bee
P_{2k,2k}&=&\frac{l_B}{l}\frac{\sqrt \pi}{2^{2k}}\frac{(2k)!}{(k!)^2}\theta(k), \nonumber\\
P_{2k+1,2k}&=&P_{2k,2k+1}=-\frac{il_B}{l}\frac{\sqrt 2}{\sqrt{2k+1}}\theta(k),  \nonumber \\
P_{2k+2,2k}&=&P_{2k,2k+2}=-\frac{l_B}{l}\frac{\sqrt \pi}{2^{2k+1}}\frac{\sqrt{(2k)!(2k+2)!}}{k!(k+1)!}\theta(k),\nonumber\\
P_{2k+3,2k}&=&P_{2k,2k+3} \nonumber \\&=&\frac{il_B}{l}\frac{\sqrt 8 (k+1)}{\sqrt{(2k+1)(2k+2)(2k+3)}}\theta(k),
\label{P_large}
\eee
where $\theta(k)$ is the step function defined such that
$\theta(k)=1$ for $k\geq 0$ and $\theta(k)=0$ for negative $k$.

Using Eqs.~\eqref{w_large+}, \eqref{w_large-} and \eqref{P_large} after cumbersome
but straightforward calculations we find  Eq.~\eqref{sigmastrong} of the main text,
where the coefficient $A_{\rm{II}}(\eta)$ is different for the positive and negative $b$:
\bee
&&A_{\rm{II}, b\to \infty}(\eta)=\frac{\sqrt \pi}{4}\frac{1}{(1+\eta^2)(1-\eta+\eta^2)^2}\nonumber \\
&& \times \left[  1 +4\eta-7\eta^2-2\eta^3-\frac{25\eta^4}{4}+4\eta^5+\eta^6 \right.\nonumber \\
&& - \frac{\pi(2-3\eta+9\eta^2-20\eta^3+9\eta^4-3\eta^5+2\eta^6)}{\Gamma^4(3/4)} \nonumber \\
&& + \left.\frac{4\eta(-1+\eta-\eta^3+\eta^4)\Gamma^4(3/4)}{\pi^3} \right] ,
\eee
\bee
&&A_{\rm{II},b\to- \infty}(\eta)=\frac{\sqrt \pi}{4}\frac{1}{(1+\eta^2)(1-\eta+\eta^2)^2}\nonumber \\
&& \times \left[  1 +4\eta-\frac{25\eta^2}{4} -2\eta^3 -7\eta^4+4\eta^5+\eta^6 \right.\nonumber \\
&& - \frac{\pi(2-3\eta+9\eta^2-20\eta^3+9\eta^4-3\eta^5+2\eta^6)}{\Gamma^4(3/4)} \nonumber \\
&& -\left.\frac{4\eta(-1+\eta-\eta^3+\eta^4)\Gamma^4(3/4)}{\pi^3} \right].
\eee
Here $\Gamma(x)$ is the gamma function.

The strong-field asymptotics of the magnetoconductivity calculated in this appendix
reflects the statistics of areas and Berry phases for minimal (triangular) trajectories
contributing to the conductivity correction. The analytical expression for $\eta=0,$
\begin{equation}
A_{\rm{II},b\to\infty}(0)=A_{\rm{II},b\to- \infty}(0)=\frac{\sqrt{\pi}}{4}\left[1-\frac{2\pi}{\Gamma^4(3/4)}\right],
\end{equation}
reproduces the numerical prefactor  found for the weak localization
in a parabolic band in the absence of the Berry phase
in Refs. \onlinecite{TI72} and \onlinecite{nonback}:
$$ A_{\rm{I},b\to\infty}(0)+A_{\rm{II},b\to\infty}(0)=2A_{\rm{II},b\to\infty}(0) \simeq -\frac{4.974}{\pi}.$$


\begin{thebibliography}{120}
\bibitem{comment1}Here we implicitly assumed that  for a given Fermi energy   there is only single value of the Fermi momentum and, consequently, the single value of $\eta$ (this is the case for HgTe/CdTe wells not too far from the transition point,  where the thickness of the quantum well is close to  $d_c$).
\bibitem{comment}
In fact, this question is more subtle and the term $N=1$ requires special attention.
It results in an ultraviolet logarithmic divergency (coming from  large momenta $q > k_F \gg 1/l$) of
the return probability $W(\phi,\phi_0,0).$ This divergency leads to  ultraviolet
renormalization \cite{aleiner-efetov}  of the key problem parameters (such  as $W_0$)
and will not be discussed here. By default, we  assume that all parameters of the problem
are already renormalized and  that $q\ll k_F.$
\end{thebibliography}
\end{document}